\documentclass[referee,onecolumn]{aa}
\usepackage{multirow}
\usepackage{upgreek}
\usepackage{wasysym}
\usepackage{txfonts}
\usepackage{gensymb}
\usepackage{morefloats}
\usepackage{rotating}
\usepackage{afterpage}
\usepackage{upgreek}
\usepackage{hyperref}
\hypersetup{colorlinks=true, linkcolor=blue, filecolor=blue, urlcolor=blue, citecolor=blue}

\begin{document}

\title{Peering Into the Extended X-ray Emission on Megaparsec Scale in 3C 187}

    \author{A. Paggi\inst{1}\fnmsep\inst{2}\fnmsep\inst{3}
	\and
	F. Massaro\inst{1}\fnmsep\inst{2}\fnmsep\inst{3}\fnmsep\inst{4}
	\and
	H. A. Pe\~na-Herazo\inst{1}\fnmsep\inst{2}\fnmsep\inst{3}\fnmsep\inst{5}
	\and
	V. Missaglia\inst{1}\fnmsep\inst{2}\fnmsep\inst{3}
	\and
	F. Ricci\inst{6}\fnmsep\inst{7}
	\and
	C. Stuardi\inst{6}\fnmsep\inst{7}
	\and
	R. P. Kraft\inst{8}
	\and
	G. R. Tremblay\inst{8}
	\and
	S. A. Baum\inst{9}
	\and
	B. J. Wilkes\inst{8}
}
    \institute{
	Universit\`{a} degli Studi di Torino, via Pietro Giuria 1, I-10125 Torino, Italy\\
	\email{alessandro.paggi@unito.it}
	\and
	INFN – Istituto Nazionale di Fisica Nucleare, Sezione di Torino, via Pietro Giuria 1, I-10125 Turin, Italy
	\and
	INAF-Osservatorio Astrofisico di Torino, via Osservatorio 20, I-10025 Pino Torinese, Italy
	\and
	Consorzio Interuniversitario per la Fisica Spaziale (CIFS), via Pietro Giuria 1, I-10125, Torino, Italy
	\and
	Instituto Nacional de Astrof\'isica, \'Optica y Electr\'onica, Apartado Postal 51-216, 72000 Puebla, M\'exico
	\and
	Dipartimento di Fisica e Astronomia, Universit\`a di Bologna, via P. Gobetti 93/2, I-40129, Bologna, Italy
	\and
	Istituto di Radioastronomia, INAF, via Gobetti 101, I-40129, Bologna, Italy
	\and
	Center for Astrophysics \(|\)  Harvard \& Smithsonian, 60 Garden St. Cambridge MA 02138, USA 
	\and
	University of Manitoba, Dept of Physics and Astronomy, Winnipeg, MB R3T 2N2, Canada
}
\date{Received ...; accepted ...}

\abstract
{The diffuse X-ray emission surrounding radio galaxies is generally interpreted either as due to inverse Compton scattering of non-thermal radio-emitting electrons on the Cosmic Microwave Background (IC/CMB), or as the thermal emission arising from the hot gas of the intergalactic medium (IGM) permeating galaxy clusters hosting such galaxies, or as a combination of both. In this work we present an imaging and spectral analysis of \textit{Chandra} observations for the radio galaxy 3C 187 to investigate its diffuse X-ray emission and constrain the contribution of these different physical mechanisms.}
{The main goals of this work are: (i) to evaluate the extension of the diffuse X-ray emission from this source, (ii) to investigate the two main processes that can account for its origin - IC/CMB and thermal emission from the IGM - and (iii) to test the possibility for 3C 187 to belong to a cluster of galaxies, that can account for the observed diffuse X-ray emission.}
{To evaluate the extension of the X-ray emission around 3C 187 we extracted surface flux profiles along and across the radio axis. We also extracted X-ray spectra in the region of the radio lobes and in the cross-cone region to estimate the contribution of the non-thermal (IC/CMB) and thermal (IGM) processes to the observed emission, making use of radio (VLA and GMRT) data to investigate the multi-wavelength emission arising from the lobes. We collected Pan-STARRS photometric data to investigate the presence of a galaxy cluster hosting 3C 187, looking for the presence of a ``red sequence'' in the source field in the form of a tight clustering of the galaxies in the color space. In addition, we made use of observations performed with COSMOS spectrograph at Victor Blanco telescope to estimate the redshift of the sources in the field of 3C 187, in order to verify if they are gravitationally bound, as one would expect in a cluster of galaxies.}
{The diffuse X-ray emission around 3C 187 is found to extend in the soft \(0.3-3\text{ keV}\) band up to \(\sim 850 \text{ kpc}\) along the radio lobe direction and \(\sim 530\text{ kpc}\) in the cross-cone direction, and it appears enhanced in correspondence of the radio lobes. Spectral X-ray analysis in the cross-cones indicates a thermal origin for the emission in this region, with a temperature \(\sim 4 \text{ keV}\). In the radio lobes, the X-ray spectral analysis in combination with the radio data suggests a dominant IC/CMB radiation in these regions, however not ruling out a significant thermal contribution. Assuming that the radiation observed in the radio lobes is due the IGM, the emission from the N and S cones can in fact be interpreted as arising from hot gas with temperatures of \(\sim 3 \text{ keV}\) and \(\sim 5 \text{ keV}\), respectively, found to be in pressure equilibrium with the surrounding. Using Pan-STARRS optical data we found that 3C 187 belongs to a red-sequence of \(\sim 40\) optical sources in the field whose color distribution is significantly different from background sources. We were able to collect optical spectra only for 1 of these cluster candidates and for 22 field (i.e., non cluster candidates) sources. While the latter show stellar spectra, the former features a galactic spectrum with a redshift close to 3C 187 nucleus.}
{The diffuse X-ray emission around 3C 187 is elongated along the radio axis, and enhanced in correspondence of the radio lobes, indicating a morphological connection between the emission in the two energy bands, and thus suggesting a dominating IC/CMB mechanism in these regions. This scenario is reinforced by the multi-wavelength radio-X-ray emission, that in these regions is compatible with the IC/CMB radiation. The X-ray spectral analysis however does not rule out a significant contribution to the observed emission from thermal gas, that would be able to emit over tens of Gyr and in pressure equilibrium with the surroundings. In fact, optical data indicate that 3C 187 may belong to a cluster of galaxies, whose IGM would contribute to the X-ray emission observed around the source. Additional X-ray and optical spectroscopic observations are however to needed secure these results and get a more clear picture of the physical processes at play in 3C 187.}

\maketitle

\section{Introduction}\label{sec:intro}

In the last decades diffuse X-ray emission associated with radio sources, and extending beyond their host galaxies up to hundreds of kpc scale, has been recently revealed by \textit{Chandra} telescope at cosmological distances, i.e. up to redshift 3.8 \citep[4C 41.17,][]{2003ApJ...596..105S}. Few recent examples are 3C 294 \citep{2003MNRAS.341..729F}, 3C 191 \citep{2006MNRAS.371...29E} and 3CR 459 \citep{2018A&A...619A..75M}.
This diffuse X-ray emission is generally interpreted as due to inverse Compton (IC) scattering of non-thermal radio-emitting electrons on Cosmic Microwave Background (CMB) photons, permeating their radio lobes \citep[IC/CMB][]{1979MNRAS.188...25H, 2000ApJ...540L..69S, 2000ApJ...544L..23T}. This scenario is supported by the observational evidence that, even without a clear morphological match between the radio and X-ray emission, extended X-ray structures are generally aligned to the radio axis and/or spatially coincident with radio structures. The lack of a clear radio counterpart to X-ray structures in high redshift sources \citep{2015MNRAS.452.3457G} could be due to the increasing energy density of the cosmic microwave background with \(z\) \citep{2014MNRAS.438.2694G}, as \(U_{\text{CMB}} = 4.22 \times {10}^{-13} {\left({1 + z}\right)}^4 \text{ erg} \text{ cm}^{-3}\), corresponding to an equipartition magnetic field\footnote{We recall that, given a region containing magnetic field with an energy density \(U_B = \frac{B^2}{8\pi}\) and relativistic electrons with an energy density \(U_e\), the equipartition magnetic field is defined as that for which the total energy \(U_e+U_B\) reaches a minimum, that is, \(U_e\approx U_B\).} \(B_{\text{cmb}}=3.25\,{(1+z)}^2\, \upmu\text{G}\) \citep{1999A&A...345..769M, 2011ApJ...729L..12M}. If the average magnetic field in the lobe is less than \(B_{\text{cmb}}\), the relativistic electrons lying there will preferentially cool through IC/CMB rather than synchrotron radiation, resulting in a dimming of radio emission while boosting the X-ray one, a process known as CMB quenching \citep[see e.g.,][]{2004MNRAS.353..523C, 2017MNRAS.468..109W}.

However, IC/CMB is not the only possible interpretation for this extended X-ray emission, since high redshift radio galaxies are often used as tracers of galaxy clusters inhabiting galaxy poor or moderately rich environments \citep[see e.g,][]{2002NewAR..46..121W, 2003MNRAS.339.1163C, 2009A&ARv..17....1W}. Thus extended X-ray emission form these sources can be due, or at least partially contaminated, by the thermal emission arising from the hot gas of the intergalactic medium \citep[see e.g.,][]{2013ApJ...770..136I, 2015MNRAS.453.2682I}.

Disentangling between scenarios of non-thermal emission via IC/CMB and thermal emission from hot gas in the intergalactic medium (IGM) is however challenging. To shed a light on the nature of such diffuse X-ray emission two main tests can be performed.

The first would be to carry out radio observations at MHz frequencies, where it would be possible to observe the synchrotron radiation of the same particles in the radio lobes responsible for the IC/CMB X-ray emission and thus verify if the spectral shape of the two components is consistent. Assuming that the X-ray radiation at \(1\text{ keV}\) is due to IC/CMB and that the emission in radio lobes is unbeamed, the synchrotron radiation from the same electrons would be observed at radio frequencies
\begin{equation}\label{eq:synch}
\nu_{\text{syn}}=1.44\,(B/\upmu\text{G})\,{(1+z)}^{-1}\text{ MHz}\,,
\end{equation}
where \(z\) is the source redshift and \(B\) is the average magnetic field in the lobe \citep[e.g.,][]{1966ApJ...146..686F}, or equivalently \(\nu_{\text{syn}}=4.68\,b\,(1+z)\text{ MHz}\), where \(b\) is the ratio of \(B\) and \(B_{\text{cmb}}\). For magnetic fields of the order of tens of \(\upmu\text{G}\) \citep[e.g,][]{2003ApJ...596..105S, 2006MNRAS.371...29E}, \(\nu_{\text{syn}}\) would fall at MHz frequencies. In this case, we would expect the photon index measured in X-ray, \(\Gamma\), to be related to the spectral index measured at MHz frequencies \(\alpha\) by the relation \(\Gamma=1-\alpha\) \citep[e.g.,][]{2012ApJ...760..132S}. This test, however, cannot be directly carried out because current and future low radio frequency facilities like the Low-Frequency Array \citep[LOFAR,][]{2013A&A...556A...2V}, the Murchison Widefield Array \citep[MWA,][]{2013PASA...30...31B}, the Long Wavelength Array \citep[LWA,][]{2009IEEEP..97.1421E}, the Giant Metrewave Radio Telescope \citep[GMRT,][]{1991ASPC...19..376S}, the Karl G. Jansky Very Large Array \citep[JVLA,][]{2011ApJ...739L...1P} and the Square Kilometre Array \citep[SKA,][]{2009IEEEP..97.1482D} could only skim the desired radio frequency range. For example in a \(50 \text{ MHz}\) LOFAR observation of a \(z\sim 2\) source, to observe the synchrotron radiation from the same electrons responsible of the IC/CMB emission in X-rays would require a magnetic field in the radio lobes \(B>100\,\upmu\text{G}\), larger than those estimated as \(\sim 25 \,\upmu\text{G}\) by \citet{2006MNRAS.371...29E}, while for 4C 41.17 (\(z=3.8\)) we would need \(B>170\,\upmu\text{G}\), still larger than the maximum equipartition magnetic field estimated for this source \citep{1998ApJ...494L.143C, 2003ApJ...596..105S}.

The second is a X-ray spectral analysis to check if the spectrum at energies of a few keV shows a thermal continuum and possibly emission lines, or if instead it appears to have non-thermal origin as for example in the form of a steep power-law with photon index \(\sim 2\), as observed in lobes of radio galaxies \citep[see e. g.][]{2005ApJ...626..733C, 2006MNRAS.370.1893H, 2017MNRAS.470.2762M}. We note however that this test in general does not allow to discriminate between the different models only by comparing fit statistics, even in cases of sources with large number of counts \citep{2010MNRAS.404.2018H}.

In the last decade we carried out the Chandra snapshot survey of the Third Cambridge catalog \citep[3C,][]{1962MNRAS.125...75B} to guarantee the complete X-ray coverage of the entire 3CR catalog \citep{2010ApJ...714..589M, massaro12, massaro15}. We discovered X-ray emission associated with radio jets \citep[see e.g.,][]{2009ApJ...696..980M}, hotspots \citep[see e.g.,][]{2011ApJS..197...24M, 2012MNRAS.419.2338O} as well as diffuse X-ray emission from hot atmospheres and IGM in galaxy clusters \citep[see e.g.,][]{2010MNRAS.401.2697H, 2012MNRAS.424.1774H, 2016MNRAS.458..681D, 2018ApJ...867...35R}, extended X-ray emission aligned with the radio axis of several moderate and high redshift radio galaxies \citep[see e.g.,][]{2013ApJS..206....7M, 2018ApJS..234....7M, 2018ApJS..235...32S, 2020ApJS..250....7J}, and the presence of extended X-ray emission spatially associated with optical emission line regions not coincident with radio structures, as in 3CR 171 and 3CR 305 \citep{massaro09,2010MNRAS.401.2697H, 2012MNRAS.424.1774H, 2012A&A...545A.143B}. One of the best examples of the latter discovery is certainly 3C 187. This Fanaroff-Riley \citep[FR,][]{1974MNRAS.167P..31F} II radio galaxy is, among those pointed in our snapshot survey, the brightest one showing extended X-ray emission on hundreds of kpc. \citet{2013ApJS..206....7M} claimed that such diffuse X-ray emission in 3C 187 is consistent with the lobe radio structure. Here, on the basis of additional information collected from the literature, we present a thorough investigation of its X-ray emission and a comparison with the optical information.

Extending the analysis by \citet{2013ApJS..206....7M}, here we present surface flux profiles of the X-ray emission to evaluate its extension and estimate its properties. We perform X-ray spectral analysis coupled with a comparison with IC/CMB model to test non-thermal scenario. Moreover we investigate the presence of a galaxy cluster claimed in literature via a red sequence based procedure \citep{1977ApJ...216..214V, 1998ApJ...501..571G}. To this end we performed optical spectroscopic observation of sources in the field of 3C 187 with Victor Blanco telescope in Cerro Tololo, Chile. However, due to bad observing conditions we were able to collect spectra only for 23 bright sources.

The paper is organized as follows. A brief multi frequency description of 3C 187 is given in Sect. \ref{sec:source}. \textit{Chandra} data reduction and analysis are presented in Sect. \ref{sec:analysis}.  Results on possible X-ray emission mechanisms and the presence of a galaxy cluster are discussed in Sect. \ref{sec:discussion}, while Sect. \ref{sec:summary} is devoted to our conclusions. Unless otherwise stated we adopt cgs units for numerical results and we also assume a flat cosmology with \(H_0=69.6\text{ km}\text{ s}^{-1}\text{ Mpc}^{-1}\), \(\Omega_{M}=0.286\) and \(\Omega_{\Lambda}=0.714\) \citep{2014ApJ...794..135B}. Spectral indexes, \(\alpha\), are defined by flux density, \(F_{\nu}\propto\nu^{-\alpha}\) both in radio and in the X-rays. Pan-STARRS magnitudes are in the AB system \citep{2012ApJ...750...99T}.

\section{Source Description}\label{sec:source}

3C 187 is a radio galaxy with a flux density at \(178\text{ MHz}\) of \(8.1\text{ Jy}\) \citep{1980PASP...92..553S}, therefore sitting at the lower end of FR II observed radio powers \citep[e.g.,][]{1992MNRAS.256..186B}.

In the 3C catalog \citep{1980PASP...92..553S}, 3C 187 has been identified with an optical source at \(z=0.350\), based on the \(4000\text{ \AA}\) break and G band absorption. \citet{1988ApJS...66..361H}, using Canada France Hawaii Telescope (CFHT) data, measured for this optical source a magnitude \(m_R=20.7\), and report it to appear elliptical without any obvious structure. These authors also found that the source appears to be a member of a cluster of \(\sim 30\) similarly extended objects projected between the radio lobes, as also stated by \citet{1995ApJS...99..349N}.

VLA imaging \citep{1994A&AS..105...91B} revealed for 3C 187 a triple radio source, with a faint core and two lobes separated by \(\sim 2\farcm0\) along a position angle \(\text{P.A.} = -21{\degree}\)\footnote{P.A. are measured counterclockwise from the north direction.}. In particular \citet{1996ApJS..107..175R} report \(1.4\text{ GHz}\) flux densities of \(246\) and \(312\text{ mJy}\) for the north and south lobes, respectively. \citet{1995ApJS...99..349N} report \(1.4\text{ GHz}\) flux densities of \(4\), \(463\) and \(459\text{ mJy}\) for the core, north lobe and south lobe, respectively, and  \(5\text{ GHz}\) flux densities of \(3\), \(44\) and \(26\text{ mJy}\) for the core, north lobe and south lobe, respectively.

The optical identification of \citet{1980PASP...92..553S} has been therefore revised by \citet{1995A&A...299...17H}, which proposed the identification of 3C 187 core with an optical source  - coincident with the radio core observed in the radio maps presented by \citet{1994A&AS..105...91B} - with \(m_R = 20.0\) and a redshift of \(z=0.465\), based on [OII] and [OIII] emission lines, as well as \(4000\text{ \AA}\) break. In the following we will adopt this redshift for 3C 187\footnote{At a redshift of \(0.465\) an angular separation of \(1\arcsec\) corresponds to a projected physical scale of \(5.93\text{ kpc}\).}.

A \textit{Chandra} snapshot observation of \(\sim 12\text{ ks}\) has been performed in January	2012, and analyzed in \citet{2013ApJS..206....7M}. These authors report extended X-ray emission co-spatial with the radio structure, detected with high significance (\(>7 \sigma\)) in four regions coincident with and between the radio lobes (see their Fig. 2). In the following we will re-analyze these \textit{Chandra} data and their correlation with the radio emission.

\section{X-RAY DATA REDUCTION AND ANALYSIS}\label{sec:analysis}

\textit{Chandra} observation of 3C 187 was retrieved from \textit{Chandra} Data Archive through ChaSeR service\footnote{\href{http://cda.harvard.edu/chaser}{http://cda.harvard.edu/chaser}}. This consists of a single \(12\text{ ks}\) observation (OBSID 13875, PI: Massaro, GO 13) carried out on January 2012 in VFAINT mode. These data have been analyzed with \textsc{CIAO} \citep{2006SPIE.6270E..1VF} data analysis system version 4.12 and \textit{Chandra} calibration database CALDB version 4.9.1, adopting standard procedures. After filtering for time intervals of high background flux exceeding \(3\,\sigma\) the average level with \textsc{deflare} task, the final exposure attains to \(11.9 \text{ ks}\).

To detect point sources in the \(0.3-7\text{ keV}\) energy band with the \textsc{wavdetect} task we adopted a \(\sqrt{2}\) sequence of wavelet scales (i.e., 1, 2, 4, 8, 16 and 32 pixels) and a false-positive probability threshold of \({10}^{-6}\). We note that this procedure did not detect a point source positionally consistent with the location of the radio core.

\begin{figure}
	\centering
	\includegraphics[scale=0.7]{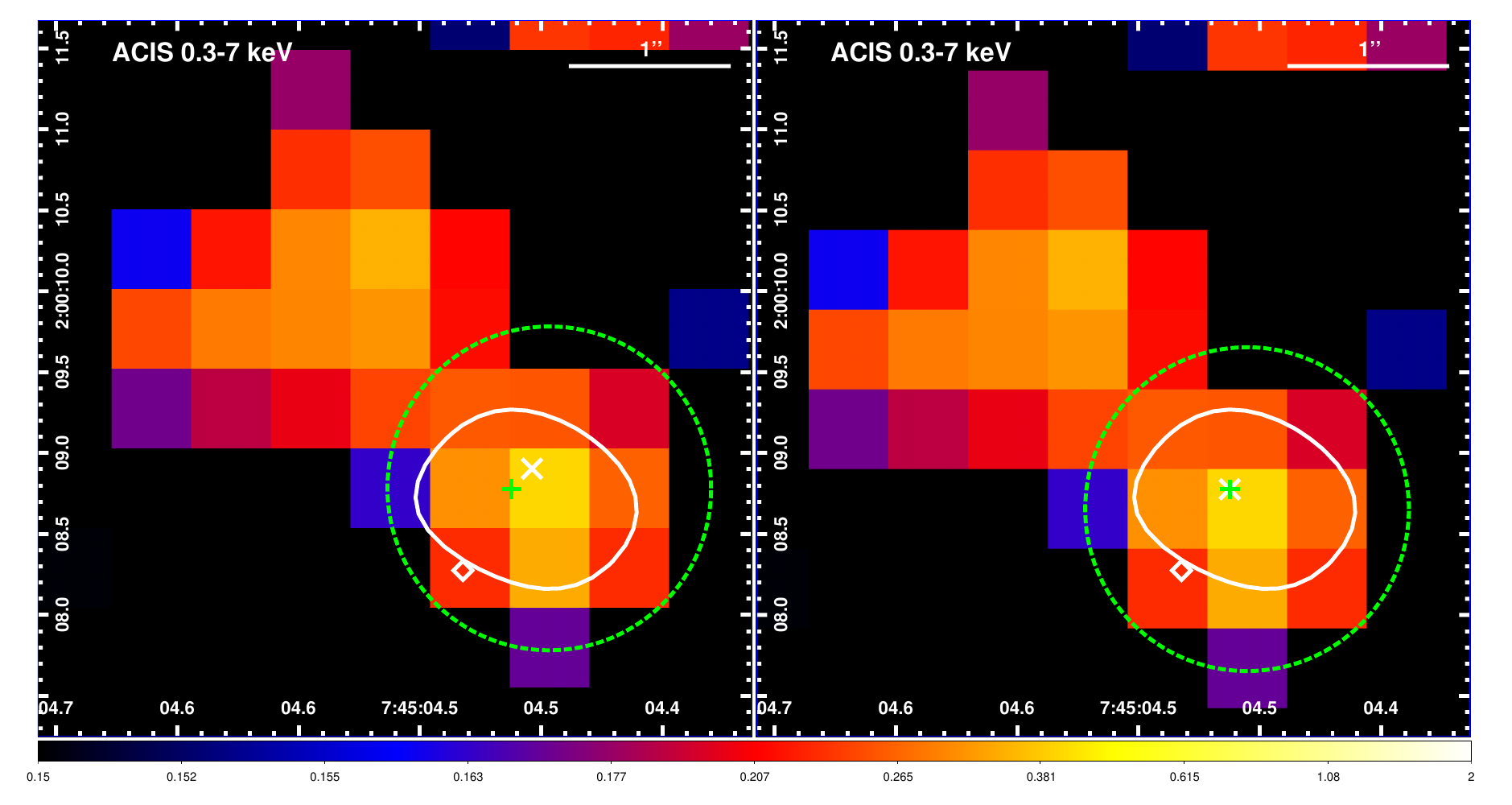}
	\caption{(Left panel) Nuclear region of 3C 187 as imaged by \textit{Chandra}-ACIS detector in the \(0.3-7 \text{ keV}\) band, smoothed with a Gaussian kernel with a \(1\times 1\) pixel (\(0\farcs 492\times 0\farcs 492\)) \(\sigma\). The white diamond indicates the optical identification by \citet{1995A&A...299...17H}, and the green cross indicates the radio core position by \citep{1995ApJS...99..349N}. The white line represents the \(1\text{ mJy}\text{ beam}^{-1}\) contour of the VLA \(4.87 \text{ GHz}\) observation presented by \citet{1994A&AS..105...91B}. The white X indicates the position of the centroid evaluated in the region represented with the green dashed line. (Right panel) Same as the left panel after the registration of the X-ray image to match the centroid position with radio core position.}\label{fig:registration}
\end{figure}

The left panel of Fig. \ref{fig:registration} shows the nuclear region (\(\sim 4\arcsec\times 4\arcsec\)) of 3C 187 as imaged by \textit{Chandra}-ACIS detector in the \(0.3-7 \text{ keV}\) band, smoothed with a Gaussian kernel with a \(1\times 1\) pixel (\(0\farcs 492\times 0\farcs 492\)) \(\sigma\). For reference in the same figure are represented the optical identification presented by \citet{1995A&A...299...17H} (white diamond), and the radio core position (green cross) at J2000 \(\text{R.A.}=07\text{h}45\text{m}04\text{s}.455\), \(\text{DEC}=+02\degree 00\arcmin08\farcs74\) \citep{1995ApJS...99..349N} together with the \(1\text{ mJy}\text{ beam}^{-1}\) contour (white full line) of the VLA \(4.87 \text{ GHz}\) (\(6 \text{ cm}\)) observation presented by \citet{1994A&AS..105...91B}. Since we do not get a clear X-ray detection of the core, we considered a circular region with a radius of \(1\arcsec\) (shown with the green dashed line) centered on the brightest pixel in this region, and evaluated the centroid of this smoothed image in this region. The centroid at J2000 \(\text{R.A.}=07\text{h}45\text{m}04\text{s}.454\), \(\text{DEC}=+02\degree 00\arcmin08\farcs90\) (indicated with a white X in the left panel of Fig. \ref{fig:registration}) is separated from the radio core position by \(0\farcs 18\), in agreement with all registration shifts reported in \citet{2011ApJS..197...24M} and compatible with the \(0\farcs8\) \textit{Chandra} absolute astrometric uncertainty\footnote{\href{https://cxc.cfa.harvard.edu/cal/ASPECT/celmon/}{https://cxc.cfa.harvard.edu/cal/ASPECT/celmon/}}. We therefore registered the X-ray images to align this centroid with the radio core, as shown in the right panel of Fig. \ref{fig:registration}. We note, however, that since we are interested in the structure of the source that is extended on arcminute scales, being hundreds kiloparsecs, small uncertainties on the X-ray map registration do not affect our analysis and results.

\subsection{X-ray Extended Emission}\label{sec:imaging}

We produced full \(0.3-7\text{ keV}\), soft (\(0.3-3\text{ keV}\)), and hard (\(3-7\text{ keV}\)) band images of \textit{Chandra} data centered on ACIS-S chip 7. We also produced PSF maps (with the \textsc{mkpsfmap} task), effective area corrected exposure maps, and flux maps using the \textsc{flux\_obs} task.

\begin{figure}
	\centering
	\includegraphics[scale=0.7]{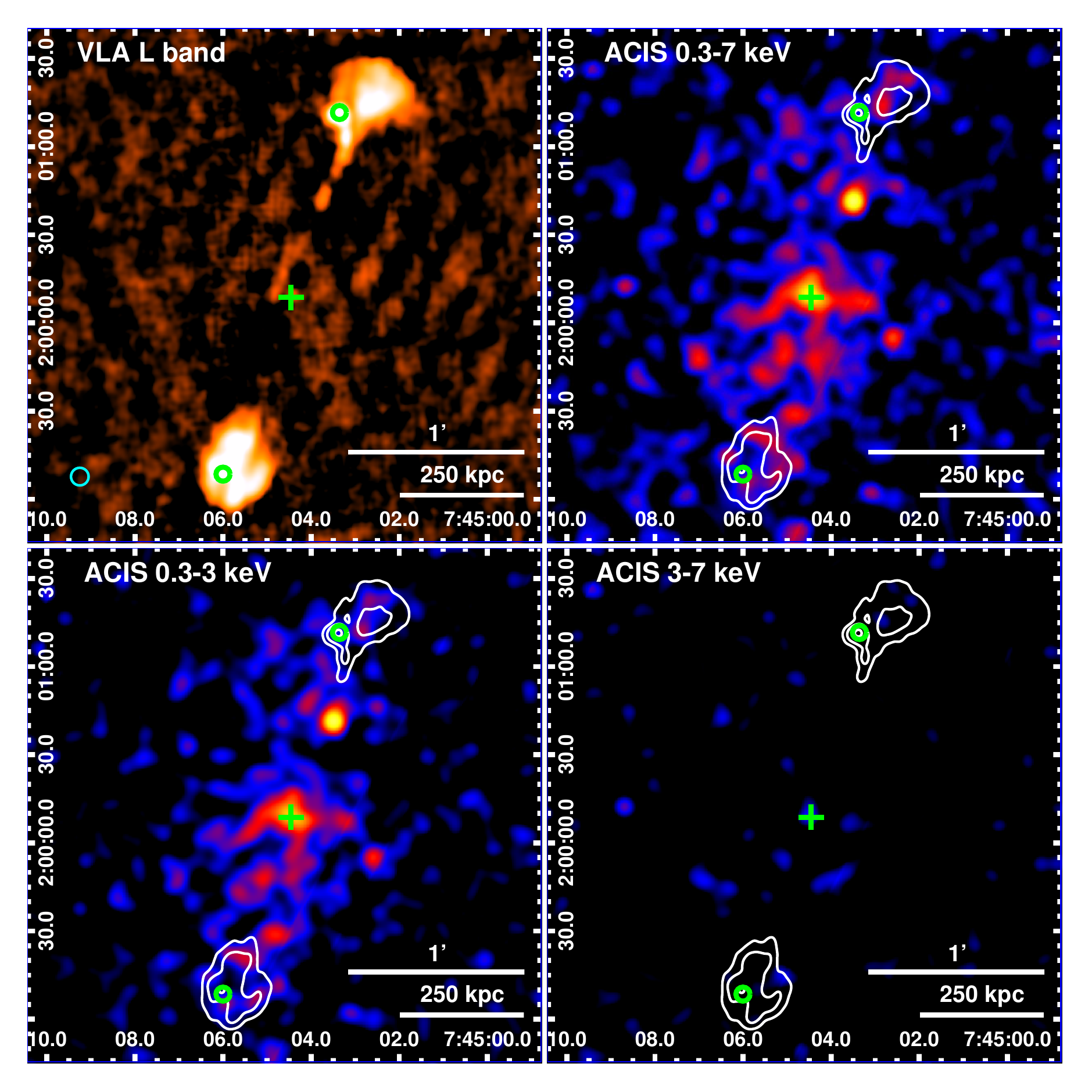}
	\caption{(Upper left panel) Archival VLA \(1.4 \text{ GHz}\) (\(20\text{ cm}\)) image of the central \(\sim 3\arcmin\) region of 3C 187 \citep{1995ApJS...99..349N} obtained via AIPS standard reduction procedure (\href{http://www.aips.nrao.edu/cook.html}{http://www.aips.nrao.edu/cook.html}). The clean beam (shown in the lower left with a cyan ellipse) is \(3\arcsec\times 3\arcsec\) with major axis \(\text{P.A.}=0{\degree}\), and the rms noise is \(87.3\,\upmu\text{Jy}\) per beam. The green circles represent the locations of the north and south lobe peaks as reported by \citet{1996ApJS..107..175R}, while the green cross marks the peak of the core emission as seen in the \(6\text{ cm}\) map \citep{1995ApJS...99..349N}. (Upper right panel) Full band \(0.3-7\text{ keV}\) ACIS-S flux image smoothed with a Gaussian kernel with a \(5\times 5\) pixel (\(2\farcs 46\times 2\farcs 46\)) \(\sigma\), with superimposed in white VLA L-band contours from the left panel image. The radio contours start at \(30\) times the rms level, increasing by factors of four. The sources marked in this panel are the same as the right panel. (Lower left and right panels) Same as the upper right panel, but in the \(0.3-3\text{ keV}\) and \(3-7\text{ keV}\) bands.}\label{fig:vla20}
\end{figure}

In Fig. \ref{fig:vla20} we present a comparison of the central \(\sim 3\arcmin\) region of 3C 187 as seen in the VLA \(1.4 \text{ GHz}\) (\(20\text{ cm}\)) map (upper left panel) and in the full (upper right panel), soft (lower left panel), and hard (lower right panel) X-ray flux maps. We can clearly see that the X-ray emission appears elongated in the same direction of the radio lobes, with enhanced emission extending up to the lobes themselves. In addition we see that X-ray emission above \(3 \text{ keV}\) is essentially compatible with the background, and therefore in the following we will consider only \textit{Chandra} images in the soft \(0.3-3 \text{ keV}\) energy range.

In Fig. \ref{fig:gmrt} we compare the inner \(\sim 5\arcmin\) region of 3C 187 as imaged by the the Giant Metrewave Radio Telescope (GMRT) \(150\text{ MHz}\) map (left panel) collected during the TIFR GMRT Sky Survey \citep[TGSS,][]{2017A&A...598A..78I}, and in the soft X-ray flux map (right panel). For comparison, we overlay on the latter both the VLA \(1.4 \text{ GHz}\) and the GMRT \(150\text{ MHz}\) contours. Again, we see the soft X-ray emission is elongated in the direction of the radio lobes as seen in the GMRT map, with most of the emission lying between these lobes.


To define regions of cone (along the radio axis) and cross-cone (perpendicular to the radio axis) emission we produced a \(0.3-3\text{ keV}\) azimuthal surface flux profile, extracting data in angular sectors centered on the brightest pixel shown in the left panel of Fig. \ref{fig:registration}, with an inner and outer radii of \(2 \arcsec\) and \(100 \arcsec\), and an arbitrary width of \(15 \degree\) (as show in the right panel of Fig. \ref{fig:azimuthal_profile}) excluding counts from detected point-like sources. We then fitted the obtained profile with a model comprising two Gaussian functions, plus a constant to take into account the background level, as shown in the right panel of Fig. \ref{fig:azimuthal_profile}. This allowed us to identify two cone regions with the angles comprised between two standard deviations from each Gaussian peak. Thus the north (N) cone is centered at a P.A. \(-13.3\degree\) and ranges between \(-42.1\degree\) and \(15.5\degree\), while the south (S) cone is centered at a P.A. \(164.0\degree\) and lies between \(128.7\degree\) and \(199.3\degree\) as shown in Fig. \ref{fig:cones}. We note that varying the angular sector width between \(10\degree\) and \(20\degree\) has little effect on the estimate of the cone regions.

To investigate the presence of enhanced X-ray emission in these regions we made use of the standard beta model \citep{1976A&A....49..137C, 1978A&A....70..677C} which is generally used to fit the surface brightness profile in relaxed galaxies or groups and clusters of galaxies,
\begin{equation}\label{eq:beta}
S_b(r)=S_0 {\left[{1+{\left({\frac{r}{r_C}}\right)}^2}\right]}^{1/2-3\beta}\,,
\end{equation}
where \(S_0\) is the central surface brightness, \(r_C\) is the core radius, and \(\beta\) is linked to the projected galaxy velocity dispersion \(\sigma_R\) and gas temperature \(T\) by \(\beta=\mu m_H {\sigma_R}^2 / k T\) (where \(\mu\) is the mean molecular weight, \(m_H\) is the mass of the
hydrogen atom, and \(k\) is the Boltzmann constant). As mentioned before, when fitting a surface flux profile rather than a surface brightness profile with a beta model the only difference will be the \(S_0\) parameter that would yield the central surface flux instead of the central surface brightness.

Then, to evaluate the extension of the emission in the cones and cross-cone direction we extracted net surface flux profiles in the directions presented in Fig. \ref{fig:cones}, excluding counts from detected sources, and extracting the background from source-free regions of chip 7 and 6. The width of the bins was adaptively determined to reach a minimum signal to noise ratio of \(3\). In the outer regions, when this ratio could not be reached, we extended the bin width up \(\sim 200\arcsec\) from the nucleus.

Surface flux profiles in the cross-cone, N, S direction are presented in Fig. \ref{fig:sb_cones}. The best fit beta models are shown with dashed lines, and residuals to these models presented on the bottom of each panel. Best fit parameters for these profiles are reported in Table \ref{tab:beta_fit}.

We see that the soft X-ray emission extends with a signal to noise ratio of at least 3 up to \(\sim 45\arcsec\) (corresponding to \(\sim 270\text{ kpc}\)) both in the W (upper left panel) and E (upper central panel) directions, respectively. The best fit beta models are shown with black dashed lines. The best fit parameters for the beta model are \(r_C=115\pm 46\text{ kpc}\) and \(\beta=0.83\pm 0.31\) for the W direction, and \(r_C=98\pm 4\text{ kpc}\) and \(\beta=0.73\pm 0.03\) for the E direction. Considering the uncertainties, larger for the W direction, the two profiles are compatible. In addition we extracted a combined surface flux profile in the W and E direction (upper right panel). For this combined profile we get best fit parameters of \(r_C=142\pm 55\text{ kpc}\) and \(\beta=0.85\pm 0.37\). Although the uncertainties for this combined profile are larger than for the W and E directions - possibly due to the mixture of gas with a different distribution - this is useful to characterize the average profile away from the enhanced emission in the direction of the radio lobes.

In the bottom panels of Fig. \ref{fig:sb_cones} we present surface flux profiles for the N and S direction. The soft X-ray emission extends with a signal to noise ratio of at least 3 up to \(\sim 75\arcsec\) (corresponding to \(\sim 450\text{ kpc}\)) and \(\sim 70\arcsec\) (corresponding to \(\sim 400\text{ kpc}\)) in the N (lower left panel) and S (lower right panel) directions, respectively. In both directions, the surface flux signal to noise ratio drops below \(3\) just after the radio lobe regions. In these panels we indicate with vertical green dashed lines and with colored areas the locations of the contour level at \(30\) times the rms level for the \(150 \text{ MHz}\) and VLA \(1.4 \text{ GHz}\) maps, respectively (see Fig. \ref{fig:cones}). The colored dashed lines represent the the best fit beta models to surface flux profiles with the exclusion of the bins falling in the colored areas, which are likely to contain enhanced X-ray flux with respect to a diffuse thermal emission described by a beta model. For the N cone the best fit \(r_C=185 \text{ kpc}\) is essentially unconstrained, while for the \(\beta\) parameter we obtain \(\beta=0.69\pm 0.30\). For the S cone instead we obtain \(r_C=292\pm 9 \text{ kpc}\) and \(\beta=0.84\pm 0.02\). These results suggest that the soft X-ray emission is significantly more extended in the S cone than in the N cone. Looking at the residuals on the bottom of each panel, we see that positive residuals tend to fall in correspondence with the colored areas that mark the VLA \(1.4 \text{ GHz}\) radio emission, more significantly in the N cone.

We also tried comparing the observed surface flux profiles in the N and S cones with that observed in the cross-cone direction. To this end we plotted the best fit beta model obtained in the W+E cones, normalizing it to match the first bin of the N and S surface flux profiles, and present it with a black dashed line in the lower panels of Fig. \ref{fig:sb_cones}. We see that the surface flux profile is above that observed in the cross-cone directions at \(R\gtrsim 200 \text{ kpc}\) in the N cone, and at all radii in the S cone. This suggests that either the thermal emission in N and S cone is more extended than in the cross-cone direction, or that there is additional emission on top of the thermal one in correspondence with radio lobes, extending in the S cone closer to the nucleus than what we observe in the N cone.

In conclusion, the X-ray emission from 3C 187 appears extended in all directions, more in the N and S cone than in the cross cone, and that it is generally compatible with a beta model distribution, with enhanced emission coincident with the radio lobes, in particular the N cone. The best fit \(\beta\) values in the W, E, S, and N cones are \(\sim 0.8\), \(\sim 0.7\), \(\sim 0.8\), and \(\sim 0.6\), although this parameter is poorly constrained. Such values are similar to those observed in groups and clusters of galaxies \citep{1996ApJ...456...80M, 1999ApJ...517..627M, 2000MNRAS.315..356H}. For the N and S cone we see enhanced emission with respect to the cross-cone in correspondence with the radio lobes, suggesting the possibility of a significant contribution to the observed X-ray flux by non-thermal process (see Sect. \ref{sec:seds}), in which case beta models do not then necessarily describe the underlying gravitational potential.

\subsection{X-ray Spectral Analysis}\label{sec:spectra}

We extracted spectra in the regions presented in Fig. \ref{fig:spectral_regions} on the basis of the surface flux profiles shown in Fig. \ref{fig:sb_cones}. In Fig. \ref{fig:spectral_regions} the cone and cross-cone direction extending up the radius where a signal to noise ratio of \(3\) is reached (see Sect. \ref{sec:imaging}) are presented in cyan (full lines for the cross-cone directions and dashed lines for the N and S directions), while the green lines represent the N and S lobes as encircled by the contour level at \(30\) times the rms level for the \(150 \text{ MHz}\) (see Fig. \ref{fig:gmrt}) maps. Since the N and S lobes almost completely encompass N and S cone regions as defined above (see Fig. \ref{fig:spectral_regions}) we decided to extract spectra in the lobes to represent the regions of enhanced radio emission. We also extracted a spectrum in the cross-cone region combining the W and E direction and excluding the overlapping regions covered by the N and S lobes.

We produced spectral response matrices weighted by the count distribution within the aperture (as appropriate for extended sources). Background spectra were extracted in the same source-free regions used for radial profile extraction (Sect. \ref{sec:imaging}).

Firstly, we subtracted background spectra from source spectra, making use of the \(\chi^2\) fit statistic, binning the spectra to obtain a minimum of \(30\) counts per bin. Spectral fitting was performed in the \(0.3-7\text{ keV}\) energy range with \textsc{Sherpa} application \citep{2001SPIE.4477...76F}, using three different models: a simple power-law (\textsc{powerlaw}) that should represent the IC/CMB emission in the non-thermal scenario, a thermal plasma (\textsc{xsapec}\footnote{\href{https://heasarc.gsfc.nasa.gov/xanadu/xspec/manual/XSmodelApec.html}{https://heasarc.gsfc.nasa.gov/xanadu/xspec/manual/XSmodelApec.html}}) with C, N, O, Ne, Mg, Al, Si, S, Ar, Ca, Fe and Ni abundances fixed to solar values, that should model the emission from the IGM, and a model that comprises both the power-law and the thermal plasma. In all the models we included photo-electric absorption by the Galactic column density along the line of sight \(N_H = 6.36\times {10}^{20}\text{ cm}^{-2}\) \citep{2016A&A...594A.116H}. The results of these fits are presented in Table \ref{tab:spectra} and in Fig. \ref{fig:spectra}. Errors correspond to the \(1\)-\(\sigma\) confidence level for one interesting parameter (\(\Delta\chi^2 = 1\)). We also tried modeling the extracted spectra with a \textsc{xsvapec} model using variable element abundances. Such model, however, yielded unconstrained abundances, with fit statistics and temperatures similar to the fixed abundance model.

In the cross-cones directions the power-law model (upper left panel of Fig. \ref{fig:spectra}) yields a photon index \(\Gamma=1.9 \pm{0.3}\), while the thermal plasma model (upper central panel) yields a poorly constrained temperature \(kT={3.9}_{-1.2}^{+2.6} \text{ keV}\), with the two models giving similar reduced \(\chi^2\) values. When applying the mixed power-law+thermal model, the power-law component is poorly constrained, and its photon index and normalization were therefore fixed to their best fit values, obtaining a reduced \(\chi^2\) similar to the previous models. The best-fit thermal component has a temperature \(kT={3.3}_{-1.0}^{+1.9} \text{ keV}\), compatible with the temperature evaluated with the thermal model alone, and as shown in the upper right panel of Fig. \ref{fig:spectra} it appears to be dominant with respect to the power-law component, which contributes to \(\sim 9\%\) of the total flux. We cannot however rule out either the thermal or the non-thermal model on a statistical base.

In the N lobe region the power-law model (central left panel) yields a photon index \(\Gamma=2.3\pm0.3\), while the thermal model (central panel) yields a temperature \(kT={2.7}_{-0.7}^{+1.1} \text{ keV}\), again with similar reduced \(\chi^2\) values. When applying the mixed power-law+thermal model, we fixed the temperature of the thermal component to the best fit value obtained in the cross-cone region to check for the presence of the same hot plasma in the N lobe region. The resulting normalization of the thermal component resulted poorly constrained, so we fixed it to its best fit value, obtaining a reduced \(\chi^2\) similar to the previous models. The best-fit power-law component has a photon index \(\Gamma={2.4}_{-0.4}^{+0.5} \text{ keV}\), compatible with the photon index evaluated with the power-law model alone, and as shown in the central right panel it appears to be dominant with respect to the thermal component, which however contributes to \(\sim 20\%\) of the total flux. Again, we cannot rule out either the thermal or the non-thermal model on a statistical base.

In the S lobe the power-law model yields a photon index \(\Gamma={1.8}\pm{0.3}\) with a reduced \(\chi^2=1.04\), while the thermal model yields an unconstrained temperature \(kT = {5.4}_{-2.0}^{+20.6}\text{ keV}\) with a reduced \(\chi^2=1.16\). This suggests that in this region the power-law model is statistically favored. The mixed model with the temperature of the thermal component fixed to the value obtained in the cross-cone region yields a zero normalization for the thermal component, and it is therefore equivalent to the power-law model, indicating that X-ray emission in the S lobe is mainly of non-thermal origin.

We then repeated the spectral fits described above, but instead of subtracting the background spectra we modeled them using the prescription given by \citet{2003ApJ...583...70M}, that is, a model comprising a thermal plasma component \citep[MEKAL;][]{kasstra1992} with solar abundances and a power law. For this analysis we binned the spectra to obtain a minimum of \(1\) count per bin, making use of the cash statistic. The results of these fits are presented in Table \ref{tab:spectra_bkg}.

The best fit parameters obtained in this way are compatible with those obtained subtracting the background spectra. However, due to the increased statistic, they are better constrained with smaller uncertainties, with the exception of the temperature in the S lobe region, which is even less constrained in this case. Another difference with respect to the previous analysis is the fit with the mixed model (power-law+thermal) in the N lobe region, where the thermal component with the temperature fixed to the value obtained in the cross-cone region yields a zero normalization, reinforcing the scenario of a dominating non-thermal emission from this region.

We note that the intrinsic (i.e., unabsorbed) X-ray luminosities \(L_{0.5-2}\approx 3\times{10}^{43}\text{ erg}/\text{s}\) inferred from the thermal models are compatible within the scatter with those expected from the \(L_X - T\) relation observed in clusters (\citealt[e.g.,][their Fig. 1]{1998ApJ...504...27M}; \citealt[e.g.,][their Fig. 2]{2016MNRAS.463..820Z}) in the N lobe and cross-cone region, while in the S cone the luminosity is significantly lower than those expected from thermal emission from thermal gas at a temperature of \(\sim 5\text{ keV}\). In addition, we note that the photon indices obtained with the power-law models are similar to those observed in lobes X-ray detected of other radio galaxies \citep{2005ApJ...626..733C}.

In conclusion, although with the statistics of the present data  we cannot rule out either the thermal or the non-thermal model, we consider the former to be favored in the cross-cone region to explain the observed X-ray emission, while the spectral fit results suggest that the X-ray emission form the N lobe is mainly due to non-thermal processes, with the presence of a hot thermal plasma contributing up to \(\sim 20\%\) of the observed flux. In the S lobe, on the other hand, a purely non-thermal emission is the favored interpretation for the observed X-ray radiation. A significant contribution from thermal emission from hot IGM in either lobe however cannot be ruled out. As a matter of fact, disentangling the origin of X-ray radiation between thermal radiation and IC/CMB emission only comparing fit statistic has proven to be challenging even with larger statistics, e.g. larger number of counts \citep[see for example][]{2010MNRAS.404.2018H}, thus additional data are needed to draw firmer conclusions on the spectral analysis of 3C 187.



\section{Discussion}\label{sec:discussion}

\subsection{Non-thermal Scenario: IC/CMB emission}\label{sec:seds}
To investigate the nature of the observed X-ray structure and its connection with the radio emission, in this section we study the multi-wavelength properties of 3C 187 and try to interpret them in the IC/CMB scenario.

\begin{figure}
	\centering
	\includegraphics[scale=0.7]{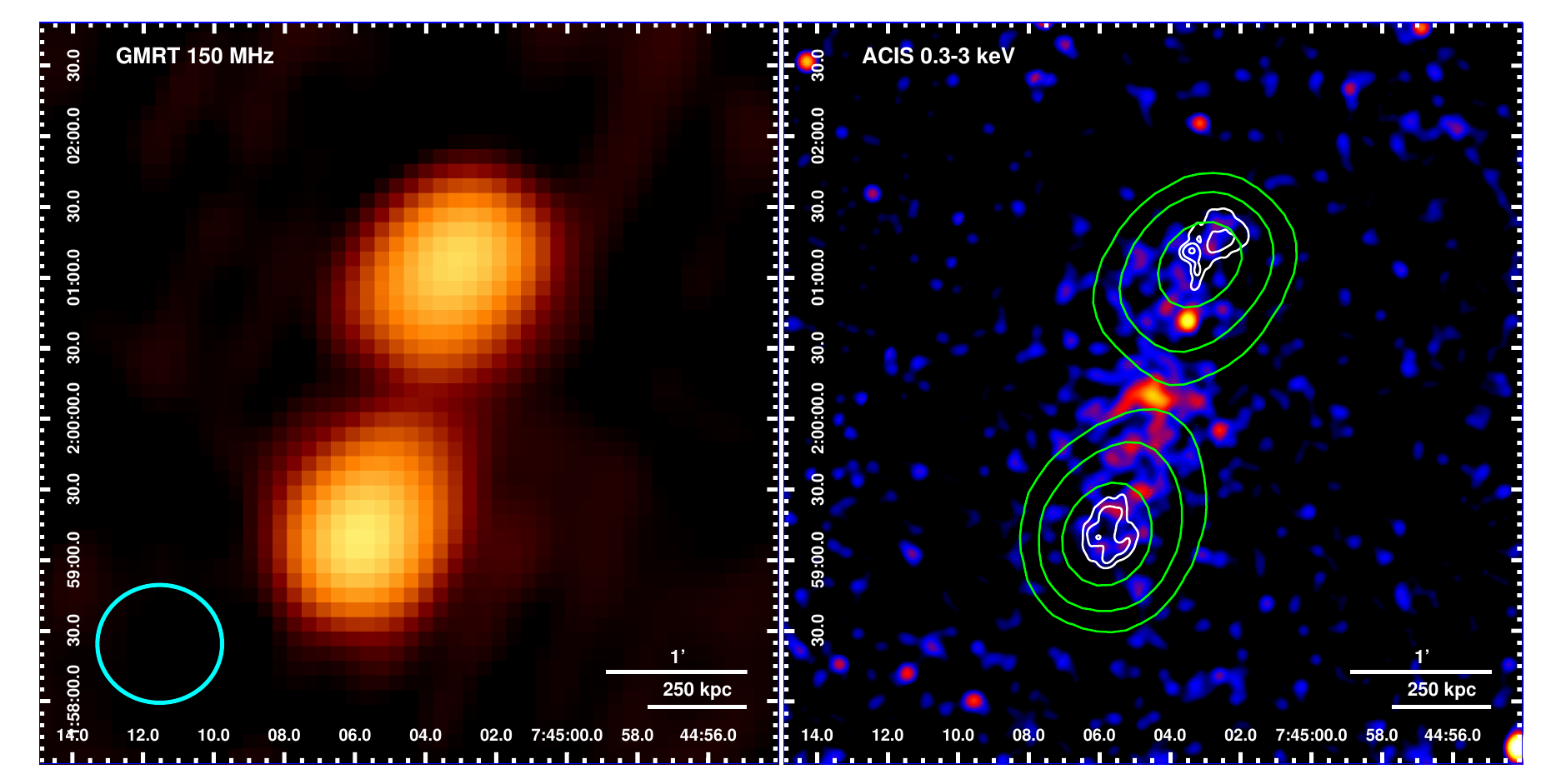}
	\caption{(Left panel) Archival GMRT \(150\text{ MHz}\) image (\href{http://tgssadr.strw.leidenuniv.nl/doku.php}{http://tgssadr.strw.leidenuniv.nl/doku.php}) of the central \(\sim 5\arcmin\) of 3C 187 obtained from the TGSS survey. The clean beam (shown in the lower left with a cyan ellipse) is \(26\farcs5\times 25\arcsec\) with major axis \(\text{P.A.}=0{\degree}\)), and the rms noise is \(2.6 \text{ mJy}\) per beam. (Right panel) Soft band \(0.3-3\text{ keV}\) ACIS-S flux image smoothed with a Gaussian kernel with a \(5\times 5\) pixel (\(2\farcs 46\times 2\farcs 46\)) \(\sigma\), with superimposed in white VLA L-band contours from the upper right panel of Fig. \ref{fig:vla20}, while green lines represent the GMRT \(150\text{ MHz}\) contours starting at \(30\) times the rms level, increasing by factors of four.}\label{fig:gmrt}
\end{figure}

\begin{figure}
	\centering
	\includegraphics[scale=0.35]{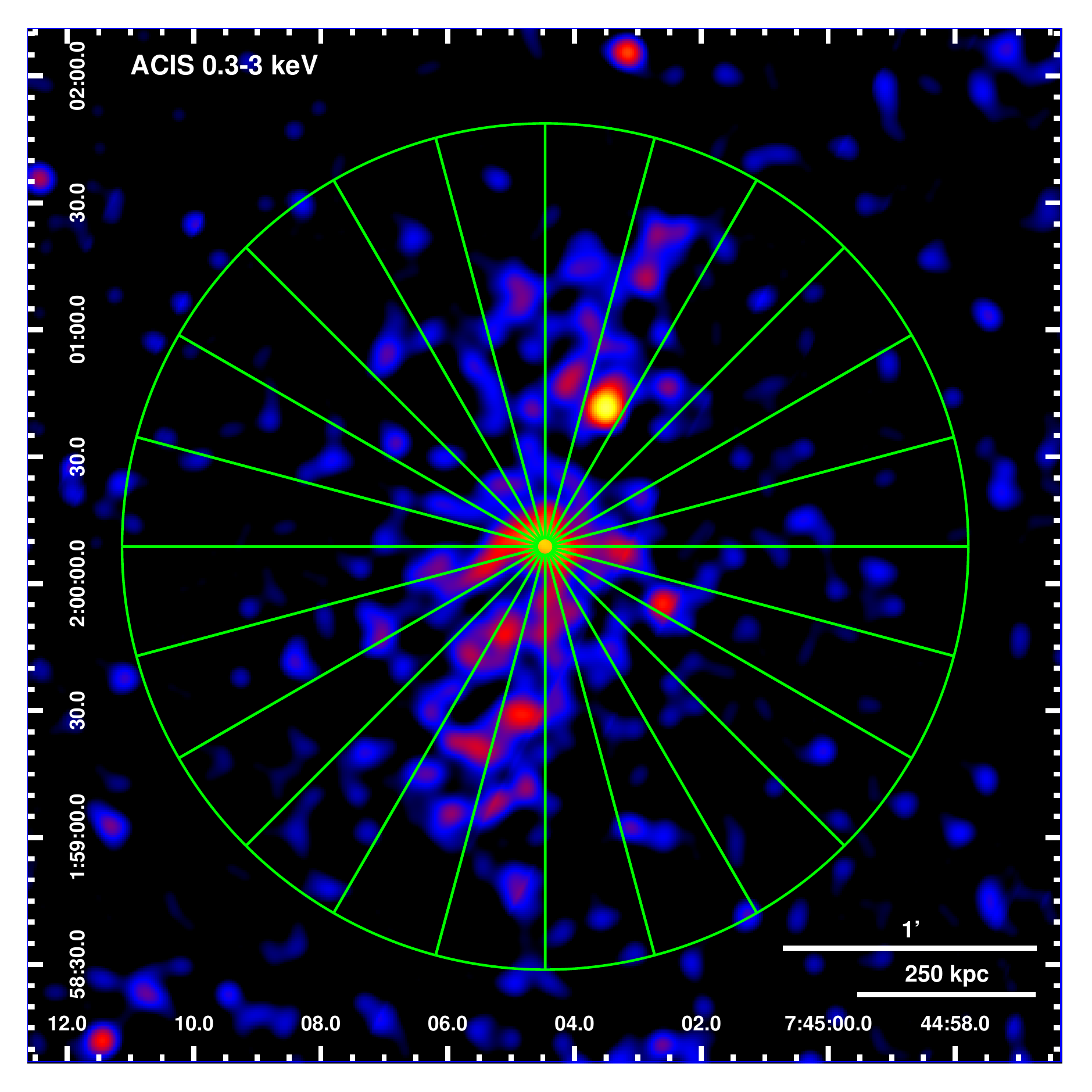}
	\includegraphics[scale=0.38]{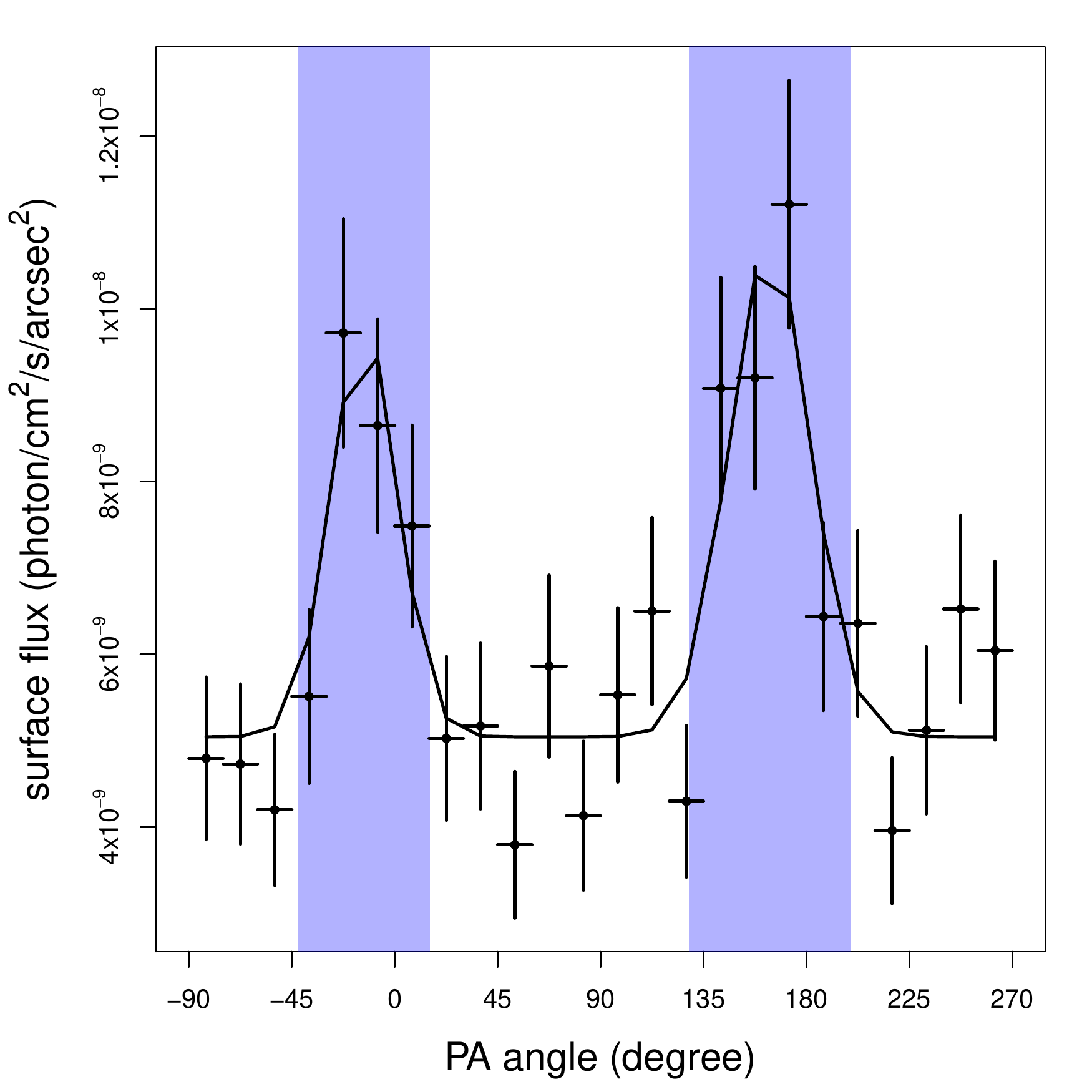}
	\caption{(Left panel) Soft band \(0.3-3\text{ keV}\) ACIS-S flux image smoothed with a Gaussian kernel with a \(5\times 5\) pixel (\(2\farcs 46\times 2\farcs 46\)) \(\sigma\), with superimposed in green the angular sectors used for the azimuthal profile extraction. (Right panel) The black points indicate the surface flux profile in the \(0.3-3 \text{ keV}\) band extracted in the sectors shown in the left panel, with the vertical black bars indicating the uncertainties on the surface flux, and the horizontal black bars indicating the bin widths. The full black line represents the best fit model - comprising two Gaussians plus a constant to account for the background level - for the profile, and the blue shaded areas represent the regions chosen for the N and S cone regions, that is, the angles comprised between two standard deviations from each Gaussian peak.}\label{fig:azimuthal_profile}
\end{figure}

\begin{figure}
\centering
\includegraphics[scale=0.7]{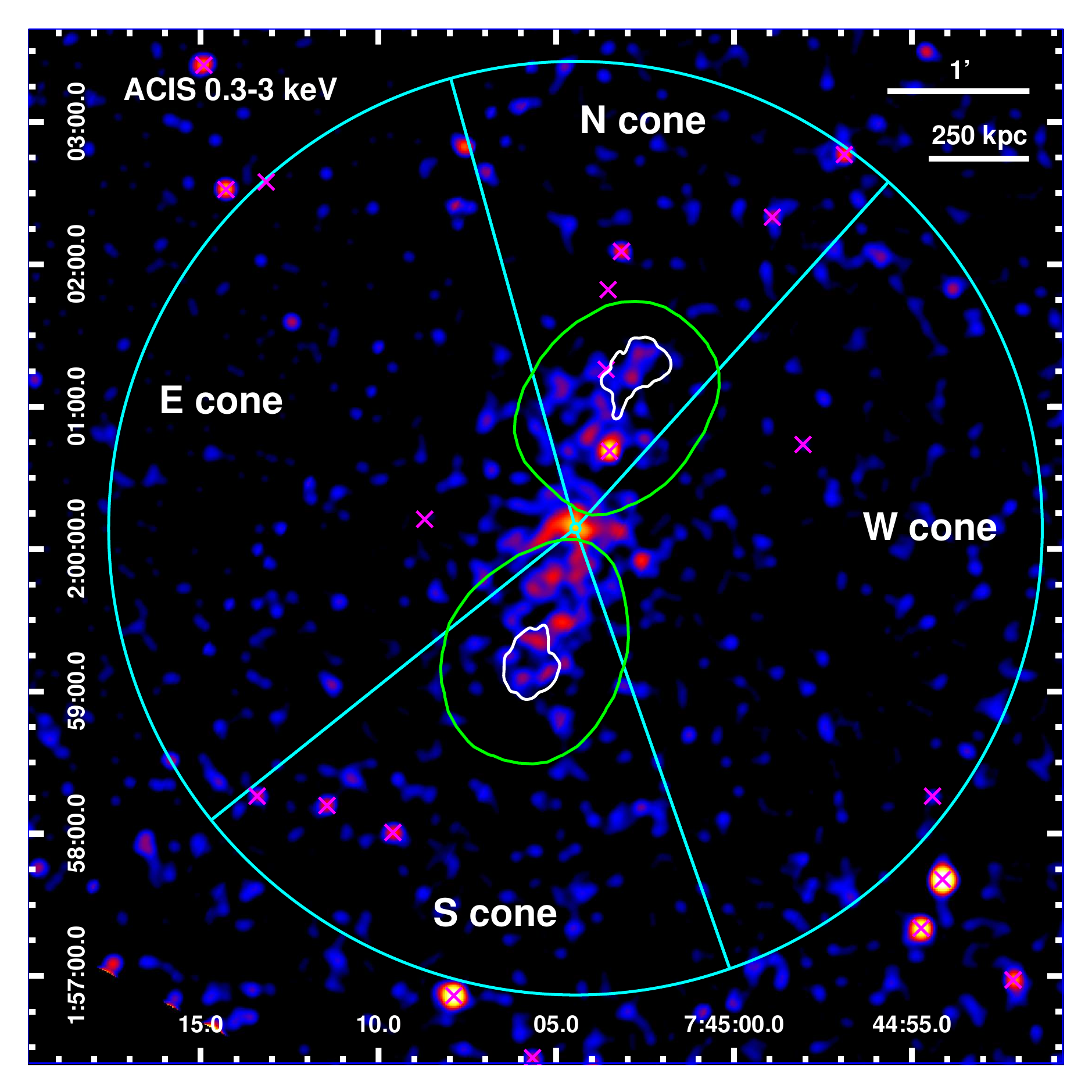}
\caption{Soft band \(0.3-3\text{ keV}\) ACIS-S flux image  smoothed with a Gaussian kernel with a \(5\times 5\) pixel (\(2\farcs46\times 2\farcs46\)) \(\sigma\), with superimposed in cyan the cone regions defined in Fig. \ref{fig:azimuthal_profile}. The white and green lines represent the contour level at \(30\) times the rms level for the VLA \(1.4 \text{ GHz}\) (see Fig. \ref{fig:vla20}) and GMRT \(150 \text{ MHz}\) (see Fig. \ref{fig:gmrt}) maps, while the magenta xs represent the detected X-ray sources.}\label{fig:cones}
\end{figure}

\begin{figure}
\centering
\includegraphics[scale=0.25]{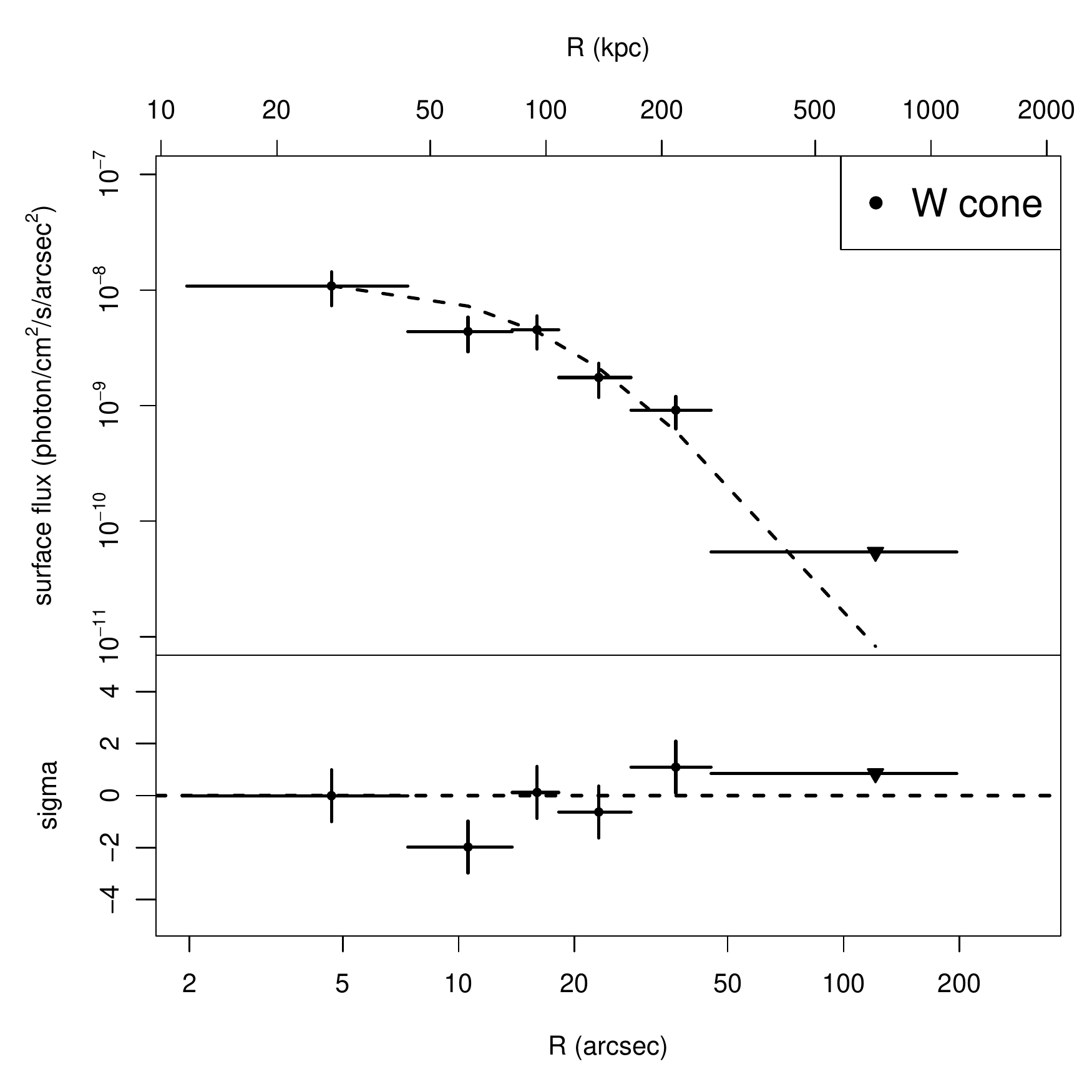}
\includegraphics[scale=0.25]{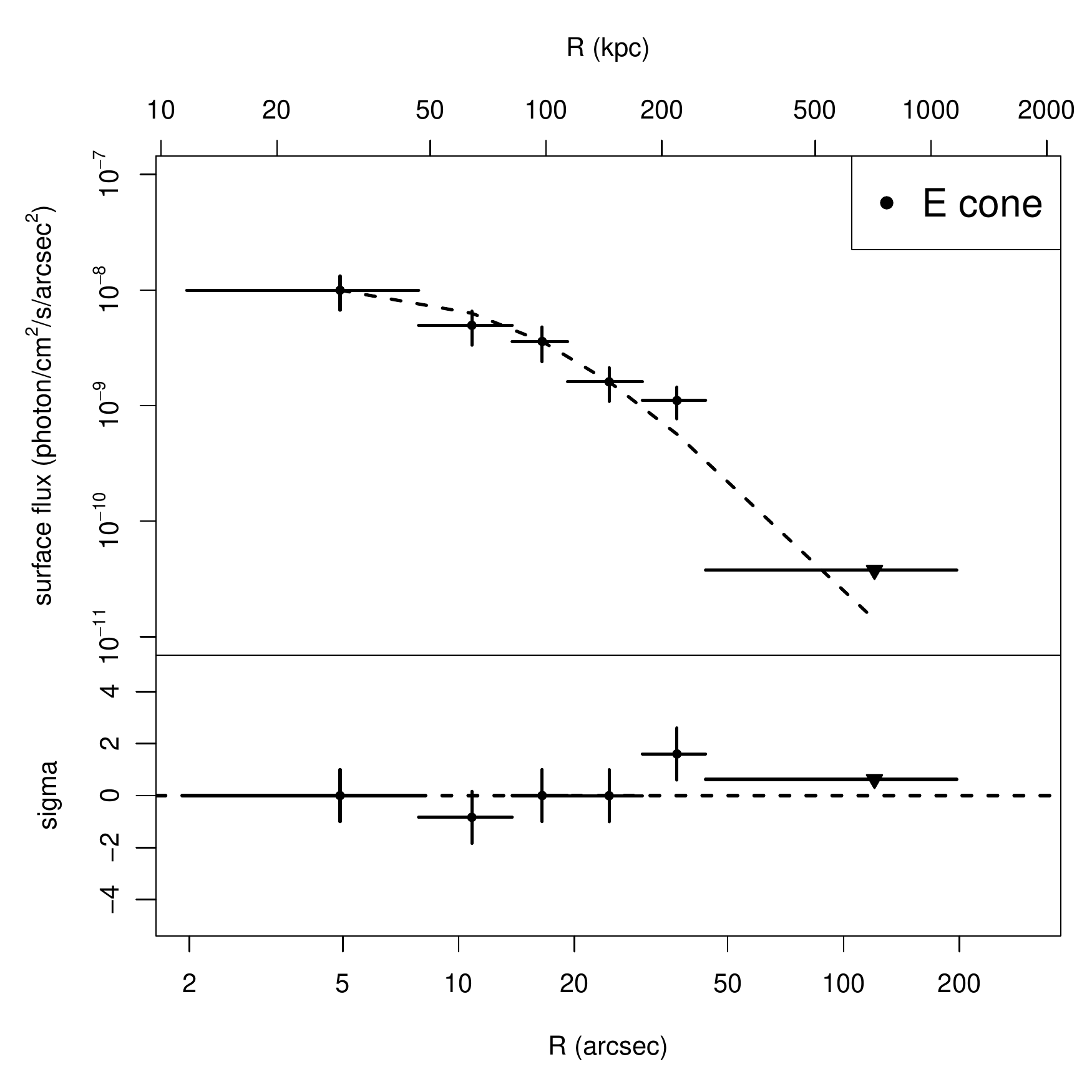}
\includegraphics[scale=0.25]{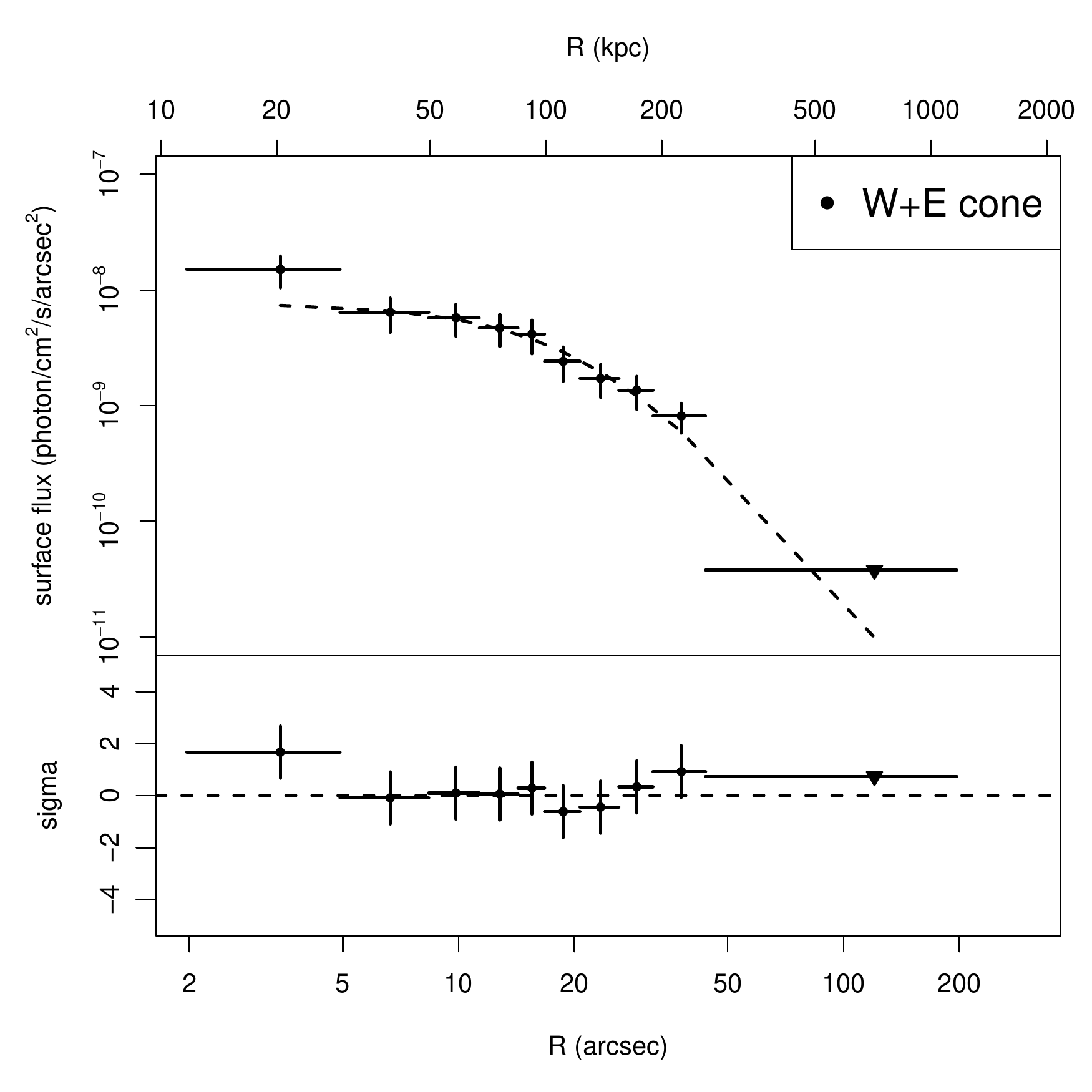}
\includegraphics[scale=0.25]{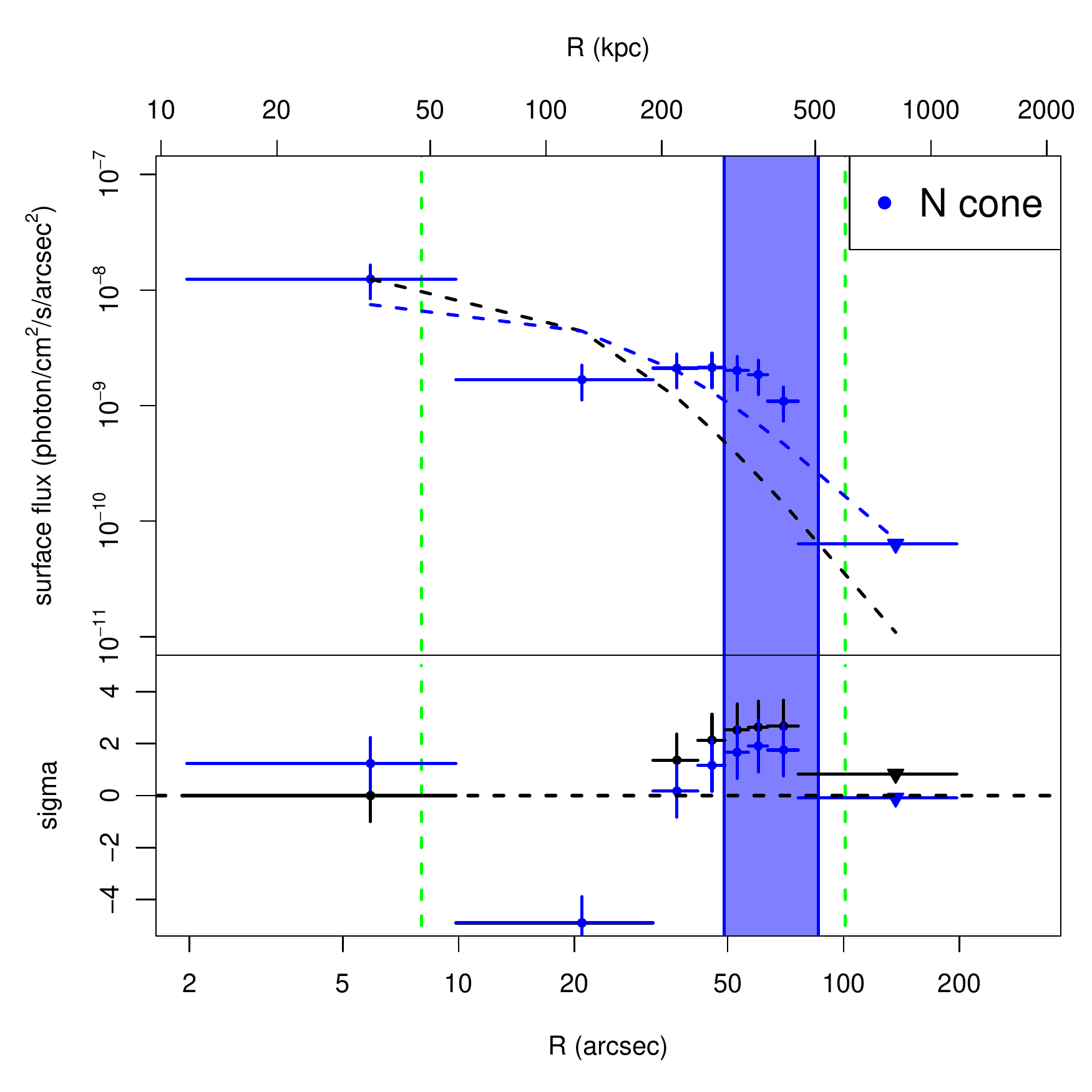}
\includegraphics[scale=0.25]{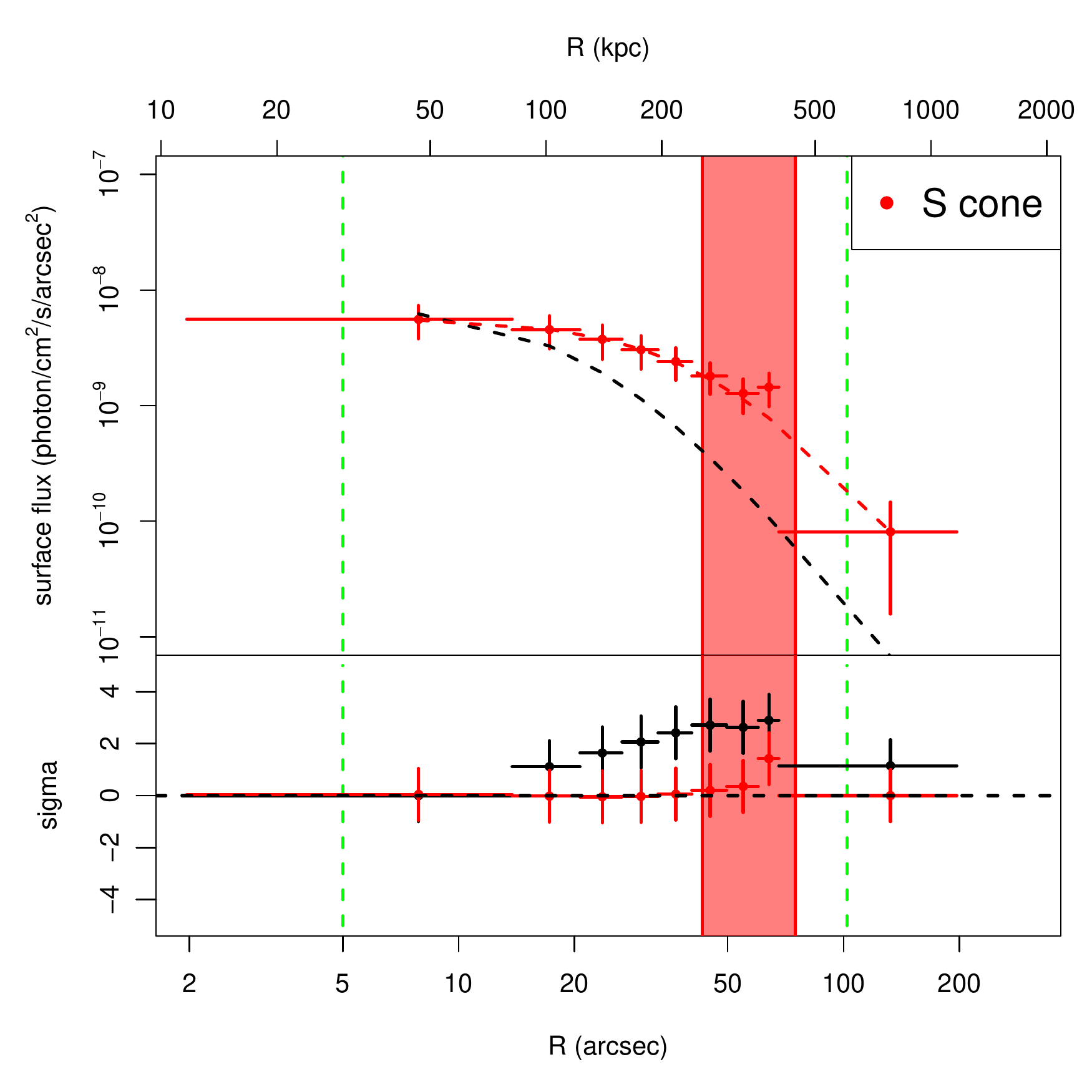}
\caption{Surface flux profiles in the \(0.3-3\text{ keV}\) band extracted in the E cone (upper left panel), W cone (upper central panel), W+E cone (upper right panel), N cone (lower left panel), and S cone (lower right panel) directions (see Fig. \ref{fig:cones}). The widths of the bins are adaptively chosen to reach a minimum signal to noise ratio of \(3\). Upper limits are indicated with downward triangles. For the upper panels the dashed black lines represent the best fit beta model the profiles. For the bottom panels, the black dashed lines represent the best fit beta model to the E+W cone, normalized to match the first bin. The vertical green dashed lines indicate the locations of the contour level at \(30\) times the rms level for \(150 \text{ MHz}\) map, while the colored vertical areas mark the locations of the contour level at \(30\) times the rms level for the VLA \(1.4 \text{ GHz}\) (see Fig. \ref{fig:cones}). The colored dashed lines represent the best fit beta model the profiles, excluding the bins in the colored vertical area. On the bottom of each panels are indicated the residuals to such fits. }\label{fig:sb_cones}
\end{figure}

\begin{figure}
	\centering
	\includegraphics[scale=0.7]{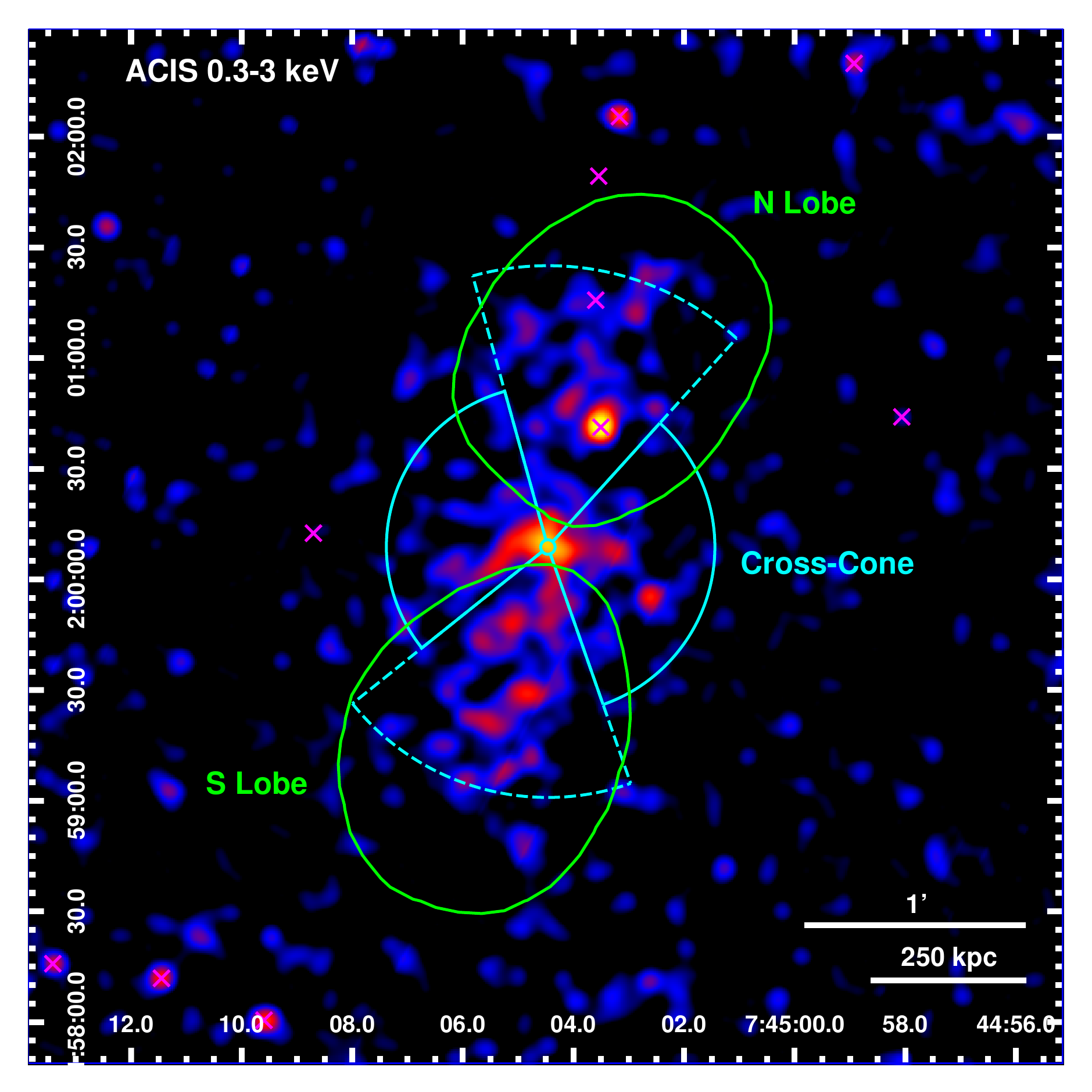}
	\caption{Soft band \(0.3-3\text{ keV}\) ACIS-S flux image smoothed with a Gaussian kernel with a \(5\times 5\) pixel (\(2\farcs46\times 2\farcs46\)) \(\sigma\), with superimposed the regions used for spectral extraction. The cyan lines represent the cones in which the soft X-ray emission extends with a signal to noise ratio of at least 3 (see Fig. \ref{fig:sb_cones}). The green lines represent the contour level at \(30\) times the rms level for the \(150 \text{ MHz}\) (see Fig. \ref{fig:gmrt}) maps. The magenta xs represent the detected X-ray sources, excluded from the spectral extraction.}\label{fig:spectral_regions}
\end{figure}

\begin{figure}
\centering
\includegraphics[scale=0.19]{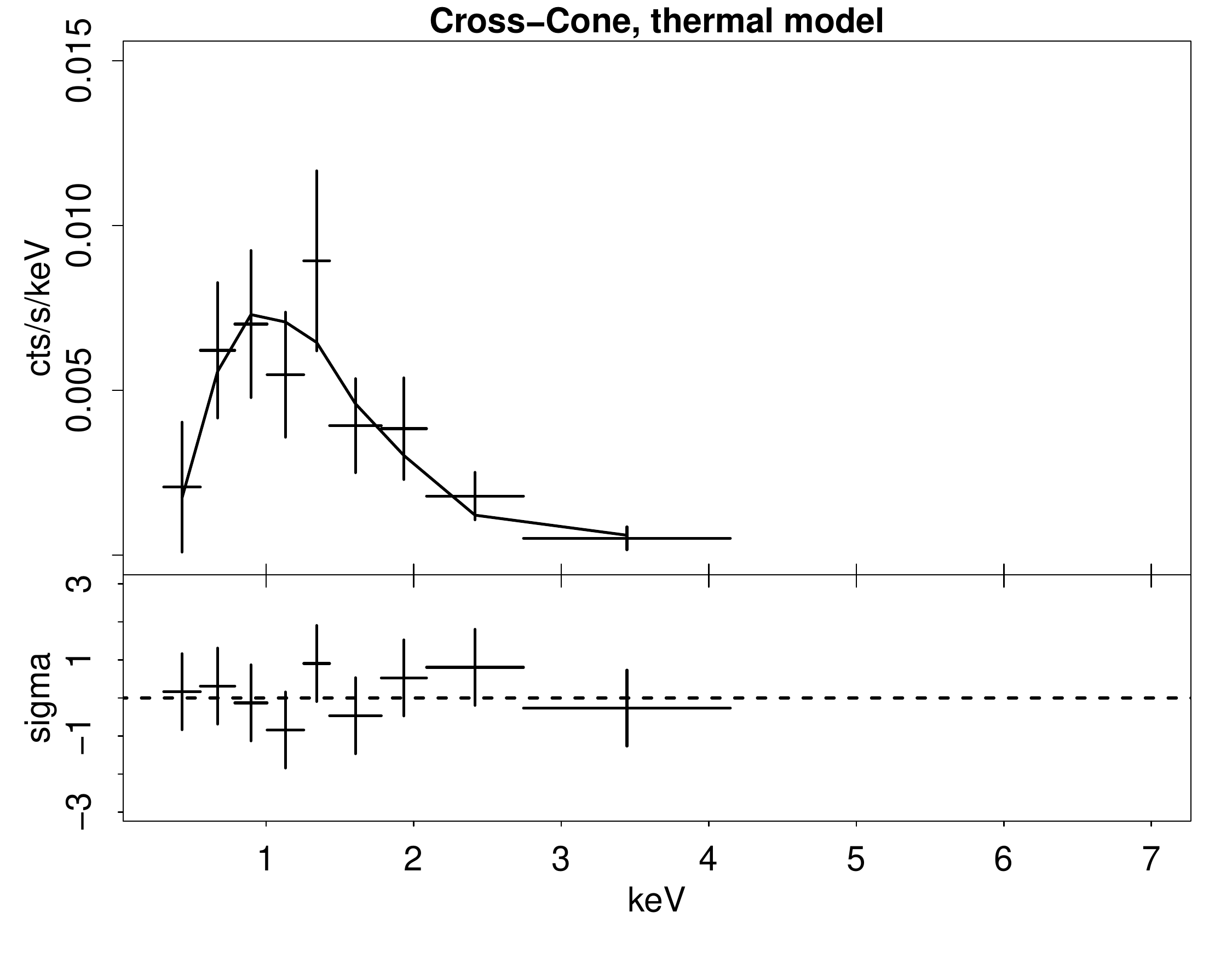}
\includegraphics[scale=0.19]{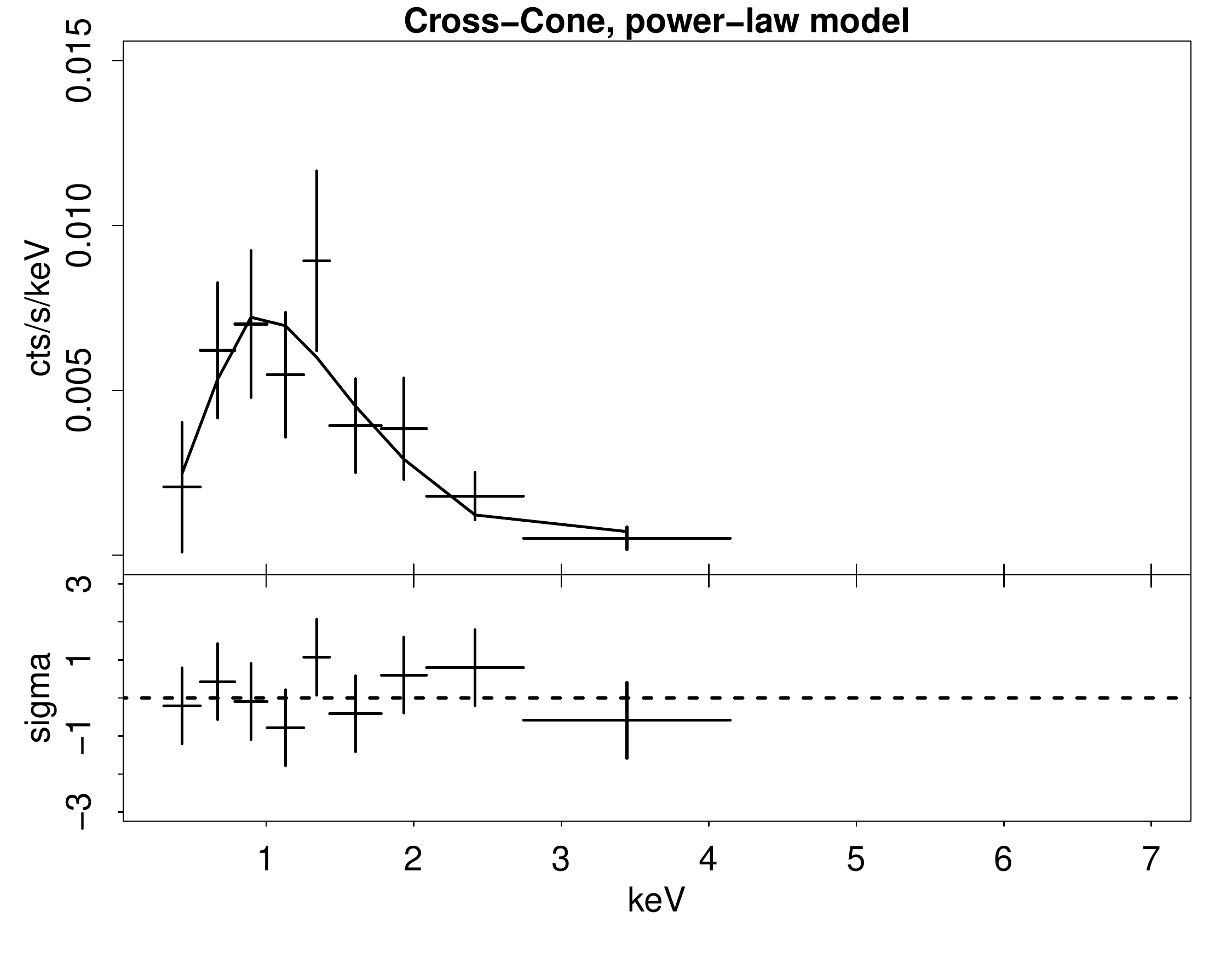}
\includegraphics[scale=0.19]{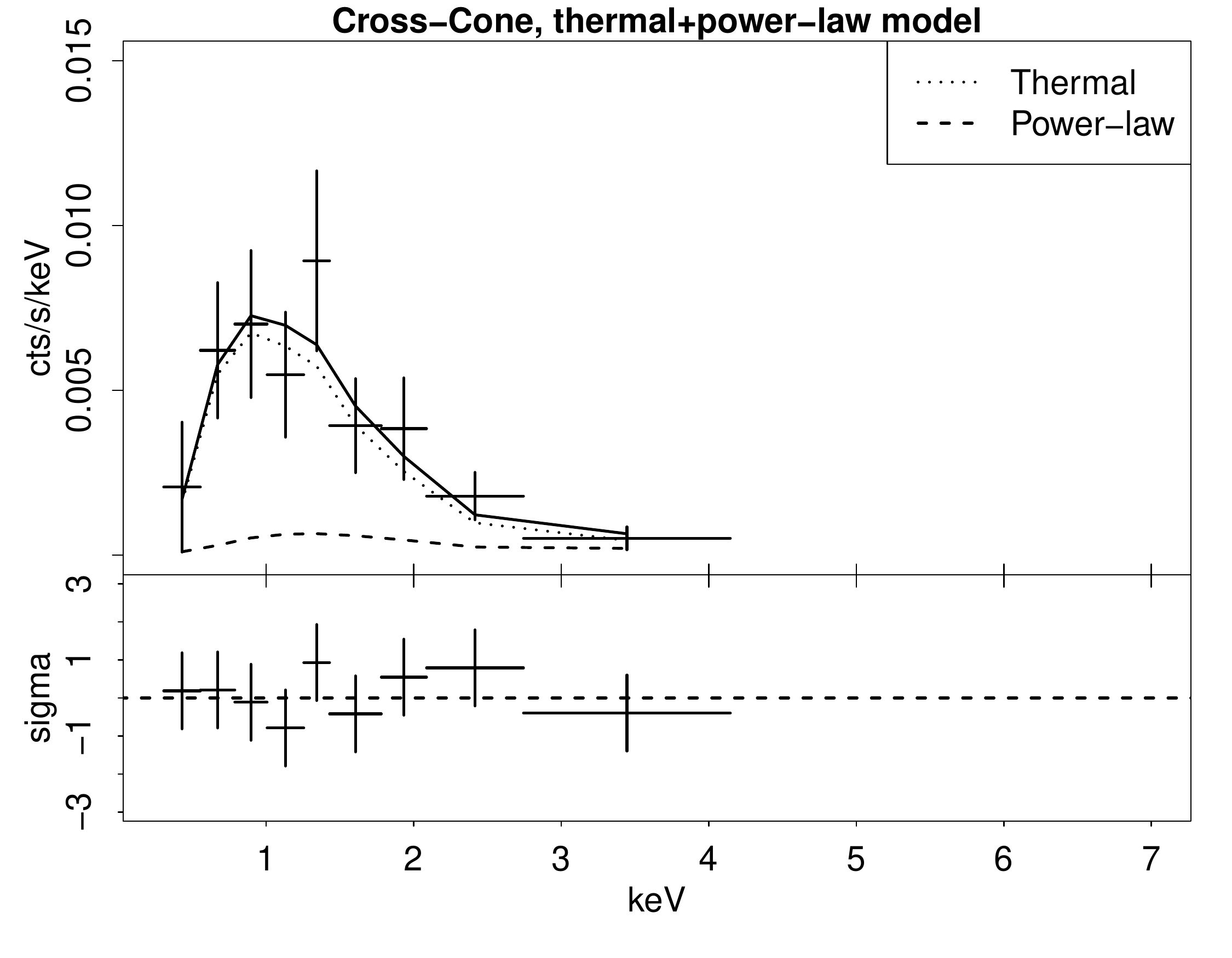}\\
\includegraphics[scale=0.19]{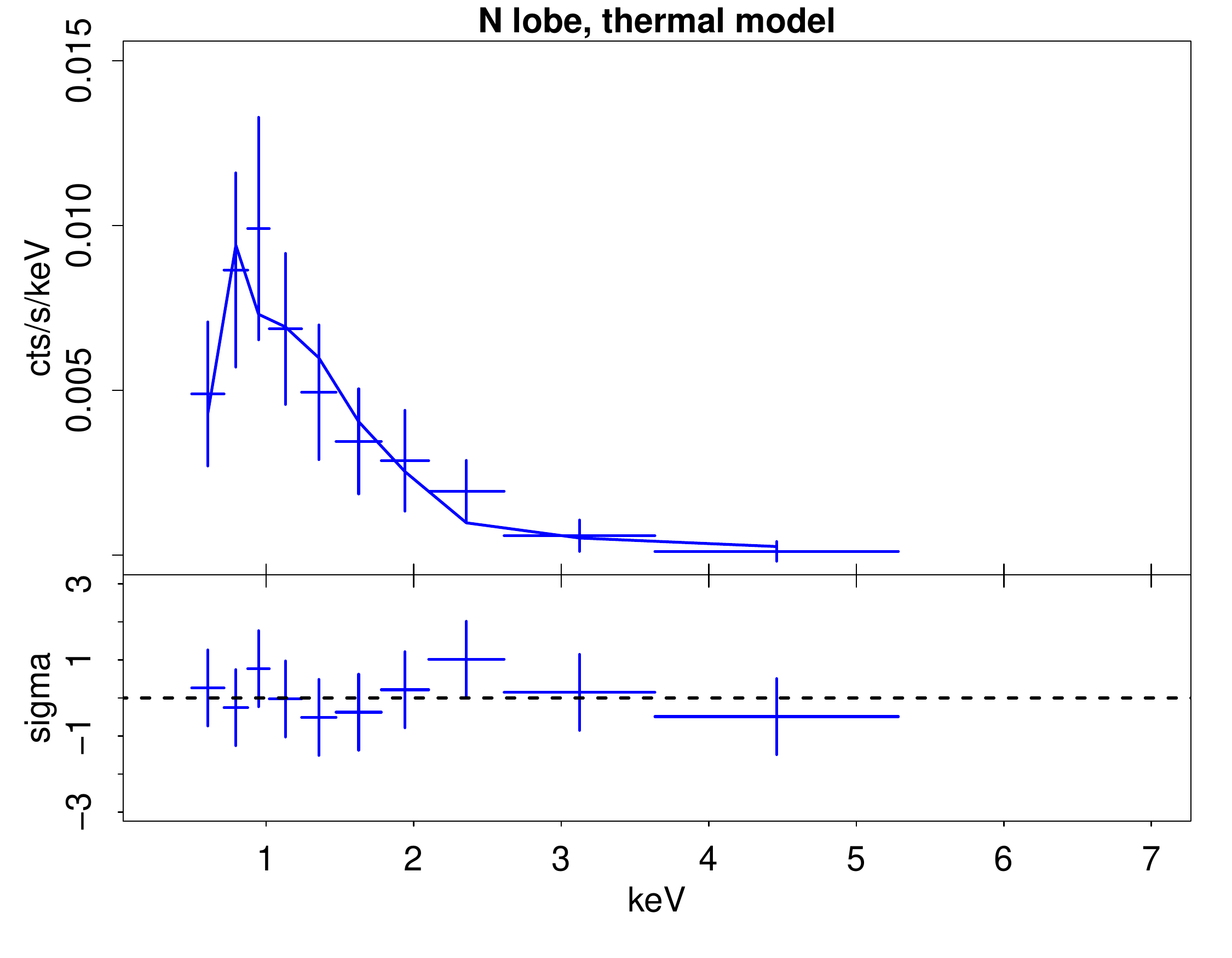}
\includegraphics[scale=0.19]{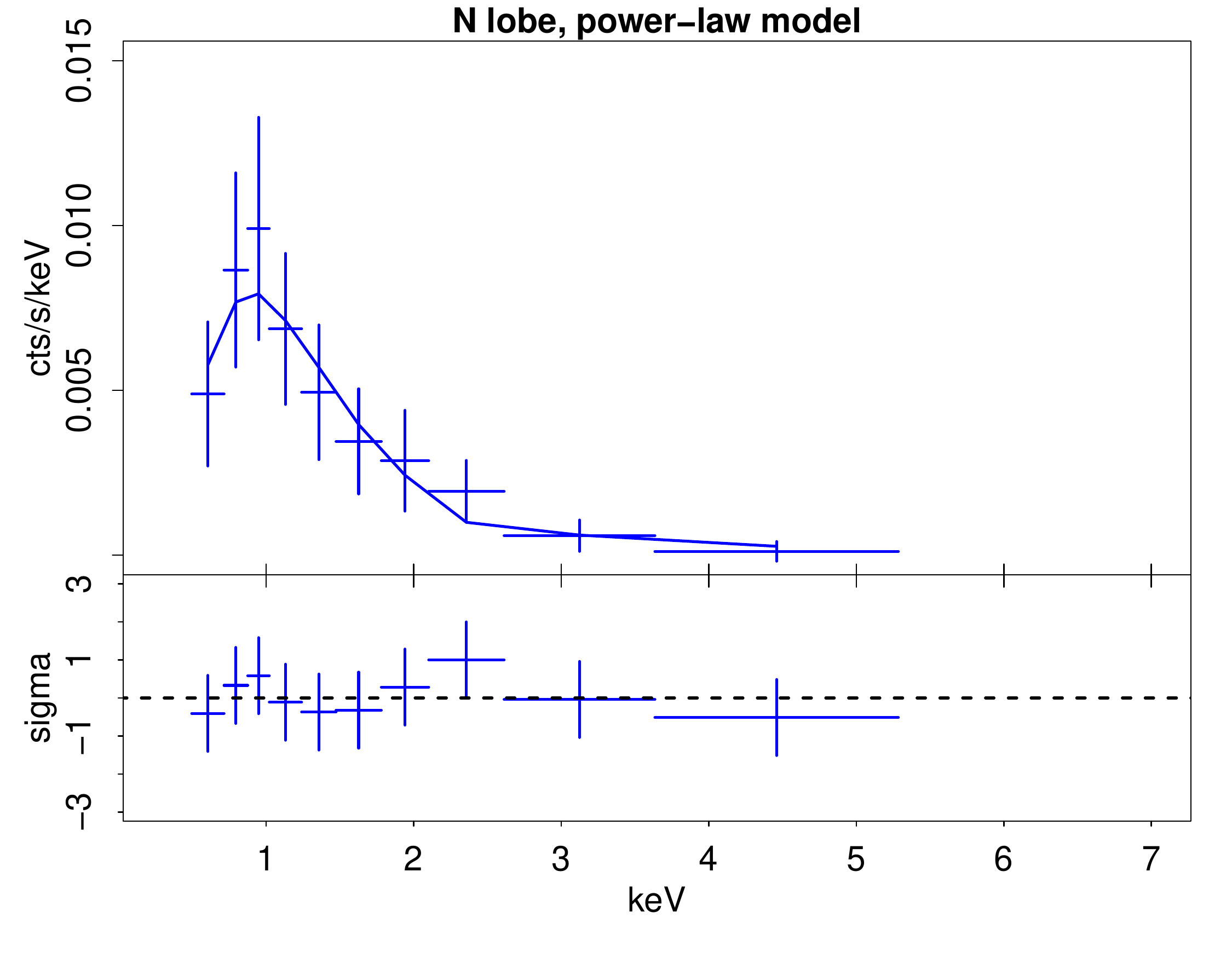}
\includegraphics[scale=0.19]{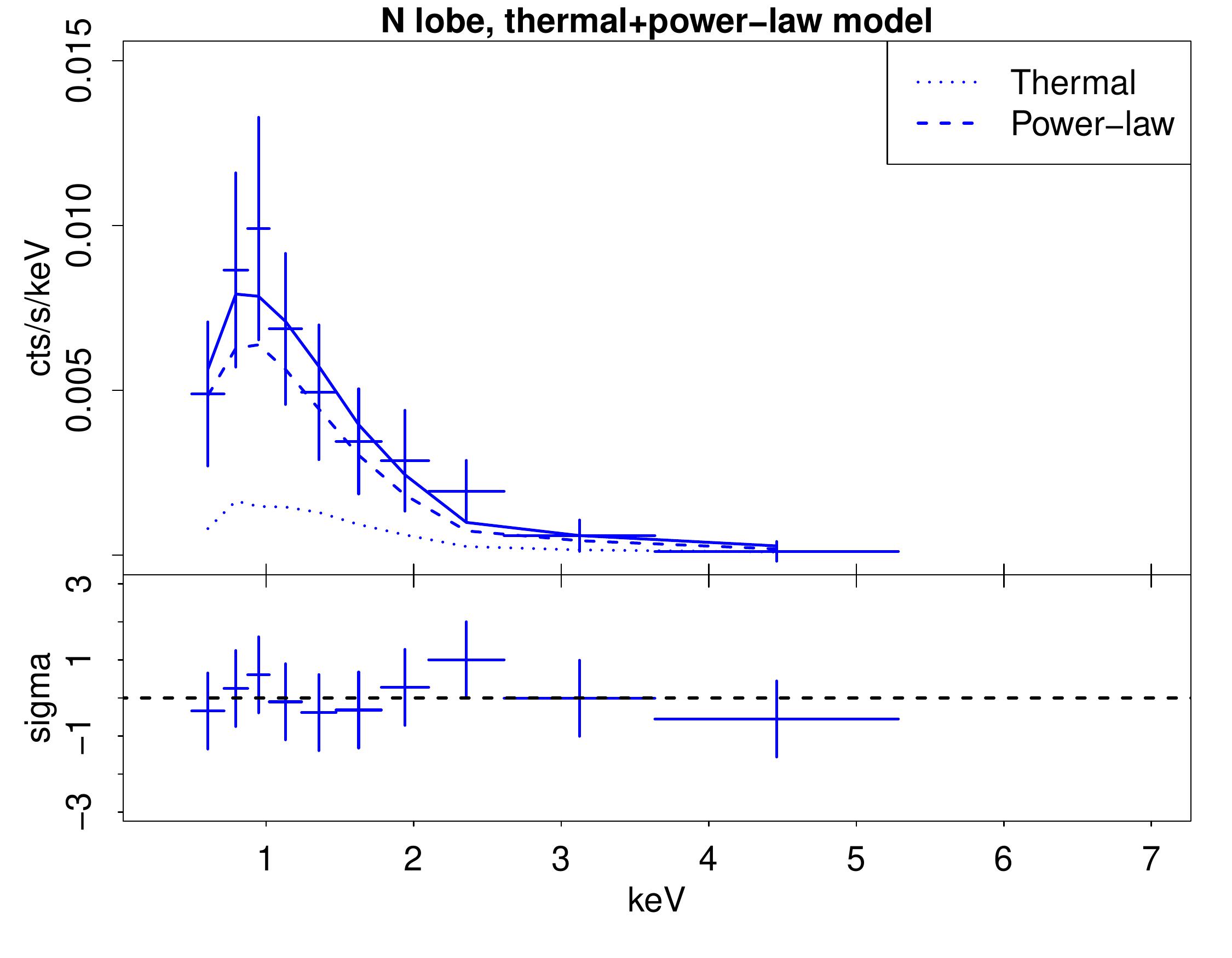}
\\
\includegraphics[scale=0.19]{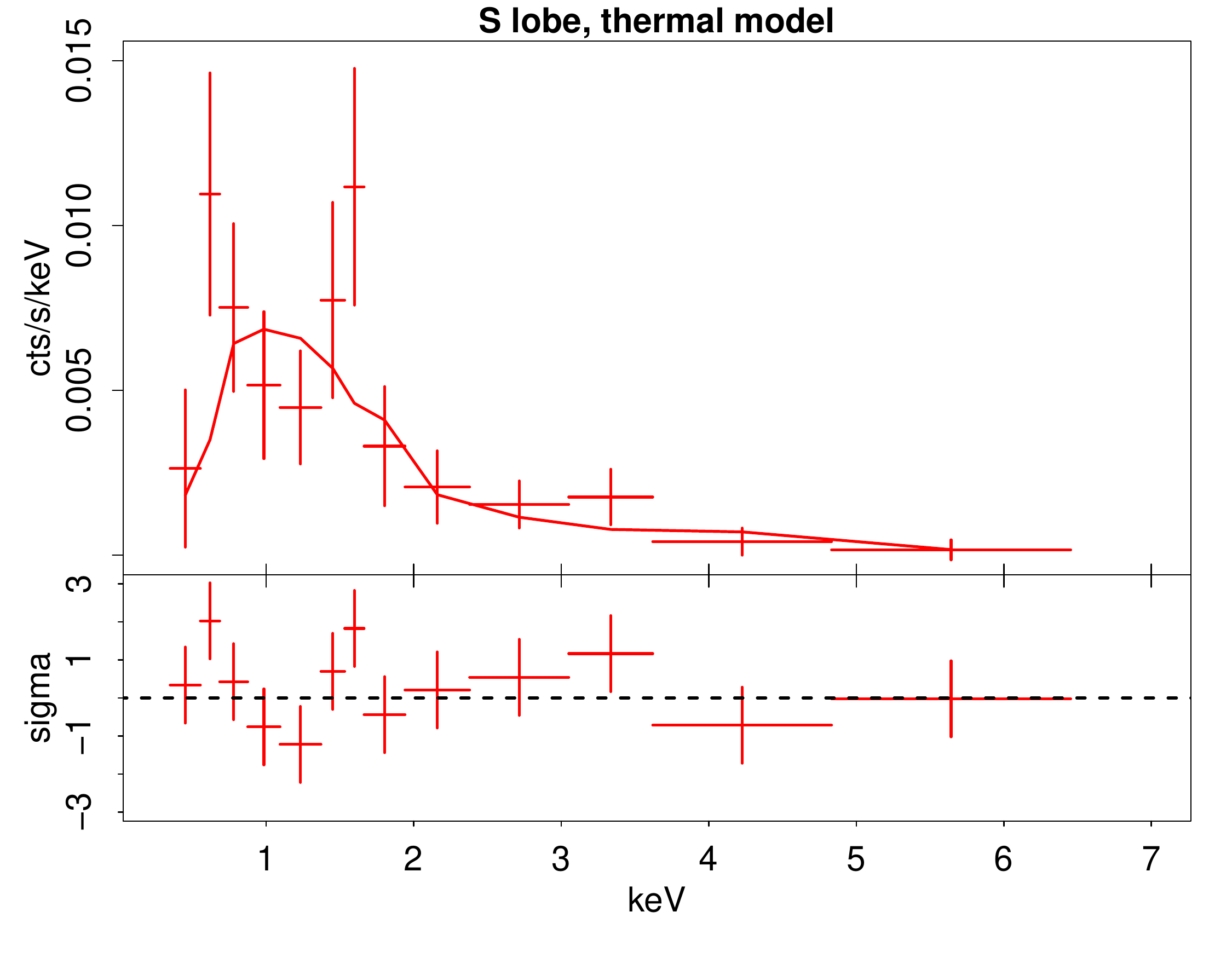}
\includegraphics[scale=0.19]{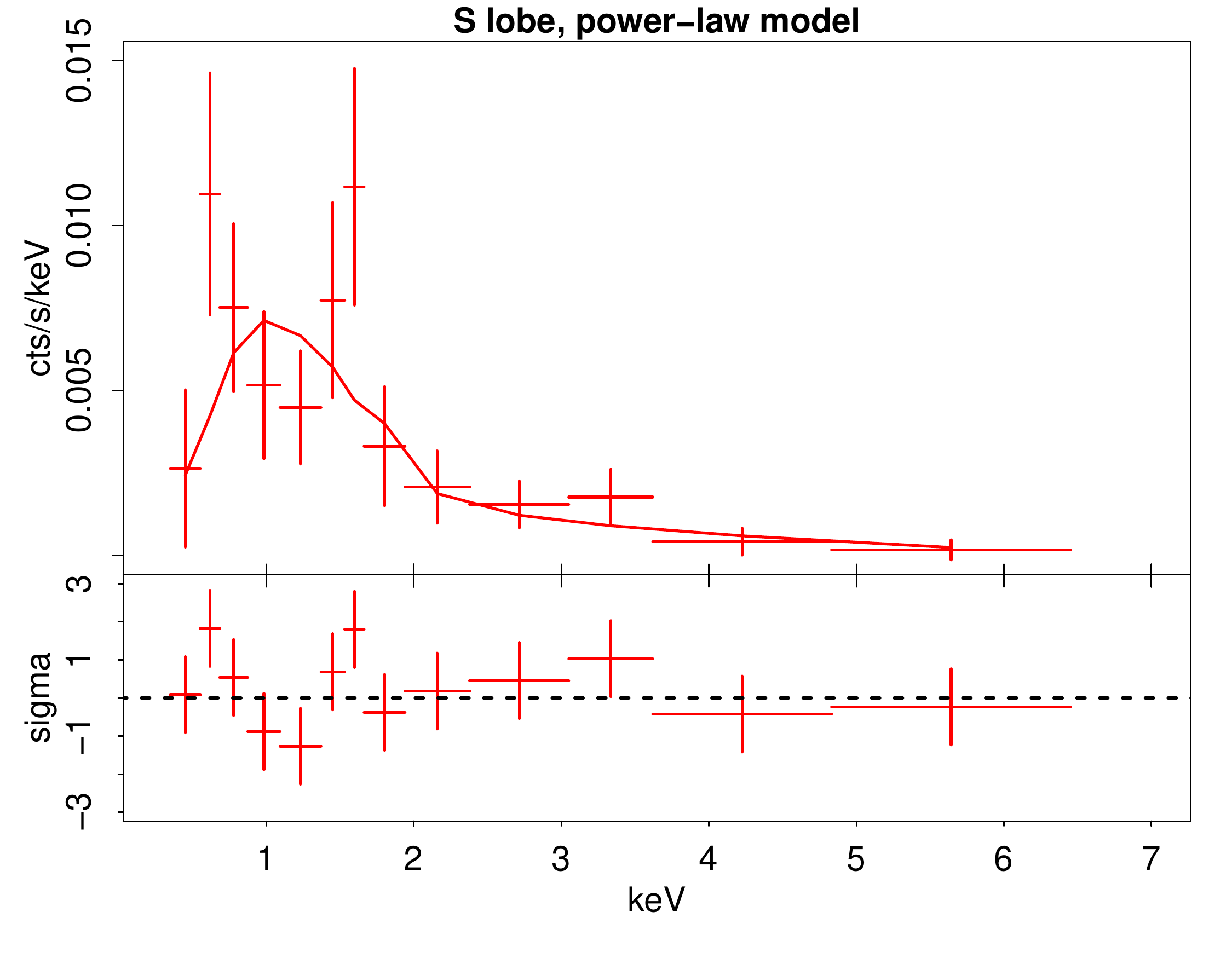}
\includegraphics[scale=0.19]{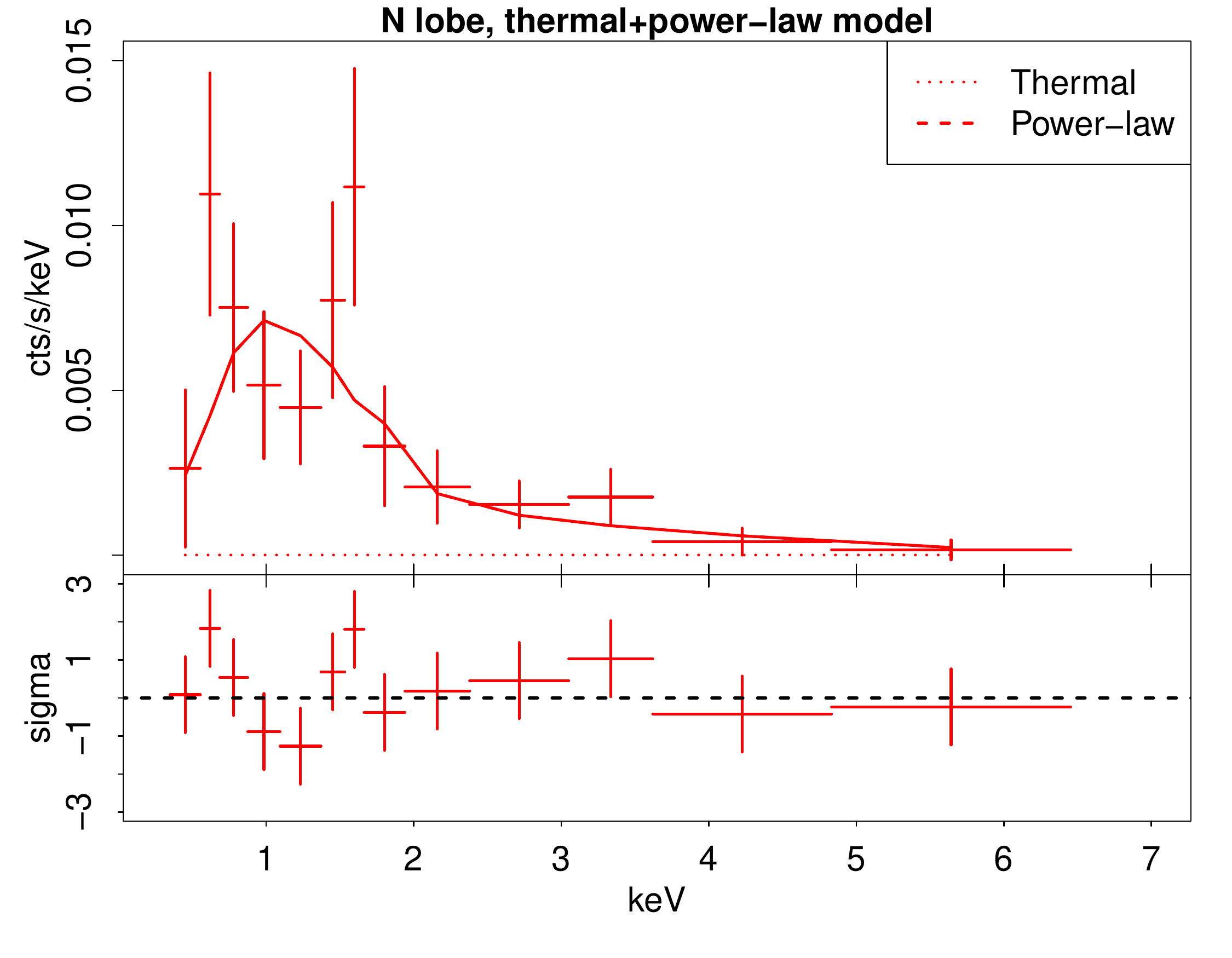}
\caption{Spectral fits in the \(0.3-7\text{ keV}\) band performed with a thermal \textsc{xsapec} (left column), power-law (central column) and mixed thermal+power-law (right column) models. From top to bottom we show spectra extracted in the cross cone, N lobe, and S lobe regions. In upper portion of each panel the full line represents the best fit model and the crosses represent the data points, while in lower portion of each panel is presented the residual distribution. For the thermal+power-law model (right column) the dotted line represents the thermal component, the dashed line represents the power-law component, and the full line represents the total model.}\label{fig:spectra}
\end{figure}

We made use of the VLA L-band map at \(1.4\text{ GHz}\) (upper left panel of Fig. \ref{fig:vla20}) and the GMRT map at \(150\text{ MHz}\) (left panel of Fig. \ref{fig:gmrt}). For both images we measured the flux densities in the regions of N and S lobes by the contour level at \(30\) times the rms level for the GMRT map (see Fig. \ref{fig:gmrt}), obtaining flux densities similar to the literature ones \citep{1995ApJS...99..349N, 1996ApJS..107..175R}.

For the X-ray emission, we made use of the spectra extracted in the N and S lobes presented in Fig. \ref{fig:spectra} convolved with the response function to obtain flux measurements. In particular, since we want to test the non-thermal scenario, for the N lobe we used the power-law component of the mixed model (center right panel of Fig.\ref{fig:spectra}), while for the S lobe we used the power-law model since it is equivalent to the mixed model. 

To model the multi-wavelength emission of 3C 187 we made use of the Jets Spectral Energy Density (SED) modeler and fitting Tool \citep[JetSeT\footnote{\href{https://jetset.readthedocs.io/en/latest/index.html}{https://jetset.readthedocs.io/en/latest/index.html}},][]{2009A&A...501..879T, 2011ApJ...739...66T} to simulate single-zone synchrotron and IC/CMB emissions in the lobes. For both lobes we adopted an electron energy distribution in the form of a power law \(n(\gamma) \propto \gamma^{-p}\) extending between a minimum and maximum energy \(\gamma_{\text{min}}\) and \(\gamma_{\text{max}}\) so that
\begin{equation}\label{eq:electron_dist}
 n_e=\int\limits_{\gamma_{\text{min}}}^{\gamma_{\text{max}}} {n(\gamma)\,d\gamma}\,,
\end{equation}
where \(n_e\) is the electron number density.

We approximated the N lobe with an ellipsoid with semi-axis \(r_l=46\arcsec\) along the radio P.A. and \(r_w=38\arcsec\) across it (the semi axis along the line of sight is taken equal to \(r_w\)) following the radio contours at \(150\text{ MHz}\). Similarly, we approximated the S radio lobe as an ellipsoid with \(r_l=50\arcsec\) and \(r_w=38\arcsec\). This yields an equivalent spherical radius \(R=\sqrt[3]{r_l\,{r_w}^2}\) of \(240\text{ kpc}\) and \(247\text{ kpc}\) for the N and S lobe, respectively, that was fixed during the fit. During the fit we assumed, \(\gamma_{\text{min}}={10}\), \(\gamma_{\text{max}}={10}^5\), a bulk Lorentz factor of \(1\) and a viewing angle of \(90\degree\). The free parameters of the model are therefore the electron energy distribution slope \(p\), the electron density \(n_e\) and the magnetic field in the lobe \(B\).

\begin{figure}
\centering
\includegraphics[scale=0.6]{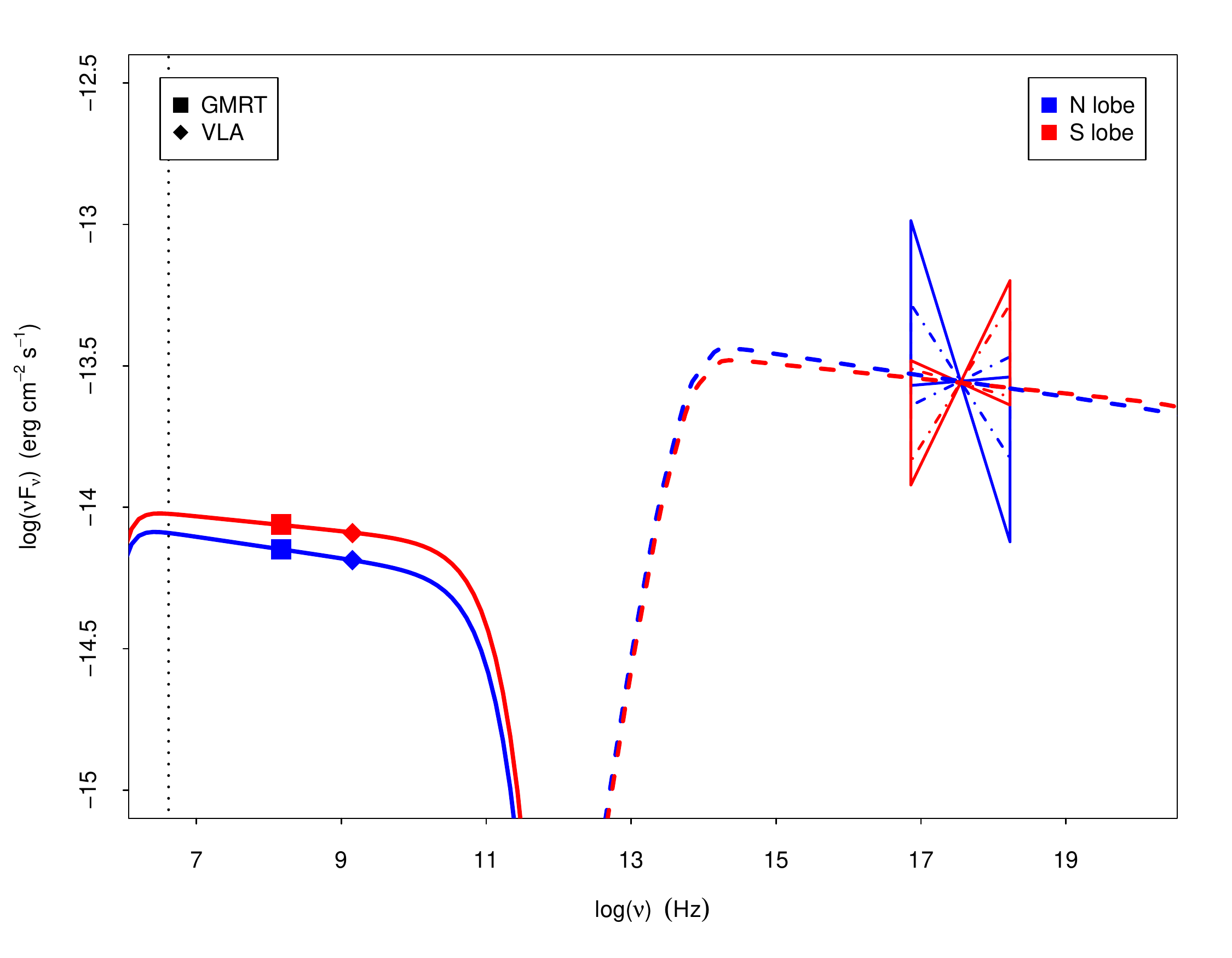}
\caption{Observed SED (\(\nu F_\nu\)) of N (in blue) and S (in red) lobes of 3C 187. Squares represent GMRT fluxes and diamonds represent VLA fluxes. The ``butterflies'' indicate the uncertainties on the spectral indices obtained from the power-law fits in the lobes (see Sect. \ref{sec:seds}). The full line butterflies represent the spectral indices obtained with the background subtraction, while the dot-dashed butterflies represent the spectral indices obtained with the background modeling. The full lines represents the simulation of the synchrotron emission, while the dashed lines represents the IC/CMB emission in the same region. The vertical dotted line marks the frequency (\(\sim 4 \text{ MHz}\)) obtained from Eq. \ref{eq:synch} at which the electrons responsible for the IC/CMB radiation at \(1 \, \text{keV}\) emit synchrotron radiation given the magnetic field values shown in Table \ref{tab:seds}.}\label{fig:seds}
\end{figure}

Results of these simulation for the N and S lobe are presented in Fig. \ref{fig:seds} using the spectral energy distribution (SED, \(\nu F_\nu\)) representation, with the full lines representing the synchrotron emission and the dashed lines representing the IC/CMB component. On the same figure we overplot the radio fluxes at their relative frequency, while for the X-ray fluxes we used ``butterfly'' plots to represent the corresponding spectral indices obtained from the lobe fittings, centered at the geometric average of \(0.3\) and \(7\text{ keV}\). The full line butterflies represent the spectral indices obtained with the background subtraction (see Table \ref{tab:spectra}), while the dot-dashed butterflies represent the spectral indices obtained with the background modeling (see Table \ref{tab:spectra_bkg}). The parameters used for reproducing the observed SED are presented in Table \ref{tab:seds}, together with the calculated total radiative luminosity \(L_{\text{rad}}\) (that is, the combined luminosity from synchrotron and IC/CMB radiation) and with the total electron energy \(E_e = V\, m_e\, c^2 \int\limits_{\gamma_{\text{min}}}^{\gamma_{\text{max}}} {\gamma \, n(\gamma)\,d\gamma}\), where \(V\) is the emitting region volume and \(m_e\) is the electron mass (we note that similar values of \(E_e\) can be evaluated with the method presented in \citealt{2006MNRAS.371...29E}). We also note that the magnetic field values obtained from the fits are similar to the equipartition value of \(\sim 7\, \upmu\text{G}\). In Fig. \ref{fig:seds} we also represent with a vertical dotted line the frequency (\(\sim 4 \text{ MHz}\)) obtained from Eq. \ref{eq:synch} at which the electrons responsible for the IC/CMB radiation at \(1 \text{ keV}\) emit synchrotron radiation given the magnetic field values shown in Table \ref{tab:seds}.

As show in Fig. \ref{fig:seds} the observed X-ray spectral slope in the radio lobes of 3C 187 is compatible (although marginally) with the scenario of the X-ray emission being due to IC/CMB radiation from the same electron population responsible for the synchrotron emission we observe at radio frequencies. 

By comparing total electron energy \(E_e\) and the radiative luminosity \(L_{\text{rad}}\) we can estimate the radiative cooling time \(\tau=E_e/L_{\text{rad}}\) as \(\tau \simeq 2.0\times{10}^{10}\text{ yr}\) and \(\tau \simeq 1.8\times{10}^{10}\text{ yr}\) for the N and S lobe, respectively. The electrons in lobes would therefore loose their energy through radiative emission on spatial scales \(\sim 6\text{ Gpc}\), far larger than the size of the observed radio structures (\(\sim 250\text{ kpc}\)).

In conclusion, the X-ray radiation from the radio lobes of 3C 187 can be explained in terms of non-thermal radiation, namely IC/CMB, with a population of energetic electrons capable of emitting the observed fluxes over tens of Gyr.

\subsection{Thermal Scenario: Hot Gas in The IGM}\label{sec:thermal}

Since the spectral analysis presented in Sect. \ref{sec:spectra} does not rule out the possibility for the diffuse X-ray emission from 3C 187 to originate from IGM hot plasma, in this section we examine the implications of this scenario. To this end, we make use of the spectra presented in Sect. \ref{sec:spectra}, in particular those obtained with the thermal \textsc{xsapec} model, that is, assuming that the observed X-ray emission is entirely due to thermal gas (left column of Fig. \ref{fig:spectra}).

From the normalization of the \textsc{xsapec} models (i.e., their emission measures \(EM\)), we can evaluate the gas proton density \(n_{\text{H}}\) in the lobes. Assuming a uniform particle density in the emitting region we have
\begin{equation}\label{eq:gas_den}
n_{\text{H}} = \sqrt{\frac{{10}^{14}\,EM\,\eta\,4 \pi\, {D_A}^2{\left({1+z}\right)}^2}{V}}\,,
\end{equation}
where \(D_A\) is the angular distance of the source, \(V\) is the emitting region volume, and \(\eta\approx 0.82\) is the ratio of proton to electron density in a fully ionized plasma. Neglecting the electron density, that contributes to the total gas density less than \(\sim 7\times{10}^{-4}\), we can estimate the thermal energy of the gas as \(E_{\text{therm}} = \frac{3\,kT\,n_{\text{H}}\,V}{2\,\mu}\), where \(k\) is the Boltzmann constant, \(T\) is the gas temperature, and \(\mu\approx 0.62 \) is the mean particle weight in units of the proton mass.

Approximating the N and S lobes with the same ellipsoids defined in Sect. \ref{sec:seds} we obtain \(E_{\text{therm,\,N}}={2.5}_{-0.8}^{+1.2}\times{10}^{61}\text{ erg}\) and \(E_{\text{therm,\,S}}={5.3}_{-2.3}^{+20.7}\times{10}^{61}\text{ erg}\). The sound-speed expansion time \(t_c=d\sqrt{\frac{\mu\, m_{\text{p}}}{5/3\, kT}}\), where \(m_{\text{p}}\) is the proton mass, and \(d\) is the distance of the center of the lobes from the source nucleus, evaluated as \(t_{\text{cs, N}} = 0.4\pm 0.1 \text{ Gyr}\) and \(t_{\text{cs, S}} = {0.3}_{-0.1}^{+0.5} \text{ Gyr}\) respectively, yields thermal powers \(L_{\text{therm}}=E_{\text{therm}}/t_{\text{cs}}\) of \(L_{\text{therm,\,N}}={2.1}_{-0.7}^{+1.1} \times {10}^{45}\text{ erg}/\text{s}\) and \(L_{\text{therm,\,S}}={6.3}_{-2.9}^{+27.5} \times {10}^{45}\text{ erg}/\text{s}\), which are \(\sim 30\) and \(\sim 80\) times larger than the X-ray luminosities estimated from the spectral fitting. In fact, we can estimate the X-ray cooling time as \(\tau=E_{\text{therm}}/L_X\) as \(\tau_{\text{N}}=12 \text{ Gyr}\) and \(\tau_{\text{S}}=21 \text{ Gyr}\) for the N and S lobe, respectively, indicating that only a small fraction of the thermal energy contained in the lobes is emitted as X-ray radiation, allowing them to shine over long times. Using the fit results obtained from background spectra modeling in the N lobe region, we obtain similar results of \(E_{\text{therm,\,N}}={2.6}_{-0.6}^{+1.0}\times{10}^{61}\text{ erg}\), \(L_{\text{therm,\,N}}={2.3}_{-0.6}^{+0.9} \times {10}^{45}\text{ erg}/\text{s}\), and \(\tau_{\text{N}}=14 \text{ Gyr}\). Since the background spectra modeling in the S lobe region yields poorly constrained, out of the band temperature, we do not consider it for these estimates.

As an additional test of the thermal origin for the diffuse X-ray emission in the lobes of 3C 187, we check if the gas in these regions happens to be in pressure equilibrium with the surrounding environment, that is, \(P\simeq P_{IGM}\), so that the IGM could contain the gas located in the lobes \citep{2002ApJ...581..948H, 2004ApJ...612..729H}. The thermal pressure of the gas can be estimated as \(P = \frac{n_{\text{H}}\, kT}{\mu}\). Approximating the cross-cone region as a double three dimensional cone, we obtain in this region a gas pressure \(P_{\text{cc}}={1.6}_{-0.6}^{+1.1}\times{10}^{-11}\text{ Ba}\), while in the N and S lobe we obtain \(P_{\text{N}}={1.0}_{-0.3}^{+0.5}\times{10}^{-11}\text{ Ba}\) and \(P_{\text{S}}={1.9}_{-0.8}^{+7.5}\times{10}^{-11}\text{ Ba}\), respectively. Again, considering the fits with background spectra modeling, we similarly obtain \(P_{\text{cc}}={1.3}_{-0.3}^{+0.4}\times{10}^{-11}\text{ Ba}\), while in the N lobe we obtain \(P_{\text{N}}={1.0}_{-0.2}^{+0.4}\times{10}^{-11}\text{ Ba}\). If we assume \(P_{IGM}=P_{\text{cc}}\), the gas thermal pressure in the lobes is compatible within the uncertainties with the pressure of the surrounding gas, indicating that the former would be contained by the latter.

In conclusion, X-ray radiation in the radio lobes of 3C 187 can be interpreted as due to thermal emission from hot gas with temperatures of \(\sim 3 \text{ keV}\) and \(\sim 5 \text{ keV}\) in the N and S lobes, respectively. The cooling times indicate that the thermal gas contained in the radio lobes emits a small fraction of its thermal energy in the form of X-ray radiation, allowing them to shine in over tens of Gyr. In addition, the X-ray emitting thermal gas in radio lobes would be in pressure equilibrium with the surroundings, thus preventing its expansion.

\subsection{Cluster Analysis}\label{sec:cluster}
As previously mentioned, \citet{1988ApJS...66..361H} and \citet{1995ApJS...99..349N} report that 3C 187 appears to be a member of a cluster of \(\sim 30\) objects projected between the radio lobes. In order to check if this source actually resides in a cluster of galaxies that could account for the extended emission we see in the \textit{Chandra} images, we checked for the presence of a ``red sequence'' in the source field (since there is no redshift measurement available for sources in the field of 3C 187, apart from the source itself). The red sequence in the form of a tight clustering of the galaxies in the color space \citep{1977ApJ...216..214V, 1999AAS...195.1202A} provides an efficient method for cluster detection \citep[e. g.][]{2003ApJ...596L.143B, 2005ApJ...623L..85M, 2007MNRAS.374..809D}.

We collected photometrical data in the field of 3C 187 from the Pan-STARRS Data Release 2 \citep[DR2,][]{2018AAS...23143601F}, namely \(g\), \(r\), \(i\), \(z\) and \(y\) magnitudes for all sources within \(2\arcmin\) (\(\sim 700\text{ kpc}\)) from the host galaxy of 3C 187. For this analysis we selected sources detected in all Pan-STARRS filters. In addition, to avoid bright stars, we restricted our selection to sources with magnitudes larger than \(14\), and considered only sources with magnitude errors smaller than \(10\%\), so obtaining a sample of 200 sources. We also collected \(\sim {50}^5\) sources in random positions of the sky, selected in the same way as for the sources in 3C 187 field. We then de-reddened the observed magnitudes of each source using the measurements of dust reddening by \citet{2011ApJ...737..103S} and the extinction model of \citet{2007ApJ...663..320F}.

\begin{figure}
	\centering
	\includegraphics[scale=0.14]{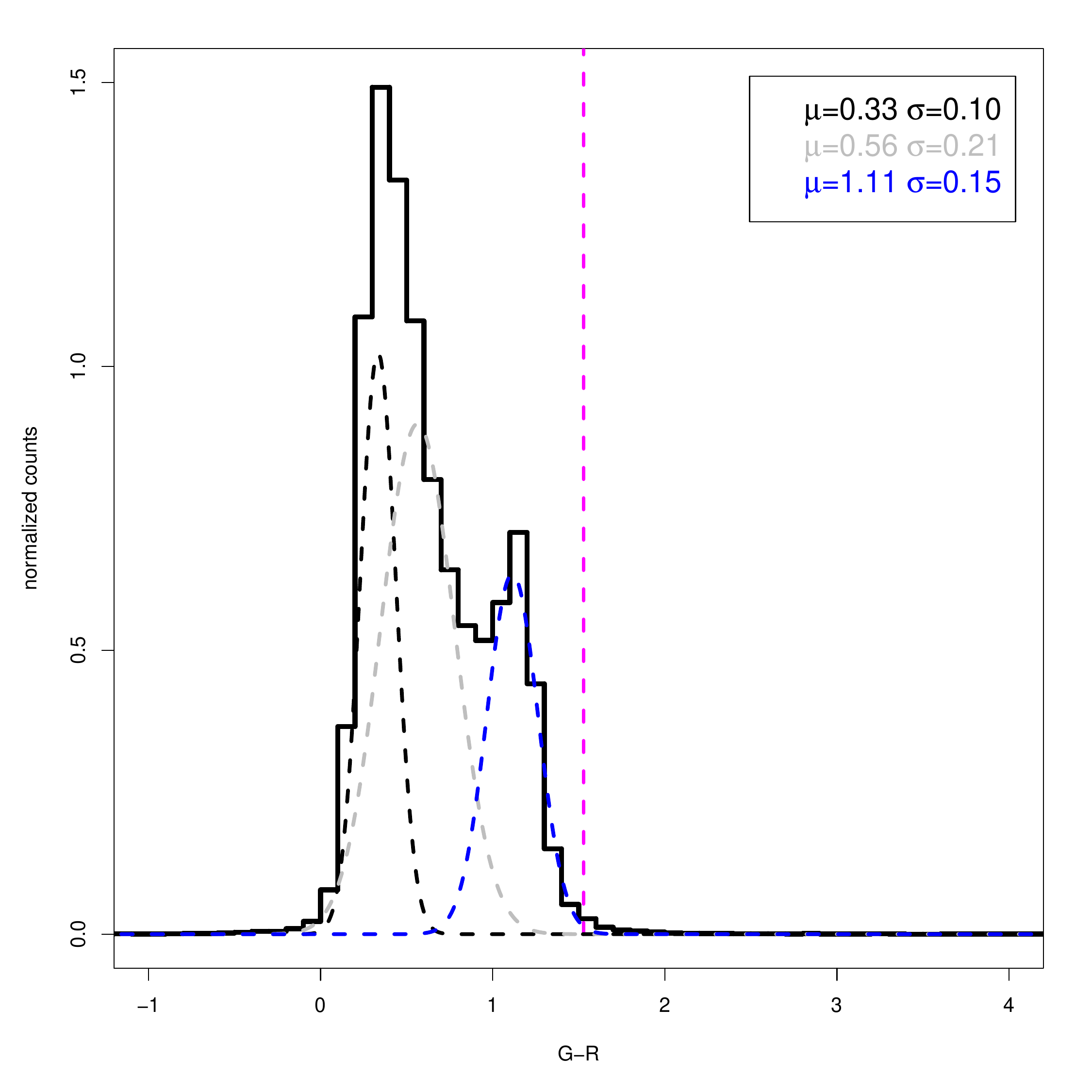}
	\includegraphics[scale=0.14]{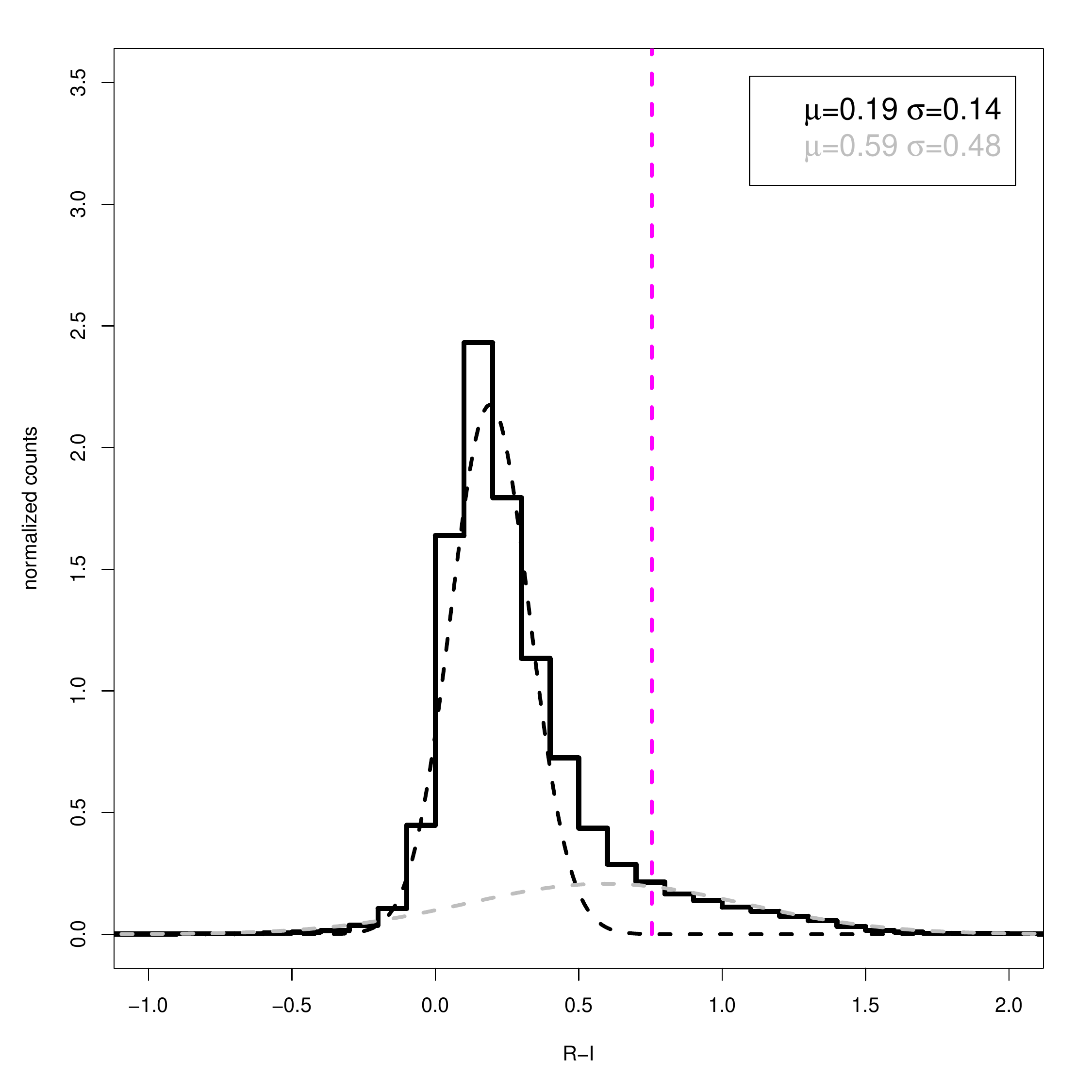}
	\includegraphics[scale=0.14]{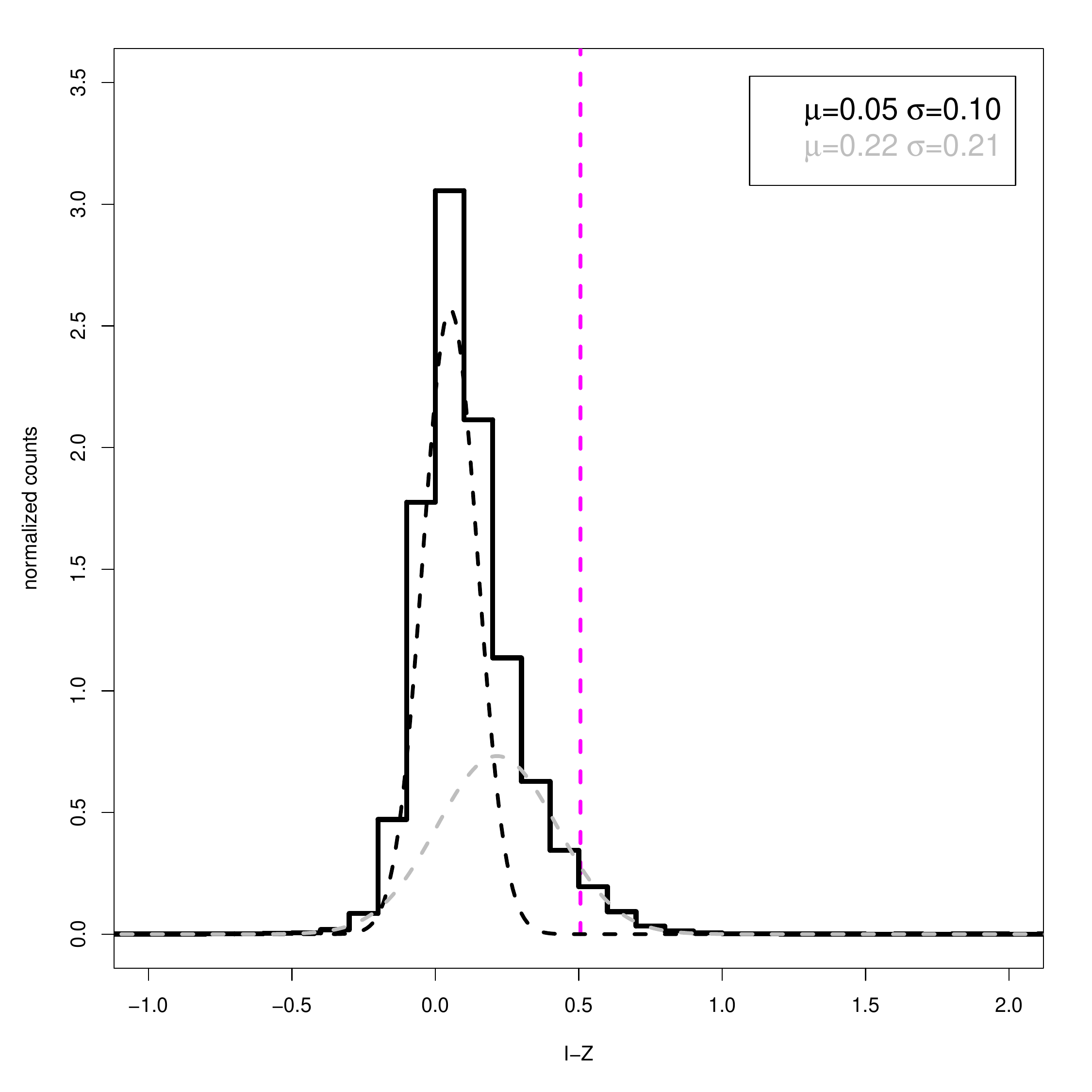}
	\includegraphics[scale=0.14]{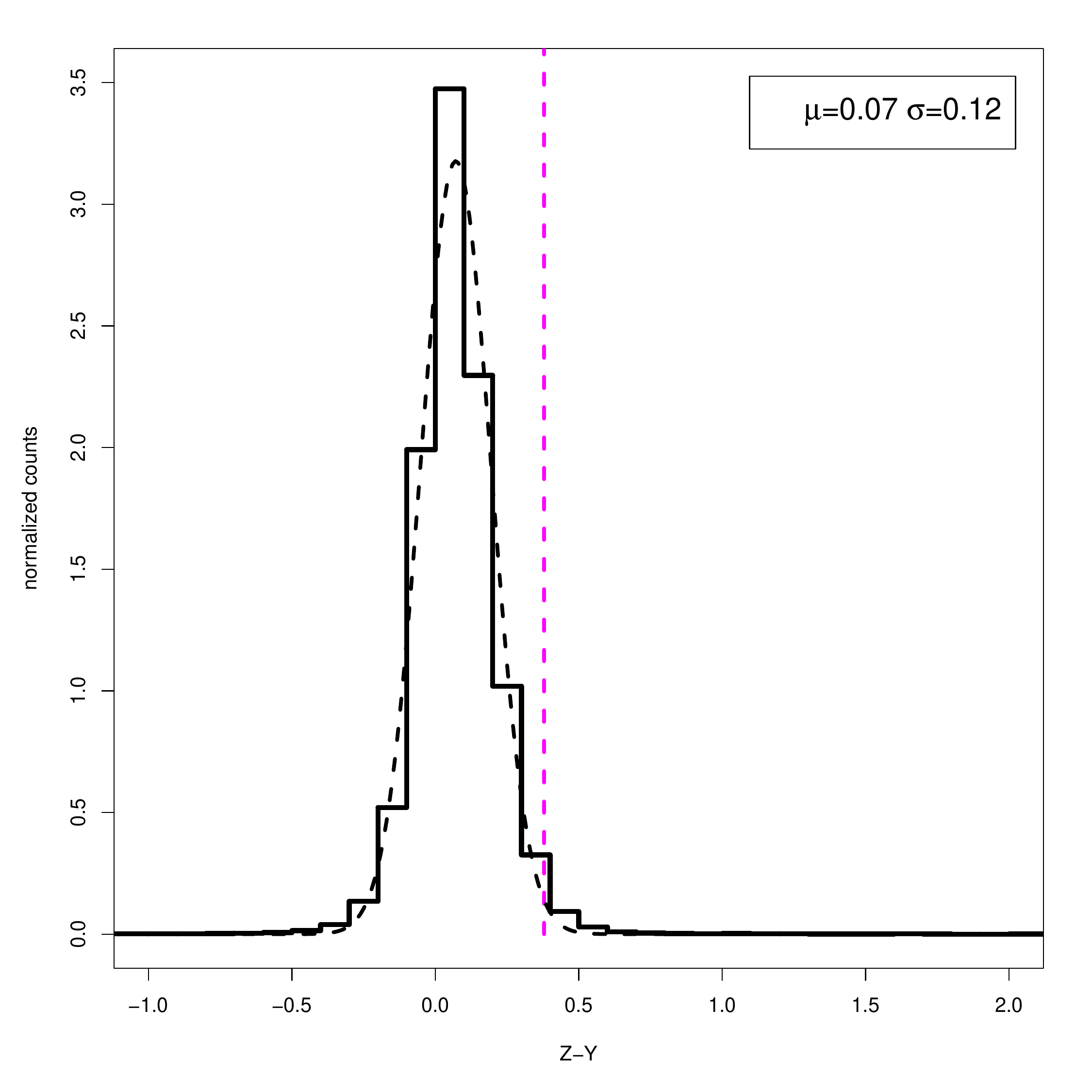}\\
	\includegraphics[scale=0.14]{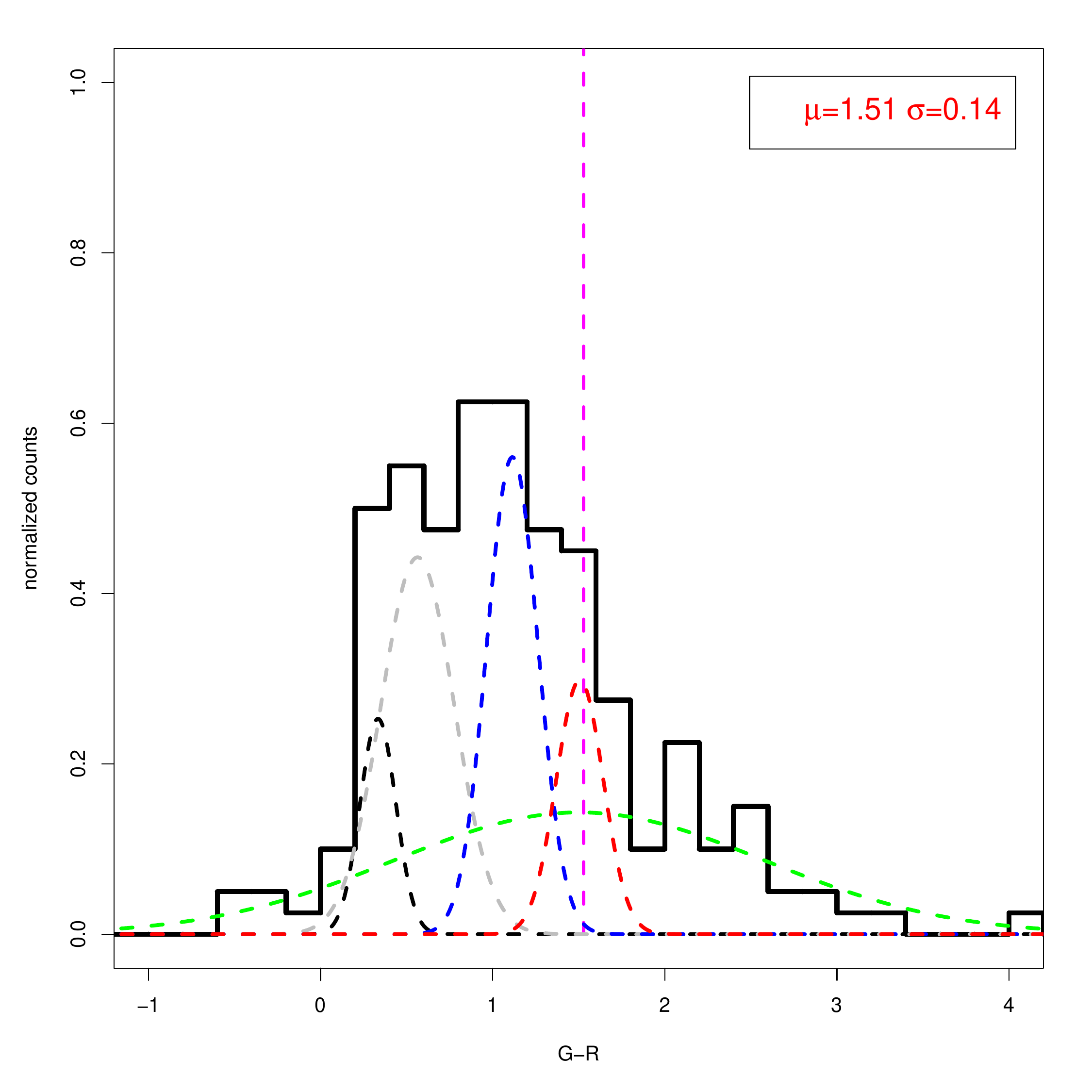}
	\includegraphics[scale=0.14]{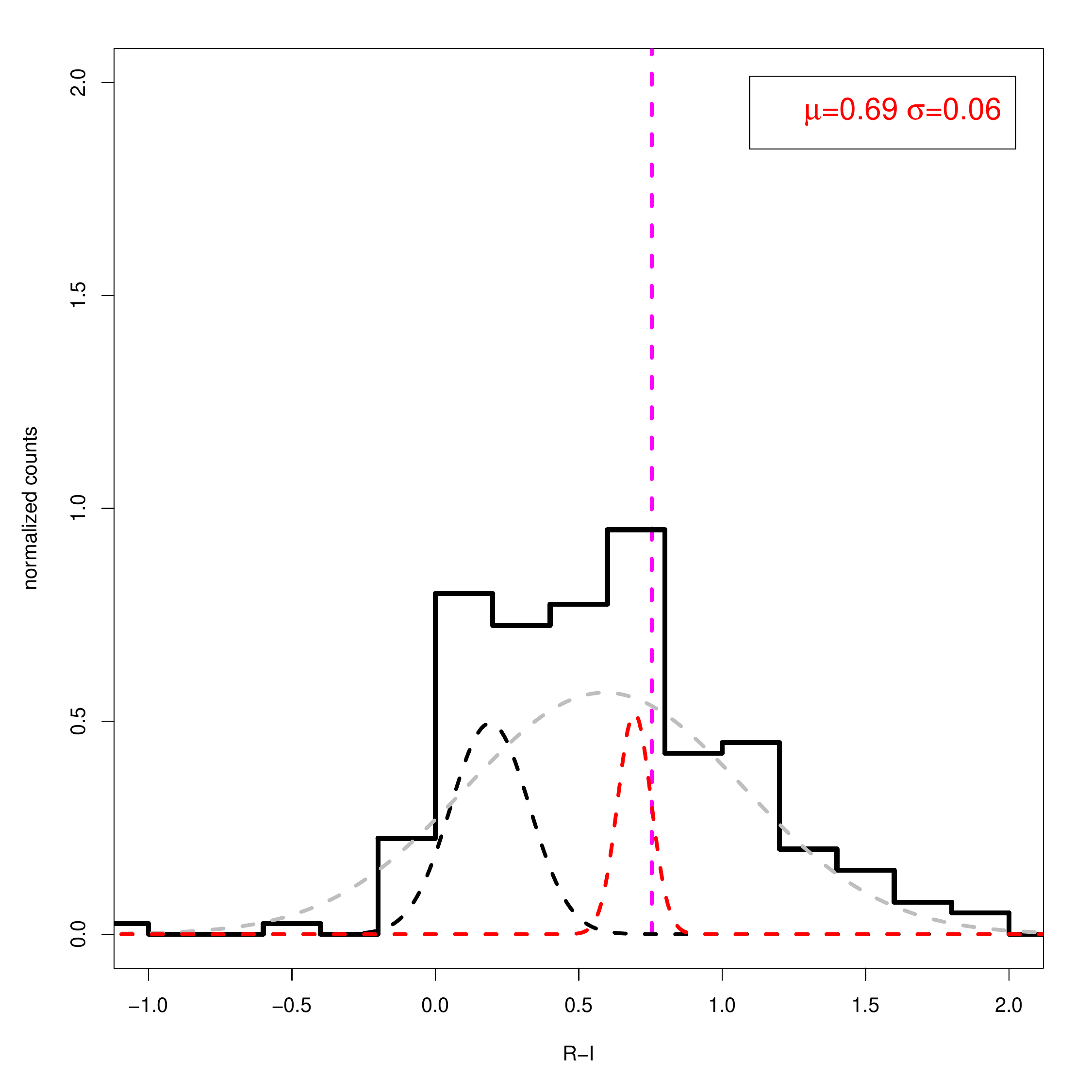}
	\includegraphics[scale=0.14]{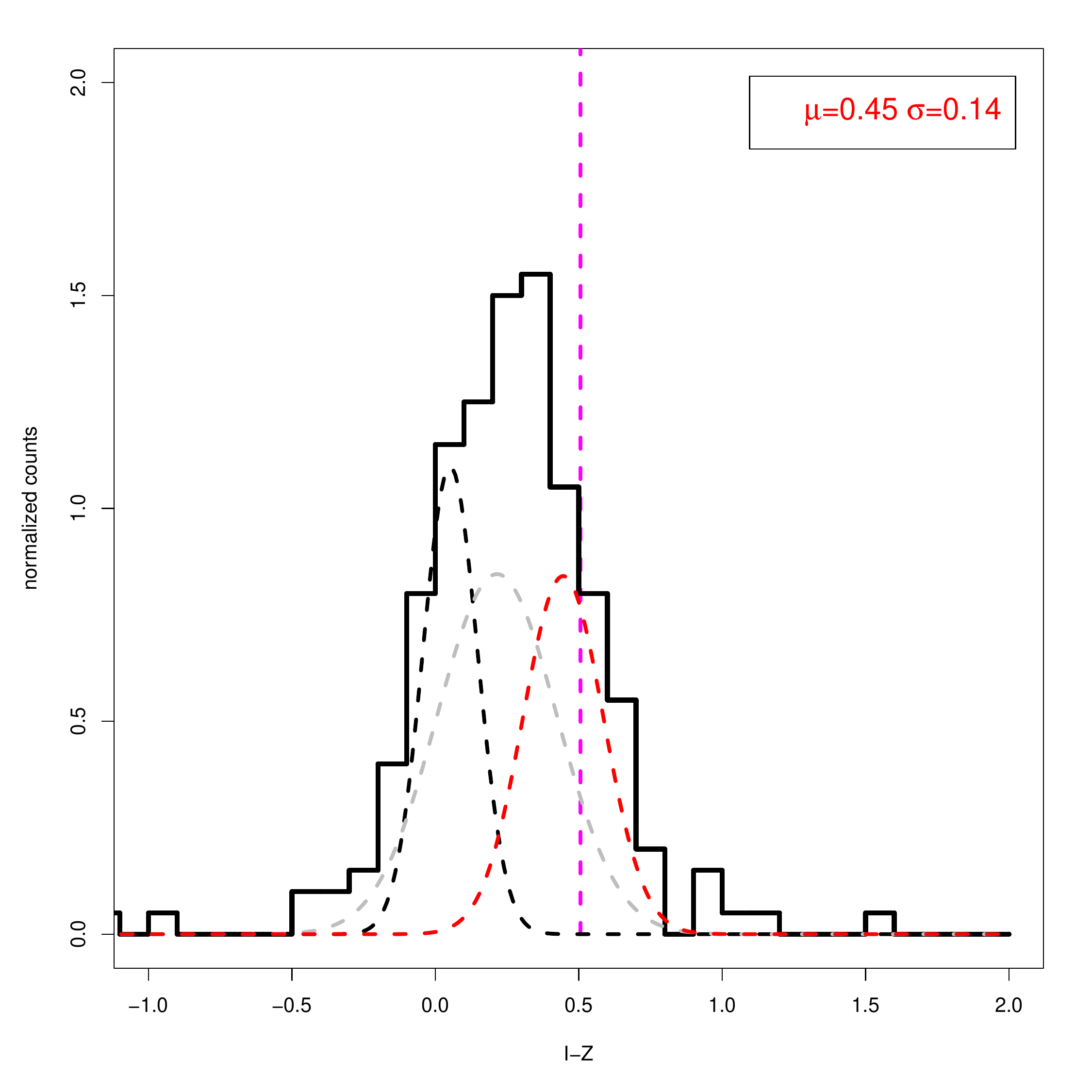}
	\includegraphics[scale=0.14]{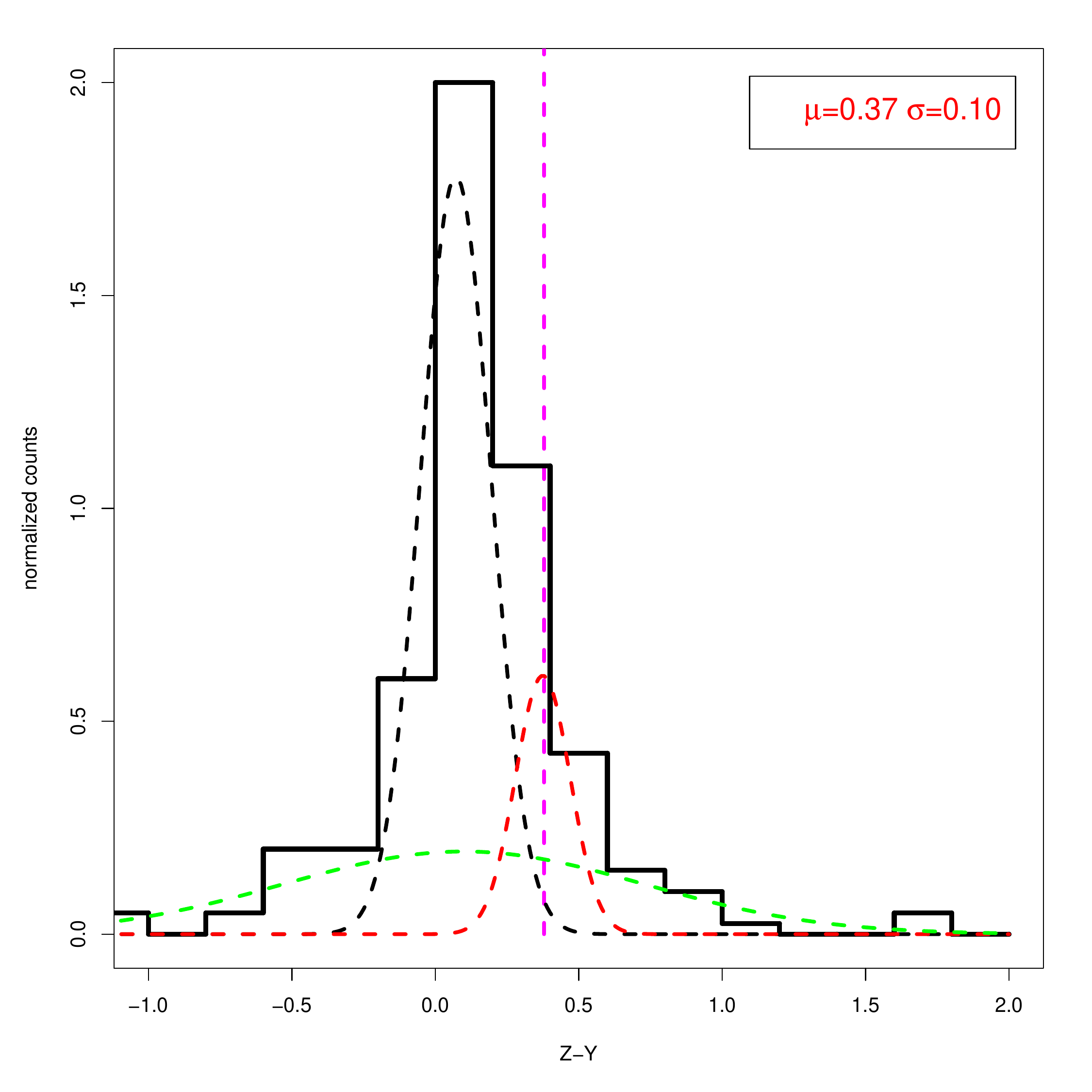}
	\caption{(Upper panels) Normalized distributions of the (de-reddened) colors for random Pan-STARRS sources. The vertical magenta line marks the color of the 3C 187 core counterpart. The best fit gaussian components are overplotted with dashed lines of different colors, with the best fit \(\mu\) and \(\sigma \) indicated in the legend. (Lower panels) Normalized distributions of the (de-reddened) colors for Pan-STARRS sources in the field of 3C 187. As in the upper panels, the vertical magenta line marks the color of the 3C 187 core counterpart. The color distributions are modeled with the same gaussians used to model the random source distribution with fixed  \(\mu\) and \(\sigma\). The additional component required to model the color distributions is indicated with a red dashed line, with the best fit \(\mu\) and \(\sigma \) indicated in the legend. The \(g-r\) and \(z-y\) colors require a further broad, sub-dominant component indicated with a green dashed line.}\label{fig:optical_sequence}
\end{figure}

To verify and confirm the existence of a red sequence in 3C 187 field we followed a procedure similar to those presented in \citet{2009MNRAS.399.1858L} and  \citet{2009ApJ...702..745H}. Following \citet{2012MNRAS.420.1861M}, we investigated the presence of red sequences in \(g-r\), \(r-i\), \(i-z\) and \(z-y\) colors. In the upper panels of Fig. \ref{fig:optical_sequence} we show the normalized distributions of the (de-reddened) colors for random Pan-STARRS sources, with the vertical purple line indicating the color of the 3C 187 core counterpart. We fitted these color distributions with a mixture of Gaussian distributions (using the R package \textsc{normalmixEM}). The best fit models are overplotted with dashed lines, with the best fit parameters (where \(\mu\) is the mean value and \(\sigma \) is the standard deviation of the Gaussian) shown in the legends. We see that the \(g-m\) distribution for the random Pan-STARRS sources is adequately represented by the combination of three narrow (\(\sigma \lesssim 0.2\)) gaussians, while for the \(r-i\) and \(i-z\) distributions we need two gaussians, one narrow and one wide (\(\sigma \sim 0.5\)) for \(r-i\) and two narrow gaussians for \(i-z\). Finally the \(z-y\) distribution is described by a single narrow gaussian. We see that the Pan-STARRS counterpart of 3C 187 core is in general redder than the random sources, and it  is fairly separated from the these in the \(g-r\) and \(z-y\) representation, while it is less separated in \(r-i\) and \(i-z\).

In the lower panels of Fig. \ref{fig:optical_sequence} we show the normalized distributions of the (de-reddened) colors for Pan-STARRS sources found in circle within \(2\) arcmin around the core of 3C 187, again with the vertical purple line indicating the color of the 3C 187 core counterpart. We modeled these distributions with the same gaussians found in the color distributions of the random sources (keeping fixed their \(\mu\) and \(\sigma\)), plus one or two additional gaussian that should describe the (non background/foreground) sources around 3C 187. These are usually described by a narrow gaussian (indicated with a red dashed line) centered around the colors of the 3C 187 core counterpart, while in the case of \(g-r\) and \(z-y\) we also find a broad, sub-dominant component (indicated with a green dashed line).

Following \citet{2009ApJ...702..745H}, to select our bona fide cluster members among the Pan-STARRS sources in the field of 3C 187 we considered the sources lying between two \(\sigma\) around the mean value of the narrow (red) component. In addition, to exclude contamination for background/foreground sources (that can be significant in \(r-i\) and \(i-z\) colors) we excluded the sources up to one \(\sigma\) from the mean value of the redder component of the random sources. As expected, this resulted in no selected sources for the \(r-i\) color, while we were able to select \(40\), \(42\) and \(60\) sources for \(g-r\), \(i-z\) and \(z-y\), respectively, including the counterpart of 3C 187 core. Since we do not assume that 3C 187 is necessarily the brightest source in the group, we further selected sources up to one magnitude fainter than the Pan-STARRS counterpart of the 3C 187 core, resulting in a final sample of
\(39\), \(41\) and \(57\) sources for \(g-r\), \(i-z\) and \(z-y\) color, respectively. A Kolmogorov–Smirnov test \citep{1933kol,1939smir} shows that the distributions of the Pan-STARRS colors cluster member candidates have p-chances \(<{10}^{-15}\) of having been randomly sampled from the population of random sources.

\begin{figure}
	\centering
	\includegraphics[scale=0.19]{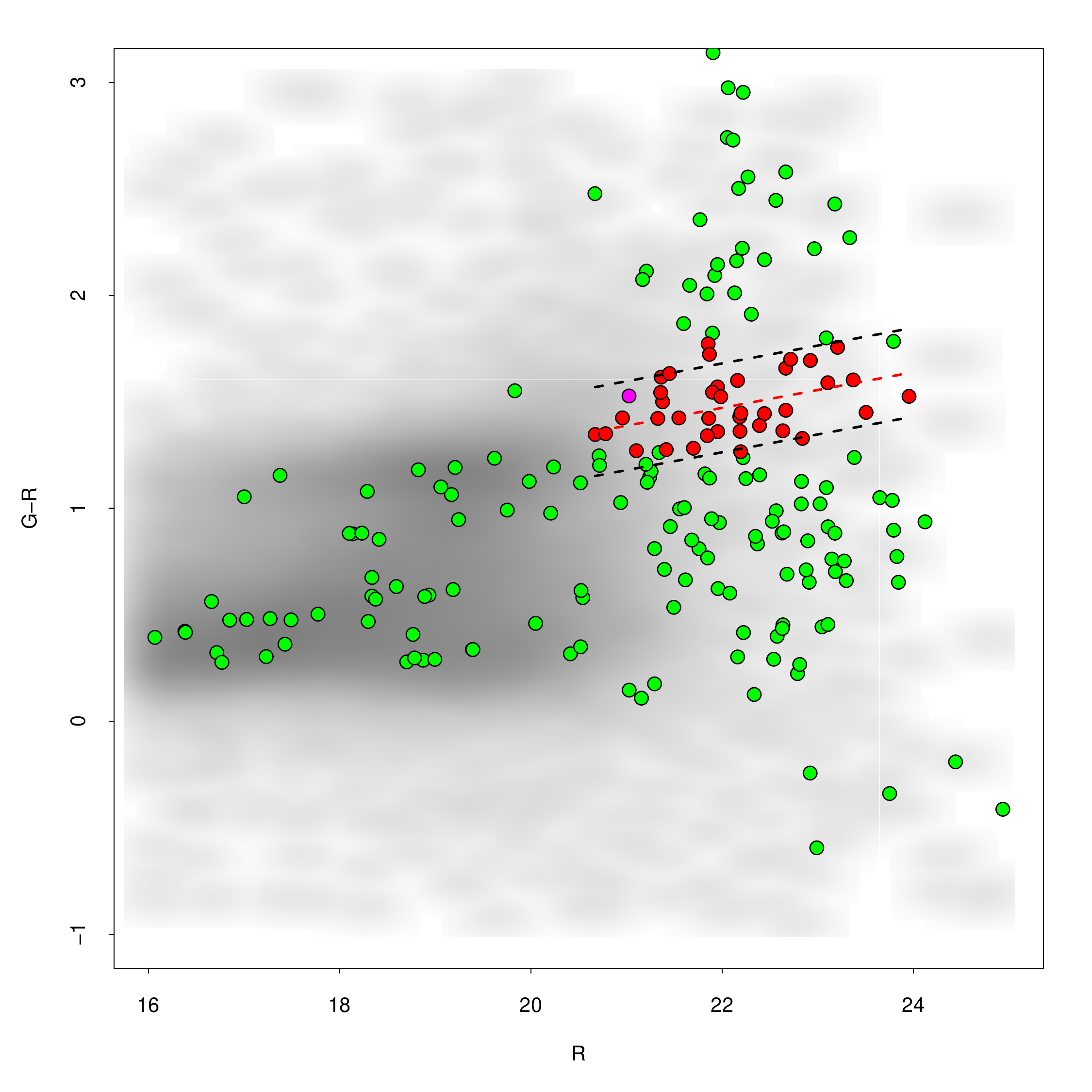}
	\includegraphics[scale=0.19]{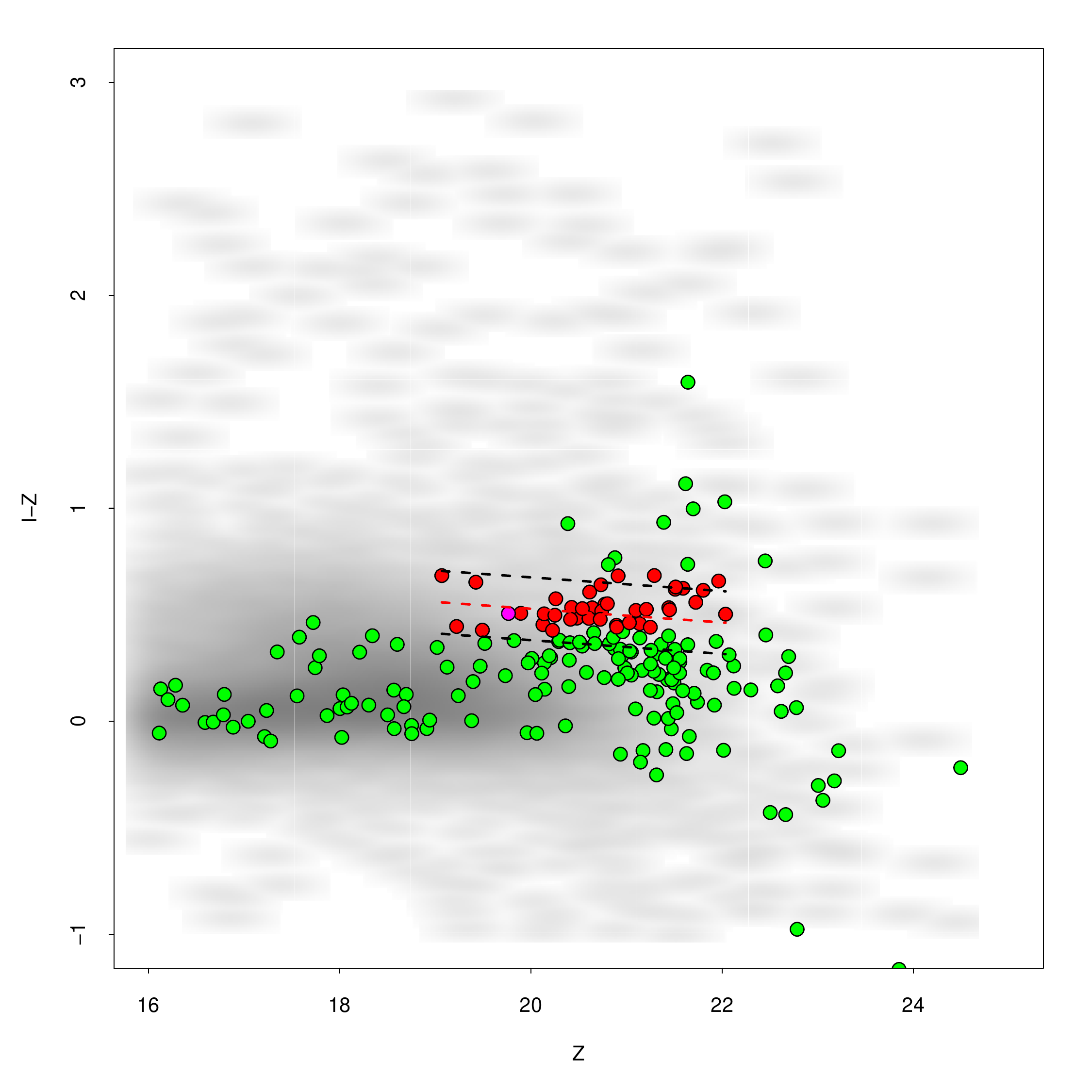}
	\includegraphics[scale=0.19]{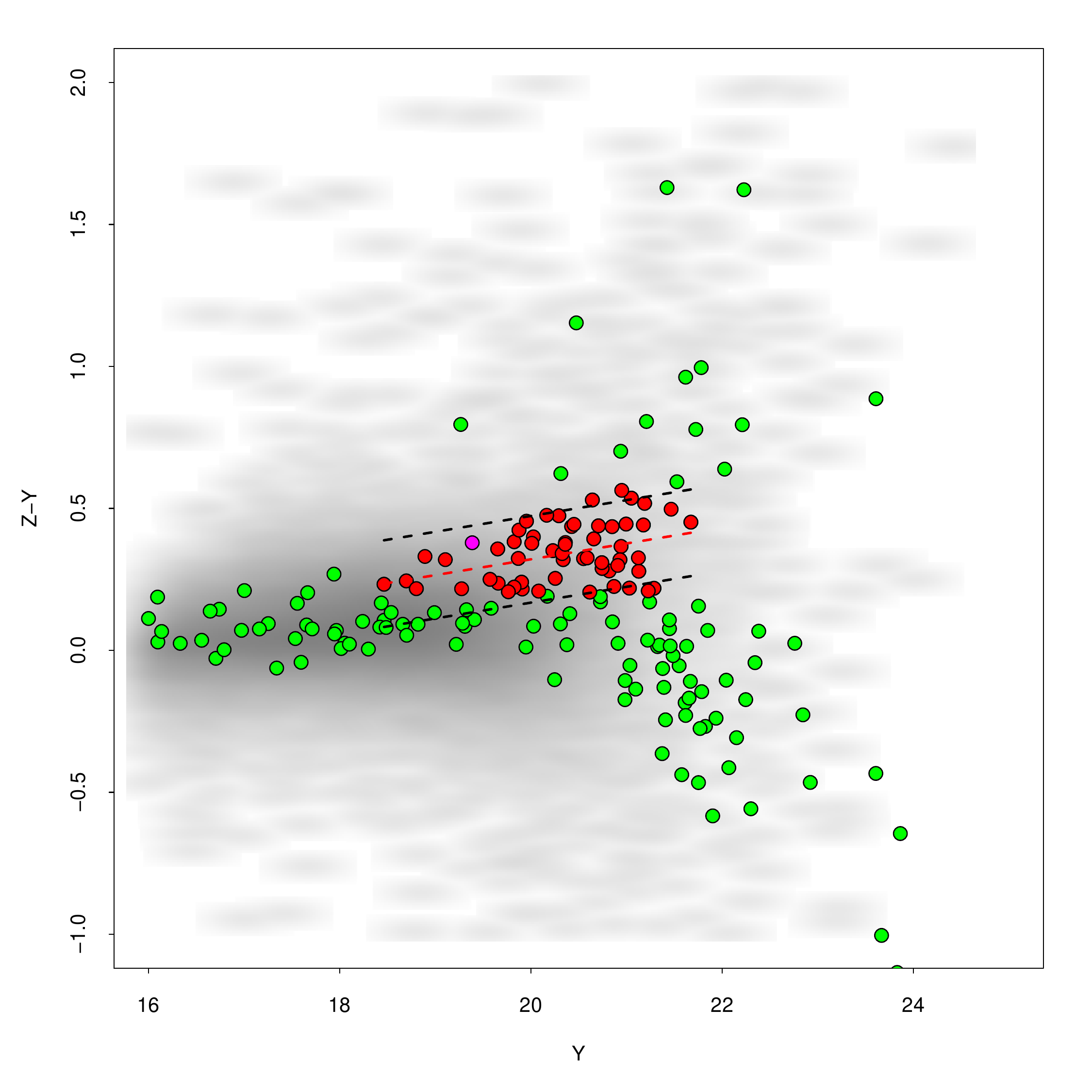}
	\caption{Pan-STARRS sources in the field of 3C 187 presented in the color-magnitude space. The cluster member candidates are reported with red circles, the other field sources are represented with green circles, while sources from random positions in the sky are plotted in the background with gray dots. The counterpart of 3C 187 core is represented with a magenta circle. The red dashed lines show a linear fit to the selected cluster member candidates, while \(90 \%\) scatters are represented with black dashed lines.}\label{fig:optical_cluster}
\end{figure}

In Fig. \ref{fig:optical_cluster} we show the Pan-STARRS sources in 3C 187 field in the color-magnitude space \(g-r\) vs \(r\), \(i-z\) vs \(z\), \(z-y\) vs \(y\) in the left, central, and right panel, respectively. The cluster member candidates are reported with red circles, the other field sources are represented with green circles, while sources from random positions in the sky are plotted in the background with gray dots. The counterpart of 3C 187 core is represented with a magenta circle. The red dashed lines show a linear fit to the cluster member candidates, weighted with the inverse square of the color errors. We obtained best fit slopes of \(0.08\pm 0.03\), \(-0.03\pm 0.02\) and \(0.06\pm 0.01\) and \(90 \%\) scatters of \(0.21\), \(0.15\) and \(0.15\) for the \(g-r\), \(i-z\) and \(z-y\) colors, respectively, represented with black dashed lines.

\begin{figure}
	\centering
	\includegraphics[scale=0.23]{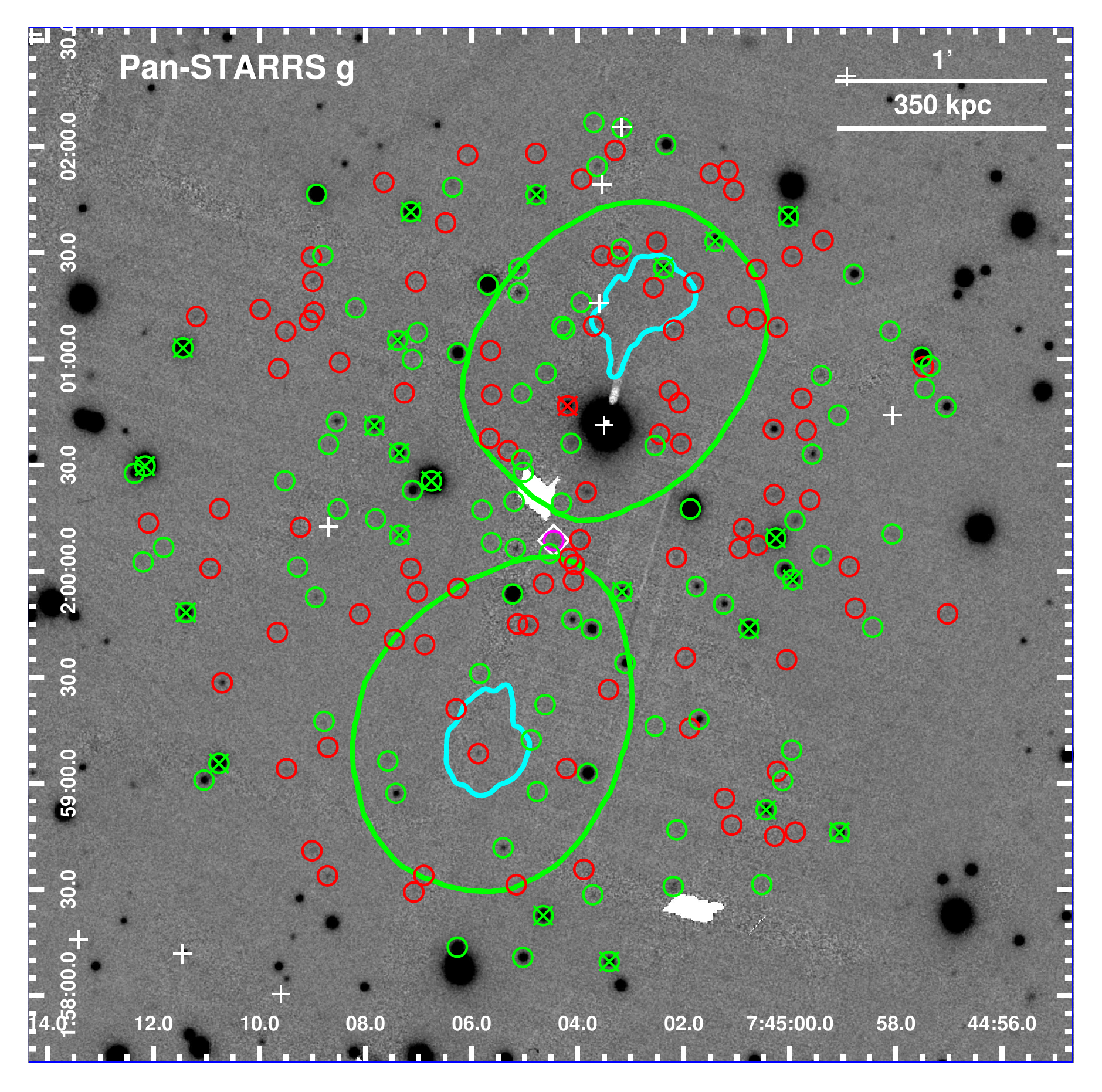}
	\includegraphics[scale=0.23]{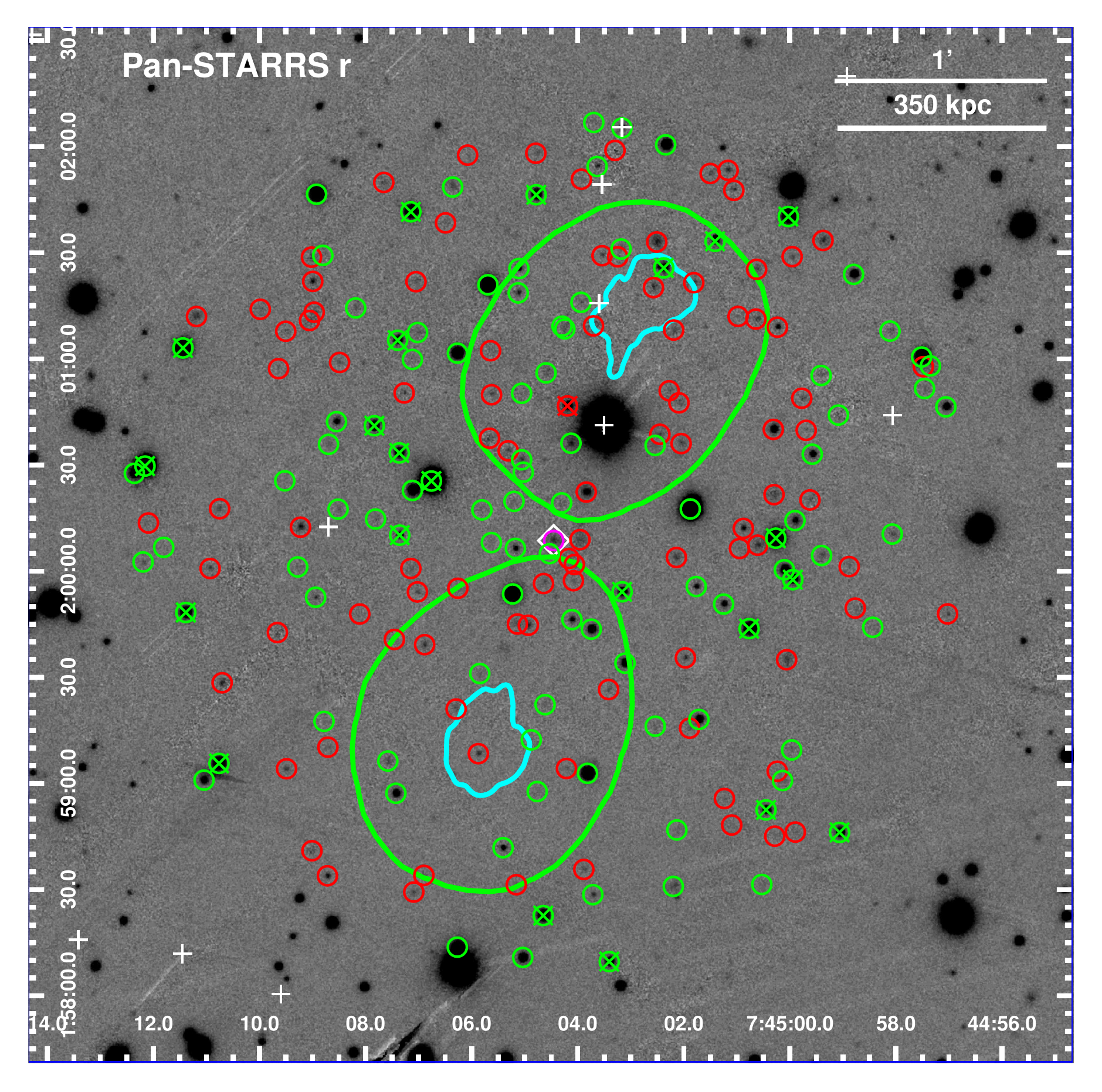}
	\includegraphics[scale=0.23]{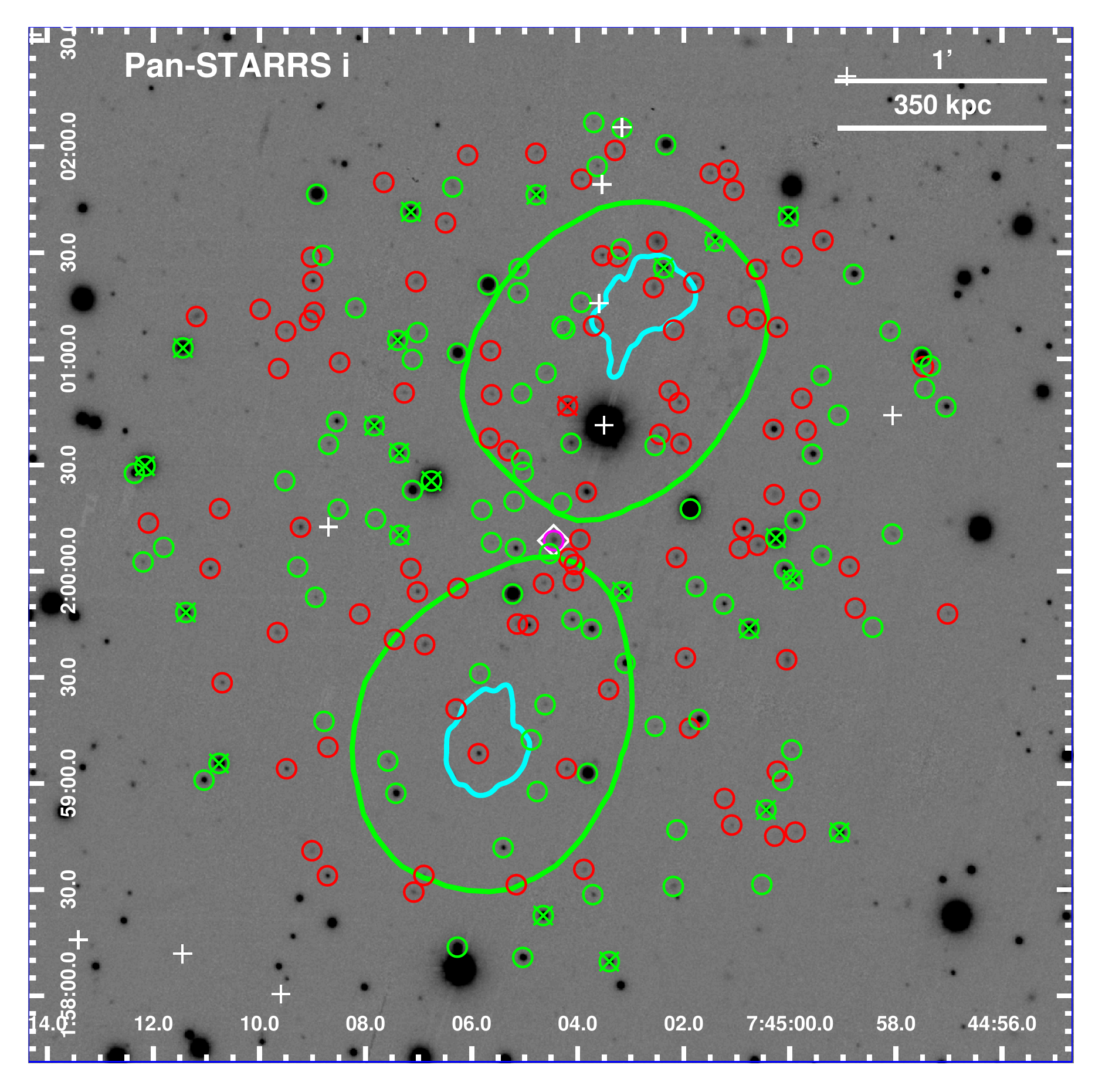}
	\includegraphics[scale=0.23]{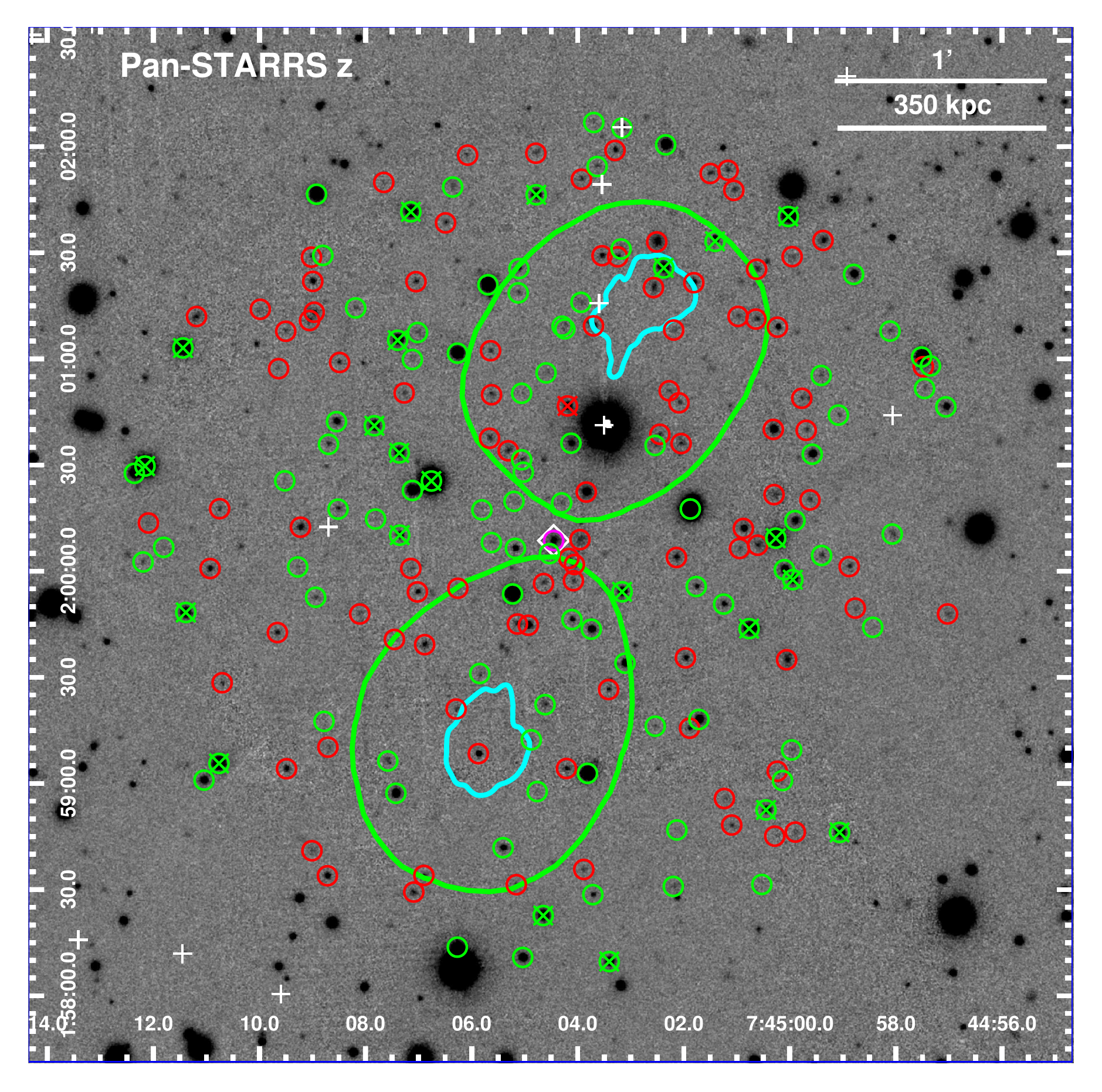}
	\includegraphics[scale=0.23]{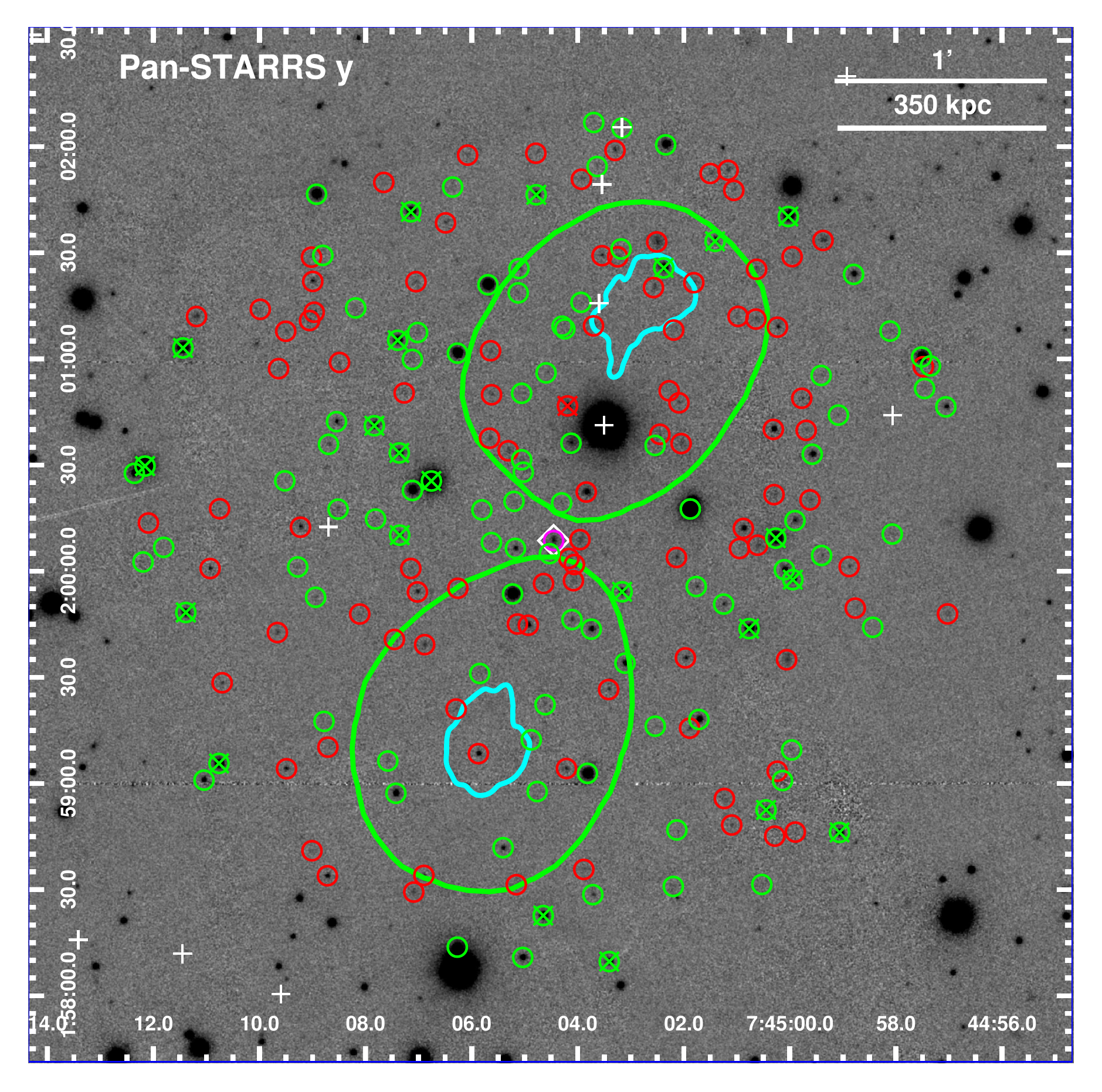}
	\caption{Pan-STARRS images of 3C 187 in the g (upper left panel), r (upper central panel), i (upper right panel), z (lower left panel) and y (lower right panel) filter with superimposed in cyan the VLA L-band contours and in green the GMRT 150 MHz contours. The red circles indicate the cluster member candidates selected in at least one color, while the green circles mark the field sources not selected as cluster member candidates in any color. The magenta circle indicates the counterpart to 3C 187 core, and the white diamond indicates the location of the optical identification by \citet{1995A&A...299...17H} for the 3C 187 nucleus. White crosses represent X-ray point sources detected in the \(0.3-7\text{ keV}\) \textit{Chandra} image. The xs indicate the sources for which we obtained optical spectra with Victor Blanco telescope. In particular, the green xs indicate sources with stellar spectra and the red x indicates the source with a galactic spectrum with redshift \(0.4662\) (see Fig. \ref{fig:spectrum}).}\label{fig:optical_cluster_coord}
\end{figure}

In Fig. \ref{fig:optical_cluster_coord} we show the locations of the Pan-STARRS sources selected as cluster member candidates in celestial coordinates, on top of the Pan-STARRS images of 3C 187 in different filters. In the same figure we superimpose in cyan the VLA L-band contours and in green the GMRT 150 MHz contours from Fig. \ref{fig:cones}. The red circles indicate the cluster member candidates selected in at least one color, while the green circles mark the field sources not selected as cluster member candidates selected in any color (\textit{field sources} hereinafter). The magenta circle indicates the counterpart to 3C 187 core, and the white diamond indicates the location of the optical identification by \citet{1995A&A...299...17H}. White crosses represent X-ray point sources detected in the \(0.3-7\text{ keV}\) \textit{Chandra} image.

From this image we can clearly see that the cluster member candidates are quite dim in the optical, with the core of 3C 187 having Pan-STARRS magnitude \(r=21.0\). On the other hand, the brightest optical source in the field is located at \(\text{R.A.}=07\text{h}45\text{m}03\text{s}.482\), \(\text{DEC}=+02\degree 00\arcmin41\farcs15\) with Pan-STARRS magnitude \(r=12.1\), coincident with the 2MASS source J07450348+0200411 and with an X-ray point source detected in \(0.3-7\text{ keV}\) \textit{Chandra} image with a flux of \({0.99}_{-0.33}^{+0.42}\times{10}^{-14} \text{ erg}/\text{cm}^2/\text{s}\), the brightest X-ray source detected in the central \(2\arcmin\) around 3C 187.

To investigate the nature of the sources in the field of 3C 187 and verify if they are gravitationally bound we performed optical spectroscopic observations of sources in the field of 3CR 187 adopting a similar procedure to that carried out for 3CR 17 \citep{madrid18}. We acquired spectral images at Victor Blanco 4m telescope in Cerro Tololo, Chile on March 10-11 2020, in remote mode. We used the COSMOS spectrograph that has a field of view of \(10\arcmin\) , and a scale of \(0\farcs 29\) per pixel. We made use of Multi-Object Slit mode, using the Red VPH Grism (r2k) for three masks with \(1\farcs 2\) slit width per source. We integrated each mask three times, with \(1200 \text{ s}\) per exposition. We acquired Hg-Ne-Ar comparison lamp spectra on each target position for the wavelength calibration. This instrumental setup gave us a spectral coverage of \(\sim 5500-9600\, \AA\) and dispersion of \(\sim 3 \AA\text{/pixel}\).  Additionally, we observed a spectro-photometric standard star for the relative flux calibration \citep[see also][]{2015AJ....149..160R, 2015AJ....149..163L, 2016Ap&SS.361..337M}. Each spectrum was reduced and extracted with standard IRAF procedures \citep{1986SPIE..627..733T, 1993ASPC...52..173T}. Unfortunately, due to bright night conditions we were able to collect spectra with a signal-to-noise ratio high enough to classify them and obtain a fair redshift estimate only for 23 bright sources in the field of 3C 187, indicated in Fig. \ref{fig:optical_cluster_coord} with xs.

\begin{figure}
	\centering
	\includegraphics[scale=0.5]{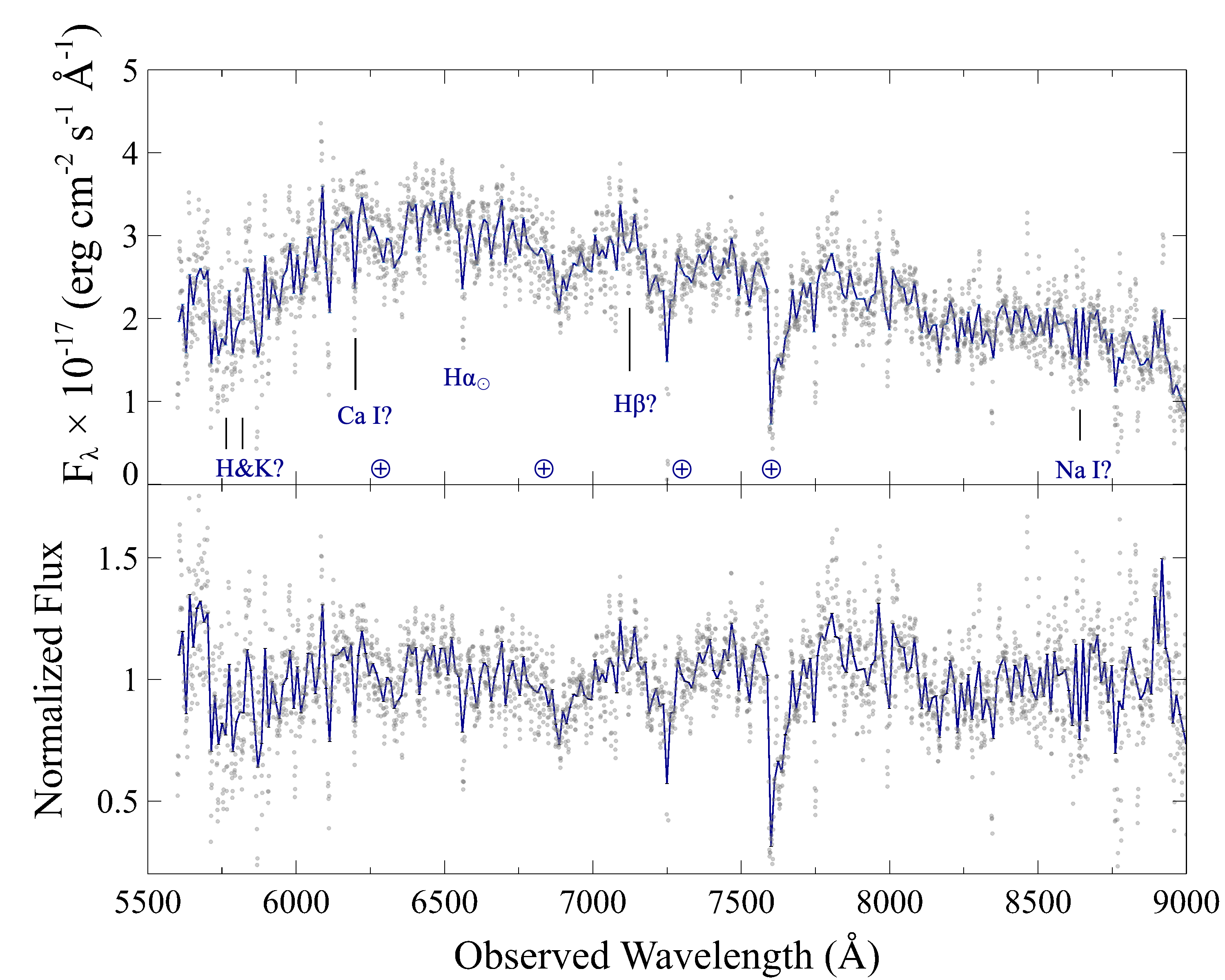}
	\caption{The spectrum of the candidate member for the cluster of 3C 187. Original data points are represented with gray background circles, while the blue line represents the best fit smoothed model. Some spectral features, H, K, Ca I, Hb, and Na I, are potentially identified at redshift \(0.466\). The \(\text{H}\upalpha_\odot\) was also detected as we observed the source during a bright night.}\label{fig:spectrum}
\end{figure}

In particular we were able to collect spectra for 22 field sources and one cluster member candidate, represented in the figure with green and red xs, respectively. All the field sources showed stellar-like spectra, while the cluster member candidate features a galactic spectrum (shown in Fig. \ref{fig:spectrum}) that allowed us to estimate a redshift of \(0.466\) based on a tentative identification of H, K, Ca I, Hb, and Na I lines. Basing on the spectral dispersion of \(\sim 3 \AA\text{/pixel}\), and considering a minimum of two pixels as the minimum resolution for a line, we estimate a redshift uncertainty of \({10}^{-3}\).
The redshift estimate of \(0.466\pm 0.001\) for this source is therefore compatible with the redshift of \(0.465\) reported by \citet{1995A&A...299...17H} for 3C 187 core, considering a maximum redshift separation \(\Delta z = 0.005\) (i.e., \(\sim 1500 \text{ km/s}\)) corresponding to the maximum velocity dispersion in groups and clusters of galaxies \citep[see, e.g.,][]{1993MNRAS.261..827M, 2004MNRAS.348..866E, 2006ApJS..167....1B}

In conclusion, on the basis of Pan-STARRS optical data, the core of 3C 187 belongs to a red-sequence of sources whose color distribution is significantly different from background sources. Optical spectroscopic observations of one cluster member candidate confirm that it is lying at the same redshift of 3C 187 core, while other field sources do not belong to the putative cluster being stars. Additional spectroscopic observations aimed at estimating the redshift of cluster member candidates are needed to reveal if they actually form a gravitationally bound cluster.

\section{Summary and Conclusions}\label{sec:summary}

We performed an imaging and spectral analysis of \textit{Chandra} data obtained for the radio galaxy 3C 187 in the context of the snapshot program for the 3C catalog \citep{2013ApJS..206....7M}. In addition to the findings of \citealt{2013ApJS..206....7M} that reported a significant X-ray emission from both the northern and the southern lobes, we preformed an analysis of the surface flux profiles in the N, S and cross cone directions, extracted X-ray spectra both in the cross-cone and in the radio lobe regions, performed a multi-wavelength analysis investigating both the thermal and non-thermal scenarios for the origin of the diffuse X-ray emission, and investigated the possibility for 3C 187 to lie in a cluster of galaxies. The main results of our analysis are:

\begin{enumerate}
\item Confirming \citet{2013ApJS..206....7M} results, we find extended X-ray emission around 3C 187 in all directions. In particular, in the soft \(0.3-3 \text{ keV}\) band the emission is extended \(\sim 850\text{ kpc}\) along the radio lobe direction with a signal to noise ratio of at least 3, while this extension reaches \(\sim 530\text{ kpc}\) in the cross cone direction. The surface flux profiles in the different directions are generally compatible with a beta model, although enhanced emission appears in correspondence with the radio lobes, more prominently in the N one. The best fit \(\beta\) values are \(0.8\), \(0.7\), \(0.7\) and \(0.8\) in the W, E, N and S cone, respectively, similar to those observed in groups and clusters of galaxies \citep[see e.g.][]{2000MNRAS.318..715X,2000MNRAS.318.1041E}.

\item The quality of the available spectral data does not allow to distinguish between a non-thermal scenario - in which the X-ray emission is of non-thermal origin with a spectrum described by a power-law - and a thermal scenario - in which the X-ray emission is of thermal origin with a spectrum described by a thermal plasma model. The two spectral models fit the X-ray spectrum extracted in the cross-cones and N lobe with similar reduced \(\chi^2\) values and residual distributions. In the S lobe, however, the thermal model yields a poorly constrained temperature with a reduced \(\chi^2\simeq 1.2\), while for the power-law model we have \(\chi^2\simeq 1.0\). Spectral fitting with a double (power-law+thermal) model seems to indicate a dominating thermal emission in the cross-cone region and a dominating non-thermal emission in the lobe regions, with photon indices similar to those observed in other radio galaxies \citep{2005ApJ...626..733C}. Similar results are obtained through modeling of the background spectra.

\item In the non-thermal scenario, we found through a combined radio-X-ray analysis that the X-ray spectral indices obtained in the lobes are compatible with IC/CMB emission from energetic electrons in these regions capable of emitting the observed fluxes over tens of Gyr.

\item In the thermal scenario X-ray radiation from the cross-cone region of 3C 187 can be interpreted as due to thermal emission from hot gas with a temperature \(\sim 4 \text{ keV}\), while in the N and S cones we find temperatures \(\sim 3 \text{ keV}\) and \(\sim 5 \text{ keV}\), respectively, with cooling times that allow the gas to shine in over tens of Gyr and in pressure equilibrium with the surroundings.

\item Using Pan-STARRS optical data we found that 3C 187 core belongs to a red-sequence of \(\sim 40\) optical sources in the field. The color distributions of these sources are significantly different from background sources, meaning that 3C 187 could reside in a cluster of galaxies, while not being the brightest one. If this is the case, the diffuse X-ray emission observed around the source could be in part due to thermal radiation from the IGM of such cluster. To confirm this result, optical spectroscopy is needed in order to robustly identify cluster members.
\end{enumerate}

To sum up, the present data yield a complex scenario for 3C 187. The diffuse X-ray emission around the source appears elongated along the radio axis, and enhanced in correspondence of the radio lobes, indicating a morphological connection between the emission in the two energy bands. The spectral analysis does not allow to rule out either the thermal or the non-thermal scenario, although the former seems favored in the cross-cone region and the latter appears dominant in the radio lobe regions. The X-ray spectral indices in the radio lobes are compatible with the IC/CMB scenario, while a thermal gas in these regions would be able to emit over tens of Gyr and in pressure equilibrium with the surroundings. We have some indications that 3C 187 may belong to cluster of galaxies, whose IGM could contribute to the X-ray emission observed around the source. However, deeper X-ray and optical spectroscopic observations are needed to get a more clear picture of this enigmatic source.

\begin{acknowledgements}
We thank the anonymous referee for their useful comments and suggestions.
This work is supported by the ``Departments of Excellence 2018 - 2022" Grant awarded by the Italian Ministry of Education, University and Research (MIUR) (L. 232/2016). This research has made use of resources provided by the Compagnia di San Paolo for the grant awarded on the BLENV project (S1618\_L1\_MASF\_01) and by the Ministry of Education, Universities and Research for the grant MASF\_FFABR\_17\_01. This investigation is supported by the National Aeronautics and Space Administration (NASA) grants GO9-20083X and GO0-21110X. A.P. acknowledges financial support from the Consorzio Interuniversitario per la Fisica Spaziale (CIS) under the agreement related to the grant MASF\_CONTR\_FIN\_18\_02. F.M. acknowledges financial contribution from the agreement ASI-INAF n.2017-14-H.0. A.P. thanks Andrea Tramacere for fruitful discussions. The Pan-STARRS1 Surveys (PS1) have been made possible through contributions of the Institute for Astronomy, the University of Hawaii, the Pan-STARRS Project Office, the Max-Planck Society and its participating institutes, the Max Planck Institute for Astronomy, Heidelberg and the Max Planck Institute for Extraterrestrial Physics, Garching, The Johns Hopkins University, Durham University, the University of Edinburgh, Queen's University Belfast, the Harvard-Smithsonian Center for Astrophysics, the Las Cumbres Observatory Global Telescope Network Incorporated, the National Central University of Taiwan, the Space Telescope Science Institute, the National Aeronautics and Space Administration under Grant No. NNX08AR22G issued through the Planetary Science Division of the NASA Science Mission Directorate, the National Science Foundation under Grant No. AST-1238877, the University of Maryland, and Eotvos Lorand University (ELTE). This research has made use of data obtained from the \textit{Chandra} Data Archive. This research has made use of software provided by the \textit{Chandra} X-ray Center (CXC) in the application packages CIAO, ChIPS, and Sherpa. This research has made use of the TOPCAT software \citep{2005ASPC..347...29T}.
\end{acknowledgements}

\newpage

\begin{table}
\caption{Results for the beta model fits to the surface flux profiles discussed in Sect. \ref{sec:imaging} and presented in Fig. \ref{fig:sb_cones}. For each extraction region the best-fit parameters are reported, with \(1-\sigma\) uncertainties indicated in parenthesis.}\label{tab:beta_fit}
\begin{center}
\begin{tabular}{|l|ccc|}
\hline
\hline
REGION  & \multicolumn{3}{c|}{beta model} \\
\hline
 & \(S_0\) & \(\beta\) & \(r_C\) \\
 & \({10}^{-8}\text{ ph}\text{ cm}^{-2}\text{ s}^{-1}  \text{ arcsec}^{-2}\) & & kpc \\
\hline
W CONE & \(1.22(0.06)\) & \(0.83(0.36)\) & \(115(46)\) \\
\hline
E CONE & \(1.15(0.01)\) & \(0.73(0.03)\) & \(98(4)\) \\
\hline
W+E CONES & \(0.77(0.05)\) & \(0.85(0.37)\) & \(142(55)\) \\
\hline
N CONE & \(0.79(0.66)\) & \(0.69(0.30)\) & \(185(172)\) \\
\hline
S CONE & \(0.58(0.01)\) & \(0.84(0.02)\) & \(292(9)\) \\
\hline
\hline
\end{tabular}
\end{center}
\end{table}

\begin{table}
\caption{Results for the spectral fits discussed in Sect. \ref{sec:spectra} with the background spectra subtraction. For the extraction regions shown in Fig. \ref{fig:spectral_regions} the \(0.3-7 \text{ keV}\) full band net counts are reported with errors in parenthesis. For the power-law model, the spectral index \(\Gamma\) and the normalization are listed, together with the intrinsic (i.e., unabsorbed) \(0.5-2\text{ keV}\) luminosity inferred from this model, and the reduced \(\chi^2\) (with degrees of freedom reported in parenthesis). For the thermal model, we report the temperature \(kT\) and the normalization, together with the intrinsic \(0.5-2\text{ keV}\) luminosity inferred from this model, the reduced \(\chi^2\), and the degrees of freedom (d.o.f.). For the mixed power-law+thermal model we list the spectral index and the normalization of the power-law component, the temperature and the normalization of thermal component, the intrinsic \(0.5-2\text{ keV}\) luminosity inferred from the total model, and the reduced \(\chi^2\). Quantities marked with an asterisk are fixed during the fit to the indicated value.}\label{tab:spectra}
\begin{center}
\resizebox{\textwidth}{!}{
\begin{tabular}{|l|c|cccc|cccc||cccccc|}
\hline
\hline
REGION & NET COUNTS (0.3-7) & \multicolumn{4}{c|}{POWER-LAW}
 & \multicolumn{4}{c|}{APEC} & \multicolumn{6}{c|}{POWER-LAW+APEC} \\
\hline
 & & \(\Gamma\) & \(\text{Norm}\) & \(L_{0.5-2}\) & \(\chi^2(\text{d.o.f.})\) & \(\text{kT}\) & \(\text{Norm}\) & \(L_{0.5-2}\) & \(\chi^2(\text{d.o.f.})\) & \(\Gamma\) & \(\text{Norm}\) & \(\text{kT}\) & \(\text{Norm}\) & \(L_{0.5-2}\) & \(\chi^2(\text{d.o.f.})\)\\
  & & & \({10}^{-5}\text{ cm}^{-2}\text{ s}^{-1}\) & \({10}^{43}\text{ erg}/\text{s}\) & & \(\text{keV}\) & \({10}^{-5}\text{ cm}^{-5}\) & \({10}^{43}\text{ erg}/\text{s}\) & & & \({10}^{-5}\text{ cm}^{-2}\text{ s}^{-1}\) & \(\text{keV}\) & \({10}^{-5}\text{ cm}^{-5}\) & \({10}^{43}\text{ erg}/\text{s}\) & \\
\hline
CROSS CONES & 125(15) &\({1.94}_{-0.28}^{+0.30}\) & \({1.95}_{-0.29}^{+0.29}\) & \({3.59}_{-0.55}^{+0.53}\) & 0.50(7) & \({3.92}_{-1.21}^{+2.59}\) & \({11.16}_{-1.46}^{+1.46}\) & \({3.57}_{-0.50}^{+0.52}\) & 0.41(7) & \(0.90^*\) & \(0.16^*\) & \({3.25}_{-0.95}^{+1.93}\) & \({9.90}_{-1.48}^{+1.48}\) & \({3.57}_{-0.53}^{+0.59}\) & 0.41(7) \\ 
\hline
N LOBE & 114(16) & \({2.29}_{-0.32}^{+0.34}\) & \({2.07}_{-0.32}^{+0.34}\) & \(3.88_{-0.60}^{+0.54}\) & 0.28(8) & \({2.70}_{-0.65}^{+1.13}\) & \({10.70}_{-1.61}^{+1.56}\) & \({3.70}_{-0.57}^{+0.72}\) & 0.31(8) & \({2.38}_{-0.40}^{+0.45}\) & \({1.66}_{-0.31}^{+0.31}\) & \(3.92^*\) & \(2.30^*\) & \({3.83}_{-0.58}^{+0.64}\) & 0.27(8) \\ 
\hline
S LOBE & 152(18) &\({1.77}_{-0.30}^{+0.34}\) & \({1.93}_{-0.32}^{+0.32}\) & \({3.58}_{-0.56}^{+0.60}\) & {1.04(11)} & \({5.38}_{-1.91}^{+20.59}\) & \({11.13}_{-1.55}^{+1.54}\) & \({3.39}_{-0.45}^{+0.56}\) & 1.16(11) & \({1.77}_{-0.30}^{+0.34}\) & \({1.93}_{-0.32}^{+0.32}\) & \(3.92^*\) & \(0^*\) & \({3.58}_{-0.56}^{+0.60}\) & 1.04(11) \\ 
\hline
\hline
\end{tabular}
}
\end{center}
\end{table}

\begin{table}
	\caption{Same as Table \ref{tab:spectra}, but for spectral fits with the background spectra modeling. At variance with Table \ref{tab:spectra}, for the extraction regions shown in Fig. \ref{fig:spectral_regions} the \(0.3-7 \text{ keV}\) full band total (source+background) counts are reported. In addition, instead of the reduced \(\chi^2\), reduced cash statistic (cstat) is reported.}\label{tab:spectra_bkg}
	\begin{center}
		\resizebox{\textwidth}{!}{
			\begin{tabular}{|l|c|cccc|cccc||cccccc|}
				\hline
				\hline
				REGION & COUNTS (0.3-7) & \multicolumn{4}{c|}{POWER-LAW}
				& \multicolumn{4}{c|}{APEC} & \multicolumn{6}{c|}{POWER-LAW+APEC} \\
				\hline
				& & \(\Gamma\) & \(\text{Norm}\) & \(L_{0.5-2}\) & cstat \((\text{d.o.f.})\) & \(\text{kT}\) & \(\text{Norm}\) & \(L_{0.5-2}\) & cstat \((\text{d.o.f.})\) & \(\Gamma\) & \(\text{Norm}\) & \(\text{kT}\) & \(\text{Norm}\) & \(L_{0.5-2}\) & cstat \((\text{d.o.f.})\)\\
				& & & \({10}^{-5}\text{ cm}^{-2}\text{ s}^{-1}\) & \({10}^{43}\text{ erg}/\text{s}\) & & \(\text{keV}\) & \({10}^{-5}\text{ cm}^{-5}\) & \({10}^{43}\text{ erg}/\text{s}\) & & & \({10}^{-5}\text{ cm}^{-2}\text{ s}^{-1}\) & \(\text{keV}\) & \({10}^{-5}\text{ cm}^{-5}\) & \({10}^{43}\text{ erg}/\text{s}\) & \\
				\hline
				CROSS CONES & 230(15) &\({1.96}_{-0.20}^{+0.21}\) & \({1.92}_{-0.22}^{+0.24}\) & \({3.56}_{-0.44}^{+0.40}\) & 0.97(414) & \({3.26}_{-0.63}^{+0.87}\) & \({10.80}_{-1.16}^{+1.21}\) & \({3.55}_{-0.34}^{+0.39}\) & 0.97(414) & \(1.57^*\) & \(0.51^*\) & \({2.73}_{-0.55}^{+0.84}\) & \({7.60}_{-1.13}^{+1.19}\) & \({3.60}_{-0.46}^{+0.56}\) & 0.97(414) \\
				\hline
				N LOBE & 244(16) & \({2.13}_{-0.26}^{+0.27}\) & \({1.86}_{-0.24}^{+0.25}\) & \(3.42_{-0.42}^{+0.46}\) & 0.98(435) & \({2.91}_{-0.52}^{+0.93}\) & \({10.07}_{-1.23}^{+1.29}\) & \({3.42}_{-0.46}^{+0.43}\) & 0.97(435) & \({2.13}_{-0.26}^{+0.27}\) & \({1.86}_{-0.24}^{+0.25}\) & \(3.26^*\) & \(0^*\) & \(3.42_{-0.42}^{+0.46}\) & 0.98(435) \\ 
				\hline
				S LOBE & 288(17) & \({1.83}_{-0.23}^{+0.25}\) & \({2.05}_{-0.26}^{+0.27}\) & \({3.78}_{-0.50}^{+0.51}\) & 1.04(481) & \({10.09}_{-5.05}^{+18.64}\) & \({11.86}_{-1.40}^{+2.25}\) & \({3.34}_{-0.40}^{+0.45}\) & 1.05(481) & \({1.83}_{-0.23}^{+0.25}\) & \({2.05}_{-0.26}^{+0.27}\) & \(3.26^*\) & \(0^*\) & \({3.78}_{-0.50}^{+0.51}\) & 1.04(481) \\ 
				\hline
				\hline
			\end{tabular}
		}
	\end{center}
\end{table}

\begin{table}
\caption{Parameters used for the simulation shown in Fig. \ref{fig:seds} to reproduce radio lobe SEDs (see Sect. \ref{sec:seds}), namely the slope of the electron energy distribution \(p\), the electron density \(n_e\), the magnetic field \(B\), and the emitting region radius \(R\) (fixed during the fits). In addition the total electron rest-frame energy (\(E_e\)) and the total radiative luminosity (\(L_{\text{rad}}\)) are listed.}\label{tab:seds}
\begin{center}
\begin{tabular}{|l|cccc|cc|}
\hline
\hline
REGION & \multicolumn{4}{c|}{Model parameters}
 & \multicolumn{2}{c|}{Energetics} \\
\hline
 & \(p\) & \(n_e\) & \(B\) & \(R\) & \(E_e\) & \(L_{\text{rad}}\) \\
 & & \({10}^{-6}\text{ cm}^{-3}\) & \(\upmu\text{G}\) & \(\text {kpc}\) & \({10}^{61}\text{ erg}\) & \({10}^{43}\text{ erg} / \text{s}\) \\
\hline
N LOBE & \(3.08(0.04)\) & \(3.45(0.62)\) & \(3.86(0.24)\) & \(240^*\) & \(9.3\) & \(14.7\) \\
S LOBE & \(3.05(0.06)\) & \(2.84(0.82)\) & \(4.21(0.42)\) & \(247^*\) & \(8.4\) & \(14.8\) \\
\hline
\hline
\hline
\end{tabular}
\end{center}
\end{table}

\end{document}